\documentclass[a4paper,fleqn]{cas-sc}
\ExplSyntaxOn
\cs_set:Npn \__first_footerline: {}
\ExplSyntaxOff

\usepackage[numbers]{natbib}

\usepackage{threeparttable}
\usepackage{subcaption}
\usepackage{float}
\usepackage{placeins}
\usepackage[most]{tcolorbox}

\newtheorem{myTheo}{Theorem}
\newtheorem{myDef}{Definition}
\newtheorem{myLem}{Lemma}
\newdefinition{myRem}{Remark}
\newdefinition{myAss}{Assumption}

\newproof{myProof}{Proof}
\newproof{myPot}{Proof of Theorem \ref{thm}}

\usepackage{hyperref}
\begin{document}
\let\WriteBookmarks\relax
\def\floatpagepagefraction{1}
\def\textpagefraction{.001}

\shorttitle{}    

\shortauthors{}  

\title [mode = title]{Action-Increment-Constrained Policy Optimization for Smooth Online Adaptive Critic Flight Control with Actuator Rate Constraints and Low
Computational Complexity}  



%

\author[1]{Yifei Li}

\cormark[1]

\ead{y.li-34@tudelft.nl}


\credit{Conceptualization, Methodology, Software, Validation, Formal analysis, Investigation, Writing – original draft, Visualization, Data curation}

\affiliation[1]{organization={Section Control and Simulation, Department of Control and Operation, Delft University of Technology},
            addressline={Kluyverweg 1}, 
            city={Delft},
          citysep={}, 
            postcode={2629HS}, 
            state={South Holland},
            country={the Netherlands}}

\author[1]{Erik-Jan van Kampen}


\ead{e.vankampen@tudelft.nl}


\credit{Supervision, Methodology}


\cortext[1]{Corresponding author}
\cortext[2]{This version extends the preliminary manuscript by providing a uniform ultimate boundedness (UUB) convergence analysis of the proposed TS-IHDP with constant $\lambda$.}


\begin{abstract}
Approximate Dynamic Programming (ADP) enables online control-law adaptation through adaptive critic methods. A safety challenge in applying ADP to angle-of-attack (AoA) tracking control is that an overly sensitive neural network (NN) controller may generate oscillatory or saturated control actions under nonlinear dynamics, potentially compromising flight stability and increasing actuator stress. To address this challenge, this paper proposes two innovations: (1) the actuator rate constraints are enforced within an Action-Increment-Constrained Policy Optimization (AI-CPO) framework inspired by safe reinforcement learning (RL), in which a primal–dual approach adaptively adjusts the penalty weight. This optimization is solved online using adaptive Temporally Smoothed Incremental Model-Based Heuristic Dynamic Programming (TS-IHDP) for smooth and computationally efficient online adaptive control; and (2) a low-pass filter integrated into the cascaded NN architecture to attenuate high-frequency pitch-rate commands while preserving the Markov property during online learning by exploiting the hierarchical actor structure. Compared with the IHDP baseline, simulations using a longitudinal air vehicle dynamics model demonstrate that TS-IHDP improves control smoothness and tracking performance, reduces system oscillations, and enhances the trained policy's robustness to model uncertainties. 
\end{abstract}





\maketitle

\section{Introduction}
\subsection{Incremental model-based ADP for online adaptive control}

Nonlinear optimal control problems have been extensively studied using Approximate Dynamic Programming (ADP), a branch of Reinforcement Learning (RL) that aims to approximately solve the well-known Hamilton–Jacobi–Bellman (HJB) equation \cite{Lewis20090903IEEEMagazine}. Within the ADP framework, neural networks (NNs) are commonly employed to approximate the value function, policy function, and system dynamics, leading to the so-called actor–critic–identifier architecture \cite{Werbos1977advanced,Prokhorov1997TINN08}. 

In ADP-based online adaptive control \cite{Vamvoudakis2010878,Bhasin201382}, incremental models have been adopted to provide local linear approximations of nonlinear system dynamics while reducing the computational cost and complexity of online system identification, thereby improving sample efficiency. To obtain this model, a first-order Taylor expansion is used to locally linearize nonlinear dynamics \cite{Khalil2002NonlinearSystems}. The resulting incremental model can then be identified online from data using an adaptive recursive least squares (RLS) algorithm \cite{Niederlinski1995TAC}. Examples include incremental model-based heuristic dynamic programming (IHDP) \cite{ZHOUAST2020105}, incremental model-based dual heuristic programming (IDHP) \cite{ZHOUCEP201813}, and incremental model-based global dual heuristic programming (IGDHP) \cite{SunJGCD2021}. The combination of incremental modeling and RLS identification offers several advantages for online adaptive critic control: (1) learning a local linear model can be more sample-efficient and computationally efficient than learning a NN model, as in conventional HDP, DHP, and GDHP. Their performance has been compared in \cite{ZHOUCEP201813,ZhouIMAV2016,Sun2020103425}; (2) the control effectiveness can be obtained directly from the incremental model, which can be identified without explicit knowledge of the system's internal dynamics, as in \cite{Lin201738}, thereby facilitating a more model-free implementation of ADP; (3) the local linear approximation can be improved through high-frequency sampling when the system dynamics vary slowly; and (4) a representative approach, IHDP \cite{ZHOUAST2020105}, constructs the policy objective by using the incremental model and $V$-function network, rather than the more complex state-action ($Q$-) function network with additional action inputs, as used in Action-Dependent HDP (ADHDP) \cite{Sokolov20159}, thereby reducing learning complexity, as demonstrated in autolanding control \cite{Prokhorov1997TINN08} and flight control \cite{KampenAIAA2006}.

\subsection{Flight control applications and control smoothness challenges}
For high-performance air vehicles, angle-of-attack (AoA) tracking control is of particular interest because AoA strongly influences aerodynamic lift and drag, thereby affecting maneuverability and turning performance, and is an important objective in longitudinal flight control design \cite{Blakelock1991AC}. Compared to the traditional backstepping approach \cite{YuAST2020106244,SinghAST2022107653}, ADP enables a more data-driven adaptive control design with reduced reliance on an accurate system model. Applications have been demonstrated using DHP \cite{Ferrari20004JGCD}, HDP, ADHDP \cite{KampenAIAA2006}, and incremental model-based ADP methods  \cite{ZHOUAST2020105,ZHOUCEP201813, SunJGCD2021}. Commonly, a cascaded control structure is adopted to achieve separate AoA and pitch-rate tracking control in the outer-loop and inner-loop subsystems, enabling a \textit{hierarchical} control strategy. The benefit of the cascaded control structure lies in the efficient and accurate pitch-rate tracking control by explicitly incorporating pitch-rate reference \cite{ZHOUCEP201813,KampenAIAA2006}. Although pitch-rate penalization can be incorporated into a monolithic NN control structure, as commonly done in offline training with black-box aircraft system models \cite{Homola2025AIAA}, it does not directly regulate the temporal smoothness of pitch-rate reference, potentially leading to control oscillations and degraded tracking performance. Moreover, the cascaded NN architecture enhances the structural interpretability of the learned control policy by distinguishing the roles of the inner and outer control loops.


The safety of NN-based flight control systems can be compromised by overly sensitive NN controllers, which may generate aggressive control actions in response to time-varying dynamics and model uncertainties, causing system oscillations. For tracking control tasks, the NN controller may generate frequently switching actions when the tracking error oscillates near zero, as seen in online adaptive flight control using IGDHP for F-16 nonlinear model \cite{SunJGCD2021}, offline-trained quadrotor flight control using proximal policy optimization (PPO) \cite{Mysore2021ICRA}, and offline-trained spacecraft proximity maneuvers and docking policies using Bayesian actor-critic algorithm (DBAC) \cite{Du2024AST109474}. To improve control smoothness and limit policy behavior, regularizing policy smoothness is necessary in training these NN controllers. This arises the first research question: 

\begin{tcolorbox}[
    colback=gray!5,
    colframe=black!60,
    title=Research Question 1,
    fonttitle=\bfseries,
    arc=1mm,
    boxrule=0.5pt
]
How can control smoothness be regularized in online adaptive critic flight control with low computation cost and implementation complexity?
\end{tcolorbox}



\subsection{Solutions to improving control smoothness}
\textit{Lipschitz constant regularization.}
The Lipschitz constant has been considered as a smoothness measure of a NN controller. In \cite{KobayashiIROS2022}, the local Lipschitz continuity constraint is considered for training a policy network to promote local policy smoothness, a more efficient way compared to regularizing global Lipschitz constant. In \cite{Song2023PMLR}, the Lipschitz constant is estimated and regularized as a part of the policy optimization objective for vehicle trajectory tracking. However, estimating the Lipschitz constant incurs additional computational cost for online learning and does not directly quantify temporal variations in control actions in a physically interpretable manner, motivating our use of \textit{action variations} as a more practical measure of control smoothness in flight control, given their direct relevance to actuator rate limits \cite{DudaAST1998265,MoghimiAST2021}. 

\textit{Conditioning for action policy smoothness (CAPS).}
Many studies have adopted CAPS to regularize control smoothness in offline reinforcement learning (RL)-based flight control, considering both temporal and spatial variations in control actions, as demonstrated in quadrotor control \cite{Mysore2021ICRA}, fixed-wing aircraft \cite{Li2026AST112516}, and Cessna Citation business jets \cite{Homola2025AIAA,Gavra202447JGCD}. Despite clear improvements in control smoothness and reductions in actuator load, CAPS requires sufficiently diverse samples to construct spatial and temporal smoothness losses. In particular, computing the spatial smoothness loss at each state requires generating a batch of spatially perturbed state samples from a prior distribution, increasing computational costs and training time during online policy learning. These overheads motivate excluding spatial policy smoothness regularization from the online learning framework. In contrast, temporal smoothness offers three advantages: (1) it can be computed efficiently using only individual training samples in model-free (action-dependent) ADP, such as ADHDP, or one-step state predictions in model-based ADP, such as IHDP, during online learning (2) unlike the Lipschitz constant, it provides an intuitive, physically meaningful measure of actuator rate, an important consideration in flight-control design with actuator rate constraints \cite{DudaAST1998265,MoghimiAST2021}; and (3) it can be used to regularize the local temporal control smoothness by using local training samples, rather than enforcing global smoothness \cite{KobayashiIROS2022}. Unfortunately, a series of studies on online adaptve critic flight control has not yet investigated temporal policy smoothness regularization as a means of mitigating control performance degradation and actuator overload caused by unsmooth policies \cite{ZHOUAST2020105,SunJGCD2021,KampenAIAA2006,Ferrari20004JGCD}.

A longstanding challenge in policy smoothness regularization methods, such as CAPS, is the selection of an appropriate smoothness penalty weight. \cite{Mysore2021ICRA,KobayashiIROS2022} adopted constant weights, but finding suitable values requires considerable effort from the practitioner. Intuitively, the weight controls a trade-off between control smoothness and control accuracy. This raises the second research question: 

\begin{tcolorbox}[
    colback=gray!5,
    colframe=black!60,
    title=Research Question 2,
    fonttitle=\bfseries,
    arc=1mm,
    boxrule=0.5pt
]
Can the penalty weight for temporal control smoothness be adaptively adjusted according to observed policy behavior,  as quantified by action increments?
\end{tcolorbox}

To this end, we consider control rate constraints in a Constrained Policy Optimization problem \cite{PaternainTAC2023}, which leads to action-increment-constrained policy optimization (AI-CPO). Previously, CPO promotes safety for cost-constrained Markov decision processes (MDPs) using a primal-dual method that adaptively adjusts the penalty weight based on constraint violations to keep the expected cumulative constraint cost below a prescribed threshold \cite{PaternainTAC2023,Hong2024PMLR}. Despite the rigorous theoretical results established for infinite-horizon optimal control \cite{PaternainTAC2023,Hong2024PMLR} and for finite-horizon optimal control \cite{Wei2025Auto112029}, the primal–dual method has not yet been applied to address the practical engineering control problem of temporal policy smoothness regularization, which constitutes the focus of this paper.

\textit{Low-pass filter.}
The low-pass filter is another approach to improve control smoothness for cascaded flight control systems, as demonstrated in command-filtered backstepping \cite{LunguAST2020105526,ZhangAST2024109552}. In the frequency-domain sense, high-frequency components in the control signal, such as those induced by sensor noise, can be effectively attenuated, at the cost of phase delay and signal biasing. However, low-pass filter alone does not explicitly regulate temporal variations of the control signal, thereby does not address the critical oscillations induced by the unregularized policy network during online learning. This motivates the study of a hybrid approach that combines a low-pass filter with temporal smoothness regularization to enable smooth online learning-based flight control, which has not yet been investigated. \cite{Mysore2021ICRA} has shown that filtering a post-training policy may cause nonvanishing oscillations. This brings up the third research question:


\begin{tcolorbox}[
    colback=gray!5,
    colframe=black!60,
    title=Research Question 3,
    fonttitle=\bfseries,
    arc=1mm,
    boxrule=0.5pt
]
How can a low-pass filter improve control smoothness of a cascaded NN actor during online learning while preserving the Markov property of the training samples?
\end{tcolorbox}

\subsection{Contributions}

The main contributions are summarized as follows.

\begin{itemize}
    \item Temporally Smoothed Incremental Model-based Heuristic Dynamic Programming (TS-IHDP) is developed for smooth online learning-based flight control. This method regularizes local temporal control smoothness with low computational overhead and readily available training samples. Compared with IHDP \cite{ZHOUAST2020105}, it shows clear improvements on system stability and control accuracy for AoA tracking control.
    \item The primal–dual method is applied for the first time to adaptively adjust the penalty weight associated with the local temporal smoothness loss in the AI-CPO formulation. The resulting optimization problem is solved online using TS-IHDP. By deriving the action-increment constraint from actuator or command-rate limits, the proposed approach reduces the need for manual penalty-weight tuning.
    
    \item 
    Unlike post-training filtering, which can degrade control performance \cite{Mysore2021ICRA}, the proposed method incorporates a low-pass filter directly into online learning. Moreover, the hierarchical control structure avoids the need to augment the NNs' inputs with filter states, as may be necessary to preserve the Markov property when training a monolithic actor \cite{WangICLR2024}.
    
\end{itemize}

The remainder of this paper is organized as follows. Section \ref{section_preliminaries} introduces the incremental model and the adaptive RLS identification algorithm. Section \ref{section_TS_IHDP} illustrates the TS-IHDP method, with weights updated using gradient descent. Section \ref{section_primal_dual_optimization} formulates the AI-CPO framework and presents a primal–dual method for updating the penalty weight. Section \ref{section_online_learning_flight_control_design} designs the cascaded online learning flight control system. Section \ref{section_simulation} presents the simulation results. Section \ref{section_conclusion} concludes this paper.

\section{Incremental model and system identification}\label{section_preliminaries}



\subsection{Incremental model}

Consider a continuous-time nonlinear system:

\begin{equation}\label{continuous_system}
    \dot{x}(t) = f(x(t),u(t))
\end{equation}
where $f: \mathbb{R}^{n}\times \mathbb{R}^{m} \rightarrow \mathbb{R}^{n}$ is a smooth nonlinear function, with $x \in \mathcal{X} \subseteq \mathbb{R}^{n}$ and $u \in \mathcal{U} \subseteq \mathbb{R}^{m}$. The sets $\mathcal{X}$ and $\mathcal{U}$ denote the state and input spaces, with $n,m \in \mathbb{N}_{+}$ being their respective dimensions. $\mathbb{N}_{+}$ denotes the set of positive integers, i.e., $\mathbb{N}_{+} = \{1,2,3,\dots\}$.

Taking the first-order Taylor expansion of \eqref{continuous_system} around an arbitrary time instant $t_0$:

\small
\begin{equation}\label{Taylor_expansion}
\begin{aligned}
&f(x(t),u(t))
= f(x(t_{0}),u(t_{0})) 
+ F(t_{0})\big(x(t)-x(t_{0})\big) \\
&+ G(t_{0})\big(u(t)-u(t_{0})\big)
+ \mathcal{O}\!\left(\|x(t)-x(t_{0})\|^{2} + \|u(t)-u(t_{0})\|^{2}\right)
\end{aligned}
\end{equation}
where $F(t_{0})$ and $G(t_{0})$ denote the Jacobian matrices of $f(x,u)$ with respect to $x$ and $u$, respectively, evaluated at $(x(t_{0}),u(t_{0}))$, i.e.,
\begin{equation}
\begin{aligned}
&F(t_{0}) \triangleq \left.\frac{\partial f(x,u)}{\partial x}\right|_{(x,u)=(x(t_{0}),u(t_{0}))} \in \mathbb{R}^{n \times n},\\
&G(t_{0}) \triangleq \left.\frac{\partial f(x,u)}{\partial u}\right|_{(x,u)=(x(t_{0}),u(t_{0}))} \in \mathbb{R}^{n \times m}.        
\end{aligned}
\end{equation}
\normalsize
The term $\mathcal{O}(\cdot)$ represents second and higher-order terms. $\Vert \cdot \Vert$ is the Euclidean norm. Since $\mathcal{O}(\cdot)$  has a relatively small influence on the transition from the state $x(t)$ to $x(t+\Delta t)$ compared with the first-order terms under a high-frequency sampling strategy, and these higher-order derivatives are difficult to identify from data, incorporating them into a simplified system model for online identification is impractical. Therefore, these higher-order terms are neglected, yielding a local linear approximation of nonlinear dynamics:

\begin{equation}\label{Taylor_expamsion_without_higher_order_terms}
\begin{aligned}
    f(x(t),u(t))&\approx  f(x(t_{0}),u(t_{0}))
    + F(t_{0})(x(t)-x(t_{0}))\\
    &+ G(t_{0})(u(t)-u(t_{0}))
\end{aligned}
\end{equation}

Use Equation \eqref{continuous_system}:

\begin{equation}\label{taylor_expansion_x}
\dot{x}(t) \approx \dot{x}(t_{0}) + F(t_{0})\left(x(t)-x(t_{0})\right)+ G(t_{0})\left(u(t)-u(t_{0})\right)
\end{equation}

The time derivatives $\dot{x}(t),\dot{x}(t_{0})$ are approximated using a forward Euler scheme: $\dot{x}(t)\approx\frac{x(t+ T)-x(t)}{T}$ and $\dot{x}(t_{0})\approx\frac{x(t_{0}+ T)-x(t_{0})}{T}$, where $T$ denotes a constant sampling period. Then one has


\begin{equation}\label{discretized_model}
\begin{aligned}
   \frac{x(t+ T)-x(t)}{T} \approx& \frac{x(t_{0}+ T)-x(t_{0})}{T} + F(t_{0})\left(x(t)-x(t_{0})\right)\\
   &+ G(t_{0})\left(u(t)-u(t_{0})\right) 
\end{aligned}
\end{equation}

A discretization is introduced with $T$ by setting $t_{0}=(k-1)T$, where $x(t_{0}) = x((k-1)T) \triangleq x(k-1)$. The index $k \in \mathbb{N}_{+}$ denotes the discrete-time step. Let $x(t)$ denote the state at the next sampling instant following $x(t_{0})$, i.e., $x(t) = x(t_{0}+T) \triangleq x(k)$. Then, a discrete-time representation of the system \eqref{continuous_system} is given by





\begin{equation}\label{incremental_model}
\Delta x(k+1) \approx \Delta x(k) +  T\Big( F(k-1)\Delta x(k)+ G(k-1)\Delta u(k)\Big)
\end{equation}\\
where $\Delta x(k)=x(k)-x(k-1)$, $\Delta u(k)=u(k)-u(k-1)$. 

A common practice in controller design is to develop controllers based on simplified models and subsequently validate their performance, including robustness, under more complex operating conditions. Accordingly, we need to introduce several commonly used assumptions to simplify the control problem for the convenience of NN-based control design and analysis. To deploy the control methods in practical applications, these assumptions must be carefully assessed. A detailed verification on these assumptions are not the main scope of this paper.

\begin{myAss}\label{assumption_sampling}
    The nonlinear system \eqref{continuous_system} is locally controllable over each interval $[k-1,k]$, $\forall k\in\mathbb{N}_{+}$. Consequently, the local linear dynamics, denoted by $(F(k-1),G(k-1))$, preserve controllability \cite{SunJGCD2021}.
\end{myAss}


\begin{myAss} \label{Assumption_high_frequency_sampling}
The sampling period $T$ is sufficiently short, and the system dynamics $f(\cdot)$ are slow-varying over the interval $[k-1,k]$, which make the local linear dynamics $F(k-1)$ and $G(k-1)$ remain approximately constant over $[k-1,k]$. This assumption can be understood from the Taylor expansion \eqref{Taylor_expansion}.
\end{myAss}


\begin{myAss}\label{Assumption_persistent_excitation}
The control input satisfies the persistent excitation (PE) condition for critic and actor network training, as well as for incremental model identification \cite{Vamvoudakis2010878}.
\end{myAss}

\begin{myAss}\label{Assumption_noise_free_measurements}
The measurements of the full system states are available and these measurements are noise-free.
\end{myAss}

\begin{myRem}
Assumption \ref{assumption_sampling} is made to enable effective controller design for the local system dynamics. Assumption \ref{Assumption_high_frequency_sampling} is introduced to minimize the effects of discretization on controller performance. Assumption \ref{Assumption_persistent_excitation} is assumed to be satisfied for our simulated flight control system by using a time-varying reference and a probe noise added to the control input (Section \ref{section_simulation}). Assumption \ref{Assumption_noise_free_measurements} is made to promote accurate estimation of the system states through system identification and ease the analysis of critic and actor networks behaviors. Practical applications may consider sensor noise, measurement delays \cite{KleinMorelli2006}, and partial observability of the states \cite{ZHOUAST2020105}, which degrade the state estimation. Therefore, Assumption \ref{Assumption_noise_free_measurements} is a ideal assumption made for a fully observable nonlinear dynamics model that does not account for these effects. The benefit lies in that Assumption \ref{Assumption_noise_free_measurements} simplifies the analysis and numerical verification of the performance of ADP methods, in line with theoretical studies in \cite{Bhasin201382,Prokhorov1997TINN08,Lin201738}.
\end{myRem}

\subsection{Adaptive RLS identification} \label{subsection_RLS}

Define the augmented state as $X(k) = [\Delta x(k), \Delta u(k)]^{T}\in\mathbb{R}^{n+m}$ and the augmented incremental model parameter as $\hat{\theta}(k)=[\hat{F}(k-1),\hat{G}(k-1)]^{T}\in\mathbb{R}^{(n+m)\times n}$ in the incremental model $\Delta \hat{x}(k+1) \approx \Delta x(k) +  \hat{F}(k-1)\Delta x(k)+ \hat{G}(k-1)\Delta u(k)$. Notably, $\hat{\theta}(k)$ is the construction of the local linear dynamics multiplied with $T$: $\theta^{*}(k)=[TF(k-1),TG(k-1)]^{T}$. The shift of the index $k$ from $\theta^{*}(k)$ to $[\hat{F}(k-1),\hat{G}(k-1)]$ is from the Taylor expansion \eqref{incremental_model}. We consider the parameter update at $k$, when the state $X(k)$ is obtained. The predicted state is $\Delta \hat{X}(k)^{T} = X(k-1)^{T} \hat{\theta}(k-1)$. The error between the actual state and predicted state is $\epsilon(k) = \Delta X(k)^{T} - \Delta \hat{X}(k)^{T}$. The parameter of the incremental model is updated by \cite{Niederlinski1995TAC}


\begin{equation}\label{equation.8}
     \hat{\theta}(k) = \hat{\theta}(k-1) + P(k)X(k-1)\epsilon(k)
\end{equation}

\begin{equation}\label{P_inversion_update}
\begin{aligned}
    P^{-1}(k) &= \rho P^{-1}(k-1) + X(k-1)X(k-1)^{T}\\
\end{aligned}
\end{equation}

The initial auxiliary matrix is set as $P(0)=I/\lambda_{\text{min}}(0)$, where $0<\lambda_{\text{min}}(0)\ll1$ is the minimum eigenvalue of $P^{-1}(0)$ \cite{Niederlinski1995TAC}. The forgetting factor is $0\ll\rho<1$. $P(k)(k=1,2,\cdots)$ is the covariance matrix, which quantifies the confidence in the  parameter estimate at $k$. It is updated recursively to reflect the amount of information available from incoming data, and its evolution determines the adaptive gain of the algorithm, thereby influencing the effective step size of parameter updates.

\begin{myRem}
    During adaptive RLS identification, the estimated control effectiveness $\hat{G}(k-1)$ is initialized as a zero vector and may have signs opposite to those of actual values $G(k-1)$ in the initial identification stage. These sign mismatches can lead to an opposite direction of the policy network update \cite{Tamimi2008TSMC38} (see policy gradient equations \eqref{equation_action_e1_w2_update},\eqref{equation_action_e1_w1_update}). As a result, the learned policy may produce ineffective control when applied to the actual system. Therefore, we make the following assumption:
\end{myRem}

\begin{myAss}
    Adaptive RLS identification \eqref{equation.8}, \eqref{P_inversion_update} eventually makes $\hat{G}(k-1)$ having signs consistent with those of $G(k-1)$ after a period time of online identification process. This assumption ensures the converged policy performs effectively in the actual system, and will be verified via simulation in Section \ref{section_online_identification}.
\end{myAss}

\subsection{One-step convergence analysis of the identifier}

We consider one-step update of $\hat{\theta}(k-1)$ and its convergence to the local linear system $\theta^{*}(k)$ under the influence of system nonlinearity, which is not considered in \cite{Niederlinski1995TAC}. To this end, we introduce the following lemmas.


\begin{myLem}[\cite{Niederlinski1995TAC}]
    Under Assumption \ref{Assumption_noise_free_measurements}, the upper bound of the identification error for the incremental model is given by
    \begin{equation}
        \Vert \tilde{\theta}(k)\Vert_{2}\leq\rho\frac{\lambda_{\text{min}}(k-1)}{\lambda_{\text{min}}(k)}\Vert \tilde{\theta}(k-1)\Vert_{2}+\Vert\Delta\theta^{*}(k)\Vert_{2}
    \end{equation}\\
    where $\lambda_{\text{min}}(k)$ and $\lambda_{\text{min}}(k-1)$ are the minimum eigenvalues of $P^{-1}(k)$ and $P^{-1}(k-1)$. The estimation error is denoted as $\tilde{\theta}(k)=\hat{\theta}(k)-\theta^{*}(k)$. The variation of $\theta^{*}(k)$ is defined as $\Vert\Delta\theta^{*}(k)\Vert_{2}=\theta^{*}(k)-\theta^{*}(k-1)$. $\Vert\cdot\Vert$ is the Euclidean norm. 
\end{myLem}

\begin{myProof}
   From step $k-1$ to $k$, the parameter update is

    \begin{equation}\label{thetaerror}
    \begin{aligned}
    \tilde{\theta}(k)&=\hat{\theta}(k)-\theta^{*}(k)\\
    &=\hat{\theta}(k-1)+P(k)X(k)\epsilon(k) -\left[ \theta^{*}(k-1)+\Delta \theta^{*}(k)\right]\\
    &=\hat{\theta}(k-1) + P(k)X(k)\left[\Delta x(k+1)^{T}-\Delta \hat{x}(k+1)^{T}\right] -\theta^{*}(k-1) -\Delta \theta^{*}(k)\\
    &=\hat{\theta}(k-1) + P(k)X(k)\left[X(k)^{T} \theta^{*}(k-1)-X(k)^{T} \hat{\theta}(k-1)\right] -\theta^{*}(k-1)- \Delta \theta^{*}(k)\\
    &=\hat{\theta}(k-1)-\theta^{*}(k-1) + P(k)X(k)X(k)^{T} \left[\theta^{*}(k-1)-\hat{\theta}(k-1)\right] -\Delta \theta^{*}(k)\\
    &=\tilde{\theta}(k-1) - P(k)X(k)X(k)^{T} \tilde{\theta}(k-1)- \Delta \theta^{*}(k)\\
    &=\left(I- P(k)X(k)X(k)^{T}\right) \tilde{\theta}(k-1)- \Delta \theta^{*}(k)\\
    \end{aligned}
    \end{equation}


    
    Equation \eqref{P_inversion_update} indicates
    \begin{equation}
    \begin{aligned}
        P(k)P(k)^{-1} =& \rho P(k)P^{-1}(k-1) + P(k)X(k)X(k)^{T}\\
    \end{aligned}    
    \end{equation}
    
    which leads to
    \begin{equation} \label{replaceeq}
    \begin{aligned}
        I -P(k)X(k)X(k)^{T} =\rho P(k)P^{-1}(k-1)
    \end{aligned}    
    \end{equation}
    
    Substitute \eqref{replaceeq} into \eqref{thetaerror}
    
    \begin{equation} \label{theta_error_tobe_iterated}
    \begin{aligned}
       \tilde{\theta}(k)= \rho P(k)P^{-1}(k-1)\tilde{\theta}(k-1) - \Delta \theta^{*}(k)
    \end{aligned}
    \end{equation}


    Take the Euclidean norm on each side of  \eqref{theta_error_tobe_iterated}
    \begin{equation}\label{theta_error_norm}
    \begin{aligned}
        \Vert \tilde{\theta}(k) \Vert &= \Vert  \rho P(k)P^{-1}(k-1) \tilde{\theta}(k-1)-\Delta\theta^{*}(k)\Vert\\
        &\leq \Vert  \rho P(k)P^{-1}(k-1) \tilde{\theta}(k-1)\Vert+\Vert\Delta\theta^{*}(k)\Vert
    \end{aligned}
    \end{equation}

   By the submultiplicative property of the matrix 2-norm and the fact that $P(k)$ is symmetric positive definite, which indicates $\Vert P(k)\Vert=\lambda_{\text{max}}(P(k))=\frac{1}{\lambda_{\text{min}}(k)}$, one has
   
   \begin{equation}\label{lambda_rho}
   \begin{aligned}
     \Vert \rho P(k)P^{-1}(k-1) \tilde{\theta}(k-1) \Vert &\leq  \rho \Vert P(k)\Vert\Vert P^{-1}(k-1)\Vert \Vert \tilde{\theta}(k-1)\Vert\\
     &= \rho\frac{\lambda_{\text{min}}(k-1)}{\lambda_{\text{min}}(k)}\Vert \tilde{\theta}(k-1)\Vert
    \end{aligned}
    \end{equation}\\
which leads to
\begin{equation}
        \Vert \tilde{\theta}(k)\Vert\leq\rho\frac{\lambda_{\text{min}}(k-1)}{\lambda_{\text{min}}(k)}\Vert \tilde{\theta}(k-1)\Vert+\Vert\Delta\theta^{*}(k)\Vert
\end{equation}

\end{myProof}

    



\begin{myLem}\label{lemma_lamda_min}
   Following the parameter update law of $P^{-1}(k-1)$ in \eqref{P_inversion_update}, one has $\rho \lambda_{\text{min}}(k-1)\leq\lambda_{\text{min}}(k)$.
\end{myLem}

\begin{myProof}
    Using Weyl's inequality for \eqref{P_inversion_update}:
    \begin{equation}
        \rho \lambda_{\text{min}}(k-1)+\lambda_{\text{min}}(X(k)X(k)^{T})\leq\lambda_{\text{min}}(k)
    \end{equation}
    Obviously
    \begin{equation}
        \lambda_{\text{min}}(X(k)X(k)^{T})=0
    \end{equation}

    It follows that
    \begin{equation}\label{lambda_k-1_k}
        \rho \lambda_{\text{min}}(k-1)\leq\lambda_{\text{min}}(k)
    \end{equation}
\end{myProof}

\begin{myTheo}
    Following the update rules \eqref{equation.8} and \eqref{P_inversion_update} over the interval from  $k-1$ to $k$, the estimate error $\tilde{\theta}(k)$ decreases when
    \begin{equation}
    -\left(1-\rho\frac{\lambda_{\text{min}}(k-1)}{\lambda_{\text{min}}(k)}\right)\Vert\tilde{\theta}(k-1) \Vert+\Vert \Delta \theta^{*}(k)\Vert\leq 0
    \end{equation}
\end{myTheo}
\begin{myProof}
    Rewrite \eqref{theta_error_norm} as
    \begin{equation}\label{}
    \begin{aligned}
        \Vert \tilde{\theta}(k) \Vert &\leq \left\Vert \tilde{\theta}(k-1)\right\Vert-\left(1-\rho\frac{\lambda_{\text{min}}(k-1)}{\lambda_{\text{min}}(k)}\right)\left\Vert \tilde{\theta}(k-1)\right\Vert+\Vert\Delta\theta^{*}(k)\Vert
    \end{aligned}
    \end{equation}

    To enforce $ \Vert \tilde{\theta}(k) \Vert\leq  \Vert \tilde{\theta}(k-1) \Vert$, it has to be satisfied that
     \begin{equation}\label{decrease_condition}
        -\left(1-\rho\frac{\lambda_{\text{min}}(k-1)}{\lambda_{\text{min}}(k)}\right)\left\Vert \tilde{\theta}(k-1)\right\Vert+\Vert\Delta\theta^{*}(k)\Vert\leq0
    \end{equation}

\end{myProof}

\begin{myRem}
    By Lemma \ref{lemma_lamda_min}, the first term in \eqref{decrease_condition} is always non-positive, which means that the parameter update laws \eqref{equation.8} and \eqref{P_inversion_update} promote the reduction of the estimation error. The second term arises from the system nonlinearity and counteracts the decrease of the estimation error. This effect can be mitigated by increasing the sampling frequency, i.e. $\lim_{T\rightarrow0}\Vert\Delta\theta^{*}(k)\Vert=0$. 
\end{myRem}

\section{Temporally Smoothed Incremental Model-based Heuristic Dynamic Programming}\label{section_TS_IHDP}

\begin{figure}
    \centering
   \includegraphics[width=0.55\linewidth]{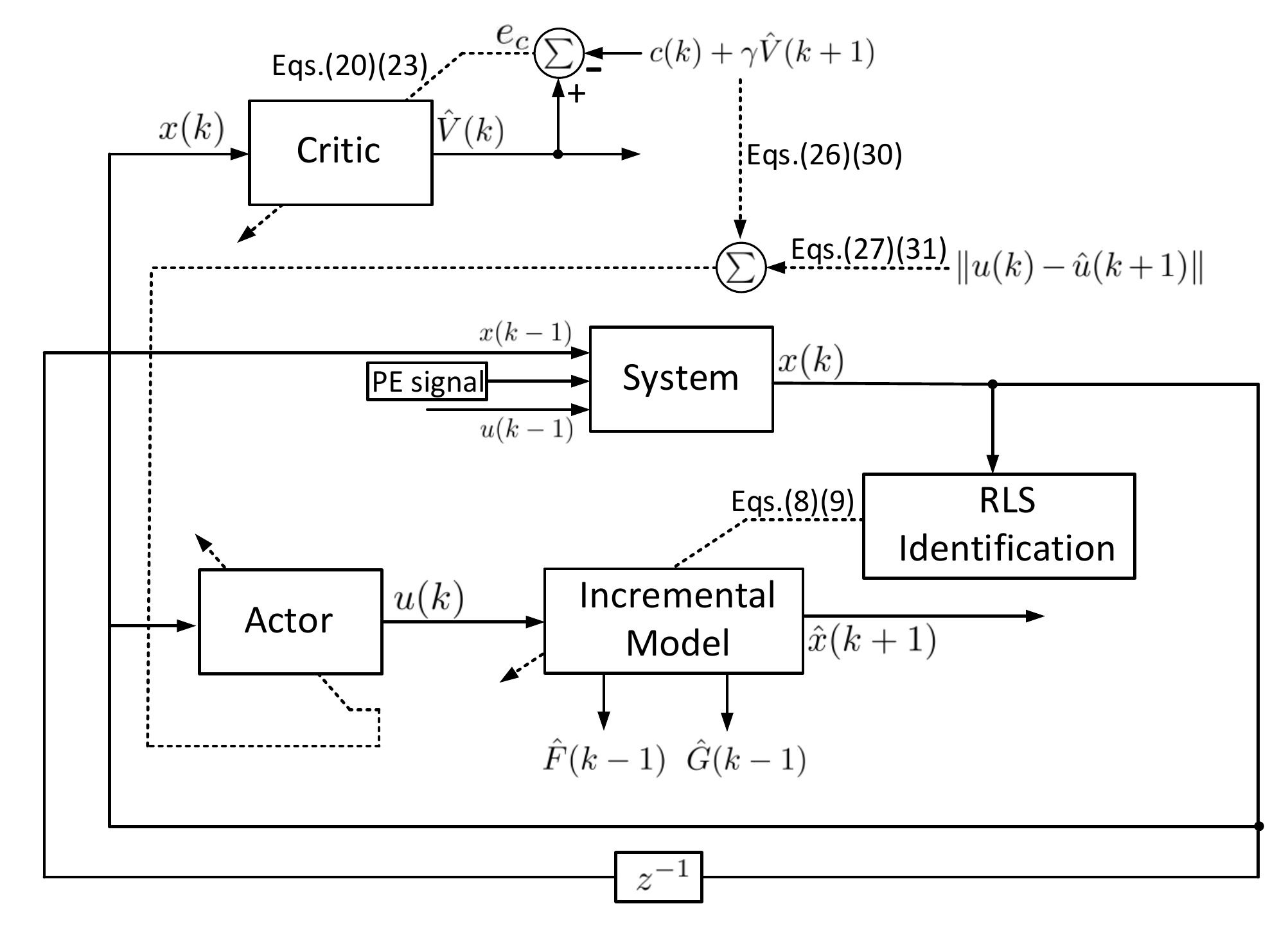}
    \caption{Schematic of the proposed TS-IHDP with an associated action increment penalty. Solid lines represent the forward propagation of signals, while dashed lines represent the backward propagation of gradients. (Adapted from \cite{ZhouTNNL34}).}\label{TS-IHDP_dirgram}
\end{figure}

This section develops the full-state feedback TS-IHDP method based on the ADHDP formulation \cite{Sokolov20159} and incorporates a temporal policy smoothness penalty into the policy optimization objective. A schematic of TS-IHDP is provided in Figure \ref{TS-IHDP_dirgram}.


\subsection{Discrete-time optimal control problem}

Consider discrete-time variables $x(k)\triangleq x(kT)$, $u(k)\triangleq u(kT)$. Using the Euler approximation $\dot{x}\approx\frac{x(k+1)-x(k)} {T}$ for the continuous-time system \eqref{continuous_system} yields $x(k+1)\approx x(k)+ T f(x(k),u(k))$.

The value function is defined as 
\begin{equation}
\begin{aligned}
    V(k) &= \sum_{i=k}^{\infty} \gamma^{\, i-k} 
    c\big(x(i), u(i)\big),\\
    &= c(x(k),u(k))+\gamma V(f(x(k),u(k))
    \label{equation_cost}    
\end{aligned}
\end{equation}
where $V(k) \triangleq V(x(k))$, $\gamma\in(0,1]$ is the discount factor, $c(x(k), u(k))=x(k)^{T}Qx(k)+u(k)^{T}Ru(k)$ is the one-step quadratic cost function. We require
$c(k)\triangleq c(x(k),u(k))$ to be positive semidefinite with respect to $(x,u)=(0,0)$, i.e. $c(0,0)=0, c(x,u)\geq 0, \forall (x,u)$, and bounded for all admissible states and input function of the
state and control. One can obtain from \eqref{equation_cost} that $0 = V(k) - c(k) -\gamma V(k+1)$.

\subsection{Critic network}
\label{subsec:critic_network}

The structure of the critic network is a fully connected single-hidden-layer multilayer perceptron (MLP) \cite{Haykin2009NeuralNetworks} that takes the input as $y(k)=[x_1(k),\ldots,x_n(k)]^{T}$ and the output as approximated value function $\hat{V}(k)$. The number of neurons in hidden layer is $N_{h_{c}}$. The weight between input neuron $i$ and hidden neuron $j$ is $\hat{w}^{(1)}_{c,ji}(k)$. The weight between hidden neuron $j$ and output neuron is $\hat{w}^{(2)}_{c,j}(k)$. 

The input to hidden neuron $j$ is
\begin{equation}
    \sigma_{c,j}(k)
    = \sum_{i=1}^{n} \hat{w}^{(1)}_{c,ji}(k)\, x_i(k)
\end{equation}

The output of hidden neuron $j$ is
\begin{equation}
    \phi_{c,j}(k)
    = \tanh(\sigma_{c,j}(k))
\end{equation}\\
where $\tanh(\cdot)$ is the nonlinear activation function.

The output of the critic is

\begin{equation}
    \hat{V}(k) = \sum_{j=1}^{N_{h_c}}
      \hat{w}^{(2)}_{c,j}(k)\, \phi_{c,j}(k),
\end{equation}

\subsection{Actor network}
\label{subsec:action_network}

The actor (policy) network is a single-hidden-layer MLP that takes the state $x(k) = [x_1(k), \ldots, x_n(k)]^{T}$ as input and outputs the action $u(k)=[u_{1}(k),\cdots,u_{m}(k)]^{T}$. The number of neurons in the hidden layer is $N_{h_a}$. The weight between input neuron $i$ and hidden neuron $j$ is denoted as $\hat{w}^{(1)}_{a,ji}(k)$. The weight between hidden neuron $j$ and output neuron $p$ is denoted as $\hat{w}^{(2)}_{a,pj}(k)$.

The input to hidden neuron $j$ is
\begin{equation}
    \sigma_{a,j}(k)
    = \sum_{i=1}^{n} \hat{w}^{(1)}_{a,ji}(k)\, x_i(k).
\end{equation}

The output of hidden neuron $j$ is:
\begin{equation}
    \phi_{a,j}(k)
    = \tanh( \sigma_{a,j}(k))
\end{equation}

The hidden to output neuron $p$ is
\begin{equation}
    \sigma_{a,p}(k) = \sum_{j=1}^{N_{h_a}}\hat{w}^{(2)}_{a,pj}(k)\, \phi_{a,j}(k)
\end{equation}

The output of the output neuron $p$ is
\begin{equation}
    u_{p}(k)=\phi_{a,p}(k)=u_{p,\max}\tanh(\sigma_{a,p}(k))
\end{equation}

\subsection{Update of the critic network}

The critic network is trained to minimize the objective function defined as $E_c(k) = \frac{1}{2} e_c^2(k)$, where $e_c(k) = \hat{V}(k) - c(k) - \gamma \hat{V}(k+1)$ is the Bellman error. In ADHDP \cite{Sokolov20159}, the definition of $e_{c}(k)$ adopts a backward-in-time formulation in order to update the policy according to the Q value function using a collected sample. In contrast, we use a forward-in-time formulation for IHDP, since the policy evaluation requires one-step state prediction using the incremental model.

The weights between the hidden layer and the output layer, $\hat{w}^{(2)}_{c}\in\mathbb{R}^{1\times N_{h_{c}}}$, are updated by

\begin{equation}\label{equation_wc2_update_vector}
    \hat{w}^{(2)}_{c}(k+1)=\hat{w}^{(2)}_{c}(k)+\Delta \hat{w}^{(2)}_{c}(k)
\end{equation}

Using gradient descent, the element-wise weight update is given by

\begin{equation}\label{wc2_element_wise_increment}
    \Delta \hat{w}^{(2)}_{c,j}(k)
    = l_c \left(- \frac{\partial E_c(k)}{\partial \hat{w}^{(2)}_{c,j}(k)}\right)
\end{equation}\\
which is obtained by using the chain rule

\begin{equation}
    \frac{\partial E_c(k)}{\partial \hat{w}^{(2)}_{c,j}(k)}
    = \frac{\partial E_c(k)}{\partial e_{c}(k)}\frac{\partial e_{c}(k)}{\partial \hat{V}(k)} \frac{\partial \hat{V}(k)}
         {\partial \hat{w}^{(2)}_{c,j}(k)}
    = e_c(k)\, \phi_{c,j}(k)
    \label{equation_critic_w2_update}
\end{equation}

The weights between the input layer and the hidden layers, $\hat{w}^{(1)}_{c}\in\mathbb{R}^{N_{h_{c}}\times m}$, are updated as

\begin{equation}
    \hat{w}^{(1)}_{c}(k+1)=\hat{w}^{(1)}_{c}(k)+\Delta \hat{w}^{(1)}_{c}(k)
\end{equation}

Using gradient descent, the element-wise weight update is given by

\begin{equation}
    \Delta \hat{w}^{(1)}_{c,ji}(k)
    = l_c \left( - \frac{\partial E_c(k)}{\partial \hat{w}^{(1)}_{c,ji}(k)}\right)
\end{equation}\\
which is obtained by using the chain rule

\begin{equation}
  \begin{aligned}
    \frac{\partial E_c(k)}{\partial \hat{w}^{(1)}_{c,ji}(k)}
    &=\frac{\partial E_c(k)}{\partial e_{c}(k)}\frac{\partial e_c(k)}{\partial \hat{V}(k)}
    \frac{\partial \hat{V}(k)}{\partial \phi_{c,j}(k)}\frac{\partial \phi_{c,j}(k)}{\partial \sigma_{c,j}(k)} \frac{\partial \sigma_{c,j}(k)}{\partial \hat{w}^{(1)}_{c,ji}(k)} \\
    &= e_c(k)\, \hat{w}^{(2)}_{c,j}(k)\phi_{c,j}^{\prime}(k) y_i(k)   \label{equation_critic_w1_update}
\end{aligned}  
\end{equation}

\subsection{Update of the actor network}\label{section_3_5}

The actor network is trained to minimize the objective function $E_a(k) =E_{a,1}(k)+\lambda E_{a,2}(k)$, where $E_{a,1}(k)=\frac{1}{2} e_a^2(k), E_{a,2}(k)=\frac{1}{2} e^{2}_{s}(k)-\varepsilon, e_a(k) = c(k)+\gamma\hat{V}(k+1)$, $e_{s}(k)=\frac{1}{2}\Vert u(k)-\hat{u}(k+1)\Vert^{2}$. $e_{s}(k)$ is the quadratic form of the estimated action increment over a sampling interval $T$, which measures temporal control smoothness. The weight $\lambda\geq 0$ controls the emphasis placed on temporal action invariance relative to the prescribed threshold $\varepsilon$ in the optimization objective \cite{Mysore2021ICRA}. $\varepsilon$ is a threshold that specifies the expected extent of variation in the estimated action, and can be determined by the actuator rate constraint or pitch-rate constraint. In \cite{Mysore2021ICRA}, $\varepsilon=0$ and the smoothness penalty has no reference criterion, so its weight is adjusted based on the observed smoothness of the control actions. We introduce $\varepsilon$ for the first time to enable $e_{s}$ approaches $\varepsilon$, thereby enforcing a prescribed level of smoothness. If $\lambda=0$, the policy update no longer incorporates the action smoothness constraint and reduces to IHDP \cite{ZHOUAST2020105}.

The weights between the hidden layer and the output layer, $\hat{w}^{(2)}_{a}\in\mathbb{R}^{n \times N_{h_{a}}}$, are updated as

\begin{equation}\label{equation_wa2_update_rule}
    \hat{w}^{(2)}_{a}(k+1)=\hat{w}^{(2)}_{a}(k)+\Delta \hat{w}^{(2)}_{a}(k)
\end{equation}\\
and the element-wise weight increment is

\begin{equation}\label{equation_wa2_increment}
    \Delta \hat{w}^{(2)}_{a,pj}(k) = l_a \left( -\frac{\partial E_{a,1}(k)}{\partial\hat{w}^{(2)}_{a,pj}(k)}-\lambda\frac{\partial E_{a,2}(k)}{\partial\hat{w}^{(2)}_{a,pj}(k)}\right)
\end{equation}

Applying the chain rule:

\footnotesize
\begin{equation}
\begin{aligned}
    &\frac{\partial E_{a,1}(k)}{\partial \hat{w}^{(2)}_{a,pj}(k)}    
    =\frac{\partial E_{a,1}(k)}{\partial e_{a}(k)}\frac{\partial e_{a}(k)}{\partial u_{p}(k)}\frac{\partial u_p(k)}{\partial\sigma_{a,p}(k)}\frac{\partial\sigma_{a,p}(k)}{\partial \hat{w}^{(2)}_{a,pj}(k)} \\ 
    =&\frac{\partial E_{a,1}(k)}{\partial e_{a}(k)}\left(\frac{\partial c(k)}{\partial u(k)}+\gamma\left(\frac{\partial \hat{V}(k+1)}{\partial \hat{x}(k+1)}\right)^{T}\frac{\partial \hat{x}(k+1)}{\partial u(k)}\right)_{p}\\
    &\qquad \quad \times \frac{\partial u_p(k)}{\partial\sigma_{a,p}(k)}\frac{\partial\sigma_{a,p}(k)}{\partial \hat{w}^{(2)}_{a,pj}(k)}\\    =&e_a(k)\left(2Ru(k)+\gamma\left(\hat{w}^{(2)}_{c}(k)\phi^{\prime}(\hat{x}(k+1))\hat{w}^{(1)}_{c}(k)\right)^{T}\hat{G}(k-1)\right)_{p}\\
    &\qquad \quad \times \phi^{\prime}_{a,p}(k)\phi_{a,j}(k)\\
    \label{equation_action_e1_w2_update}
\end{aligned}
\end{equation}
\normalsize
and compute the element-wise weight gradient from temporal smoothness loss $E_{a,2}(k)$:

\begin{equation}
\begin{aligned}
    \frac{\partial E_{a,2}(k)}{\partial \hat{w}^{(2)}_{a,pj}(k)}    
    &=\frac{\partial E_{a,2}(k)}{\partial e_{s}(k)}\frac{\partial e_{s}(k)}{\partial u_{p}(k)}\frac{\partial u_p(k)}{\partial\sigma_{a,p}(k)}\frac{\partial\sigma_{a,p}(k)}{\partial \hat{w}^{(2)}_{a,pj}(k)} \\ 
    &=e_s(k)(u_{p}(k)-\hat{u}_{p}(k+1))\phi^{\prime}_{a,p}(k)\phi_{a,j}(k)
    \label{equation_action_e2_w2_update}
\end{aligned}
\end{equation}\\
where $\phi^{\prime}_{a,p}(k)=u_{p,\max}\tanh^{\prime}(\sigma_{a,p}(k))$.

The weights between the input layer and the hidden layer $\hat{w}^{(1)}_{c}\in\mathbb{R}^{N_{h_{c}}\times m}$ are updated as

\begin{equation}\label{equation_action_wa1_update}
    \hat{w}^{(1)}_{a}(k+1)=\hat{w}^{(1)}_{a}(k)+\Delta \hat{w}^{(1)}_{a}(k)
\end{equation}\\
and the element-wise weight increment is given by

\begin{equation}
    \Delta \hat{w}^{(1)}_{a,ji}(k)
    = l_a \left(- \frac{\partial E_{a,1}(k)} {\partial\hat{w}^{(1)}_{a,ji}(k)}- \lambda\frac{\partial E_{a,2}(k)} {\partial\hat{w}^{(1)}_{a,ji}(k)}\right),
\end{equation}\label{equation_action_e1_e2_w1_update}

Applying the chain rule:
\footnotesize
\begin{equation}
\begin{aligned}
    &\frac{\partial E_{a,1}(k)}{\partial \hat{w}^{(1)}_{a,ji}(k)}
    =\frac{\partial E_{a,1}(k)}{\partial e_{a}(k)}\left(\frac{\partial e_{a}(k)}{\partial u(k)}\right)^{T}
    \frac{\partial u(k)}{\partial \sigma_{a,j}(k)}\frac{\partial \sigma_{a,j}(k)}{\partial \hat{w}^{(1)}_{a,ji}(k)} \\
    =&\frac{\partial E_{a,1}(k)}{\partial e_{a}(k)}\sum^{m}_{p=1}\frac{\partial e_{a}(k)}{\partial u_{p}(k)}
    \frac{\partial u_{p}(k)}{\partial \sigma_{a,p}(k)}\frac{\partial\sigma_{a,p}(k)}{\partial\phi_{a,j}(k)}\frac{\partial \phi_{a,j}(k)}{\partial \sigma_{a,j}(k)}\frac{\partial \sigma_{a,j}(k)}{\partial \hat{w}^{(1)}_{a,ji}(k)} \\
    =&\frac{\partial E_{a,1}(k)}{\partial e_{a}(k)}\sum^{m}_{p=1}\left(\frac{\partial c(k)}{\partial u(k)}+\gamma\left(\frac{\partial \hat{V}(k+1)}{\partial \hat{x}(k+1)}\right)^{T}\frac{\partial \hat{x}(k+1)}{\partial u(k)}\right)_{p}\\
    &\qquad \qquad \times \frac{\partial u_{p}(k)}{\partial \sigma_{a,p}(k)}\frac{\partial\sigma_{a,p}(k)}{\partial\phi_{a,j}(k)}\frac{\partial \phi_{a,j}(k)}{\partial \sigma_{a,j}(k)}\frac{\partial \sigma_{a,j}(k)}{\partial \hat{w}^{(1)}_{a,ji}(k)} \\   =&e_a(k)\sum^{m}_{p=1}\left(2Ru(k)+\left(\gamma\hat{w}^{(2)}_{c}(k)\phi^{\prime}(\hat{x}(k+1))\hat{w}^{(1)}_{c}(k)\right)^{T}\hat{G}(k-1)\right)_{p}\\
    &\qquad \quad \times \phi^{\prime}_{a,p}(k)\hat{w}^{(2)}_{a,pj}(k)\phi_{a,j}^{\prime}(k) x_i(k)
    \label{equation_action_e1_w1_update}
\end{aligned}
\end{equation}

\normalsize

and compute the element-wise weight gradient from temporal smoothness loss $E_{a,2}(k)$:
\footnotesize
\begin{equation}
\begin{aligned}
    &\frac{\partial E_{a,2}(k)}{\partial \hat{w}^{(1)}_{a,ji}(k)}=\frac{\partial E_{a,2}(k)}{\partial e_{s}(k)}\left(\frac{\partial e_{s}(k)}{\partial u(k)}\right)^{T}\frac{\partial u(k)}{\partial \sigma_{a,p}(k)}\frac{\partial\sigma_{a,p}(k)}{\partial \phi_{a,j}(k)}\\
    &\qquad \qquad \quad \quad \times \frac{\partial \phi_{a,j}(k)}{\partial \sigma_{a,j}(k)}\frac{\partial \sigma_{a,j}(k)}{\partial \hat{w}^{(1)}_{a,ji}(k)} \\
    =&\frac{\partial E_{a,2}(k)}{\partial e_{s}(k)}\sum^{m}_{p=1}\frac{\partial e_{s}(k)}{\partial u_{p}(k)}
    \frac{\partial u_{p}(k)}{\partial\sigma_{a,p}(k)}\frac{\partial \sigma_{a,p}(k)}{\partial \phi_{a,j}(k)}\frac{\partial \phi_{a,j}(k)}{\partial \sigma_{a,j}(k)}\frac{\partial \sigma_{a,j}(k)}{\partial \hat{w}^{(1)}_{a,ji}(k)} \\
    =&e_s(k)\sum^{m}_{p=1}(u_{p}(k)-\hat{u}_{p}(k+1))\phi^{\prime}_{a,p}(k)\hat{w}^{(2)}_{a,pj}(k)\phi_{a,j}^{\prime}(k) x_i(k)
\end{aligned}\label{equation_action_e2_w1_update}
\end{equation}
\normalsize
\begin{myRem}
As can be seen from the above gradient equations, the temporal smoothness loss $E_{a,2}(k)$ penalizes action increments and hence enables learning an action-increment–constrained policy under a multi-objective optimization formulation regarding $E_{a}(k)$, which is absent in standard IHDP \cite{ZhouIAC2017}. The corresponding gradient items $\frac{\partial E_{a,2}(k)}{\partial \hat{w}^{(2)}_{a,pj}(k)}$, $\frac{\partial E_{a,2}(k)}{\partial \hat{w}^{(1)}_{a,ji}(k)}$ in \eqref{equation_action_e2_w2_update}, \eqref{equation_action_e2_w1_update} regularize the actor network smoothness.
\end{myRem}

\section{Convergence Analysis}\label{section_convergence_analysis}

This section analyzes the convergence of TS-IHDP under gradient-based network updates. Using the trace analysis framework of \cite{Vamvoudakis2010878}, the uniform ultimate boundedness (UUB) of the network weight errors is established. The analysis further reveals that the temporal smoothness loss introduces tighter constraints on the actor's learning rate.

\subsection{Basics of uniformly ultimately bounded stability}

Since the system under consideration is nonlinear, the resulting value function $V(x(t))$ is non-convex. Consequently, gradient-descent–based policy optimization generally converges only to suboptimal solutions or local optima in the network weights. Denote the suboptimal weights by ${w}_c^{*}, w_a^{*}$, defined as $w_c^{*} = \arg\min_{\hat{w}_c}\frac{1}{2} \Vert \hat{V}(k) - c(k) - \gamma \hat{V}(k+1)\Vert^{2}$, and $w_a^{*} = \arg\min_{\hat{w}_a}\big(\frac{1}{2}\Vert c(k)$ $+\gamma\hat{V}(k+1)\Vert^{2}+\frac{1}{8}\lambda\Vert u(k)-\hat{u}(k+1) \Vert^{4}\big)$ if assuming \(U_c = 0\). The estimation errors of the critic and actor weights are defined by $\tilde{w}(k) := \hat{w}(k) - w^{*}$. Then, Equations \eqref{equation_critic_w2_update}, \eqref{equation_critic_w1_update}, \eqref{equation_action_e1_w2_update},\eqref{equation_action_e2_w2_update},\eqref{equation_action_e1_w1_update},\eqref{equation_action_e2_w1_update}
define a dynamical system of estimation
errors for a nonlinear function \(F\):

\begin{equation}
\tilde{w}(k+1) = \tilde{w}(k) - F\!\left(\hat{w}(k),x(k),\hat{x}(k+1)\right)
\label{equation_error-dynamics}
\end{equation}

\begin{myRem}
   The suboptimal weights $\omega^{*}_{c}$, $\omega^{*}_{a}$ are defined only for the individual state $x(k)$ during online learning, and vary across different states. Therefore, IHDP searches only for point-wise suboptimal weights at each update. This differs from the suboptimal weights defined based on the mean-square loss over a dataset in batch learning. 
\end{myRem}

\begin{myDef}[\cite{Khalil2002NonlinearSystems}]
A dynamical system is said to be uniformly ultimately bounded
(UUB) with ultimate bound \(\varepsilon>0\), if for any \(\delta>0\) and \(k_0>0\),
there exists a positive number \(N = N(\delta,\varepsilon)\) independent of \(k_0\),
such that $
\|\tilde{w}(k)\| \le \varepsilon, \text{for all } k \ge N $, whenever \(\|\tilde{w}(k_0)\| \le \delta\).
\end{myDef}

\begin{myTheo}[UUB of a discrete-time dynamical system \cite{Khalil2002NonlinearSystems}]If, for system \eqref{equation_error-dynamics}, there exists a function $L(\tilde{w}(k), k)$
such that for all $\tilde{w}(k_0)$ in a compact set $K$,
$L(\tilde{w}(k), k)$ is positive definite and the first difference
satisfies $\Delta L(\tilde{w}(k), k) < 0, \text{for} \; \|\tilde{w}(k_0)\| > \varepsilon$, for some $\varepsilon > 0$ such that the $\varepsilon$-neighborhood of $\tilde{w}(k)$
is contained in $K$, then the system is UUB and the norm of the state
is bounded within a neighborhood of $\varepsilon$.\label{UUB_theorem}
\end{myTheo}

\subsection{Network weight increments}

\begin{myAss}\label{assumption_1}
The suboptimal network weights are bounded as $\|w_c^\ast\| \le w_c^{\max}$, $\|w_a^\ast\| \le w_a^{\max}$.
\end{myAss}

\begin{myLem}\label{Lemma1}
Under Assumption \ref{assumption_1}, define $L_1(k) = \frac{1}{l_c} \operatorname{tr}\!\left[\tilde w_c^{(2)}(k)^T \tilde w_c^{(2)}(k)\right]$, and its first difference $\Delta L_{1}(k)=L_{1}(k+1)-L_{1}(k)$ is

\begin{equation}
\begin{aligned}
\Delta L_1(k)&= l_c e_{c}^2(k)\|\phi_c(k)\|^2 + \|e_{c}(k)-\tilde{w}^{(2)}_{c}(k)\phi_c(k)\|^2
- \Vert \zeta_{c}(k)\Vert^2- e^{2}_{c}(k)   
\end{aligned}  
\end{equation}
where $\zeta_c(k) = \tilde w_c^{(2)}(k)\phi_c(k)$ is the approximation error of the critic output. $\operatorname{tr}(\cdot)$ denotes the trace operator
\end{myLem}

\begin{myProof}
According to \eqref{equation_wc2_update_vector}, \eqref{wc2_element_wise_increment}, \eqref{equation_critic_w2_update} and the fact that $w_c^{ (2)*}$ is time-invariant:

\begin{equation}
\begin{aligned}\label{eq:wctupdate}
\tilde w_c^{(2)}(k+1)
&= \hat w_c^{(2)}(k+1) - w_c^{\ast (2)}= \tilde w_c^{(2)}(k)- l_c e_{c}(k) \phi_c(k)^{T}
\end{aligned}   
\end{equation}
where $\phi_{c}(k)\in\mathbb{R}^{N_{h_{c}}\times1}$, $e_{c}(k)$ is a scaler. 

The squared Frobenius norm of $\tilde{w}_c^{(2)}(k+1)$ is 

\begin{equation}
\begin{aligned}\label{eq:wcsquare}
\operatorname{tr}\left[\tilde w_c^{(2)}(k+1)^{T}\tilde w_c^{(2)}(k+1)\right]=&\operatorname{tr}\left[\left(\tilde w_c^{(2)}(k)- l_c e_{c}(k) \phi^{T}_c(k)\right)^{T}\left(\tilde w_c^{(2)}(k)- l_c e_{c}(k) \phi^{T}_c(k)\right)\right]\\ 
=&\operatorname{tr}\left[\tilde{w}_c^{(2)}(k)^{T}\tilde w_c^{(2)}(k)\right] +\operatorname{tr}\left[l^{2}_c e^{2}_{c}(k)\phi_{c}(k) \phi_c(k)^{T}\right] \\
&\quad- l_c e_{c}(k)\operatorname{tr}\left[\tilde{w}_c^{(2)}(k)^{T} \phi_c(k)^{T}\right]- l_c e_{c}(k)\operatorname{tr}\left[\phi_c(k)\tilde{w}_c^{(2)}(k)\right]\\
=&\operatorname{tr}\left[\tilde{w}_c^{(2)}(k)^{T}\tilde w_c^{(2)}(k)\right] +\operatorname{tr}\left[l^{2}_c e^{2}_{c}(k)\phi_{c}(k) \phi^{T}_c(k)\right] \\
&\quad- 2l_c e_{c}(k)\operatorname{tr}\left[\phi_c(k)\tilde{w}_c^{(2)}(k)\right]\\
=&\operatorname{tr}\left[\tilde{w}_c^{(2)}(k)^{T}\tilde w_c^{(2)}(k)\right] +l^{2}_c e^{2}_{c}(k)\Vert\phi_{c}(k)\Vert^{2}  \\
&\quad- 2l_c e_{c}(k)\tilde{w}_c^{(2)}(k)\phi_c(k)
\end{aligned}   
\end{equation}

Since $\tilde w_c^{(2)}(k)\phi_c(k)$ is a scalar, the cross term becomes
\begin{align}
-2 l_c e_{c}(k)\tilde w_c^{(2)}(k)\phi_c(k)= l_c\Bigl(\| e_{c}(k)-\tilde{w}^{(2)}_{c}(k)\phi_c(k)\|^2-\|\tilde{w}^{(2)}_{c}(k)\phi_c(k)\|^2- e_{c}^{2}(k)\Bigr)
\end{align}

The difference $\Delta L_{1}(k)$ yields

\begin{equation}
\begin{aligned}
\Delta L_1(k)
&=\frac1{l_c}\Bigr(\operatorname{tr}[\tilde{w}_c^{(2)}(k+1)^T \tilde w_c^{(2)}(k+1)]-\operatorname{tr}[\tilde w_c^{(2)}(k)^T \tilde w_c^{(2)}(k)]
\Bigr)\\
&= l_c e_{c}^2(k)\|\phi_c(k)\|^2 + \|e_{c}(k)-\tilde{w}^{(2)}_{c}(k)\phi_c(k)\|^2
- \Vert \tilde{w}^{(2)}_{c}(k)\phi_c(k)\Vert^2- e^{2}_{c}(k)\\
&= l_c e_{c}^2(k)\|\phi_c(k)\|^2 + \|e_{c}(k)-\tilde{w}^{(2)}_{c}(k)\phi_c(k)\|^2
- \Vert \zeta_{c}(k)\Vert^2- e^{2}_{c}(k)
\end{aligned}  
\end{equation}\\
where $\zeta_{c}(k)=\tilde{w}^{(2)}_{c}(k)\phi_{c}(k)$.
\end{myProof}

\begin{myLem}\label{Lemma2}
Under Assumption \ref{assumption_1}, define $L_2(k)=\frac{1}{l_ak_{1}}\operatorname{tr}\!\left[\tilde w_a^{(2)}(k)^T \tilde w_a^{(2)}(k)\right]$, and its first difference $\Delta L_{2}(k)=L_{2}(k+1)-L_{2}(k)$ is bounded by

\begin{equation}
\begin{aligned}
\Delta L_2(k)
&= \frac{1}{l_a k_{1}} \operatorname{tr}\!\left[
\tilde w^{(2)}_a(k+1)^{T} \tilde w^{(2)}_a(k+1)
- \tilde w^{(2)}_a(k)^{T} \tilde w^{(2)}_a(k)
\right] \\
&\leq \frac{1}{k_{1}} \Big(
2 l_a e_a^{2}(k)\, \|C(k)\|^{2}\, \|\phi_a(k)\|^{2}
+ 4 \|c(k)\|^{2}
+ 8 \gamma^{2} \|\zeta_c(k+1)\|^{2} \\
&\quad
+ 8 \gamma^{2} \|w_{c}^{*(2)} \phi_c(k+1)\|^{2}
+ \|C(k)\|^{2} \|\zeta_a(k)\|^{2}
- e_a^{2}(k)
\Big) \\
&\quad
+ \frac{\lambda}{k_{1}} \Big(
2 \lambda l_a e_s^{2}(k)\, \|U(k)\|^{2}\, \|\phi_a(k)\|^{2}
+ 2 e_s^{2}(k)
+ \|U^{T}(k) \zeta_a(k)\|^{2}
- e_s^{2}(k)
\Big)
\end{aligned}
\end{equation}\\
where $\zeta_a(k)=\tilde w_a^{(2)}(k)\phi_a(k)$.
\end{myLem}


\begin{myProof}
According to Equations \eqref{equation_wa2_update_rule}-\eqref{equation_action_e2_w2_update} and the fact that $w^{(2)*}_{a}$ is time-invariant:

\begin{equation}
\begin{aligned}
\tilde w^{(2)}_a(k+1)
&= \hat w^{(2)}_a(k+1) - w^{(2)*}_a = \tilde w^{(2)}_a(k)-l_ae_{a}(k)C(k)\phi_{a}(k)^{T}- \lambda l_{a} e_{s}(k) U(k)\phi_{a}(k)^{T}
\label{eq:wa2_update}
\end{aligned}
\end{equation}\\
where $\phi_{a}\in\mathbb{R}^{N_{a_{h}}\times1}$, $C(k)\in \mathbb{R}^{n\times 1}$ and its $p$-th element\\ $C_{p}(k)=\left(2Ru(k)+\left[\gamma\hat{w}^{(2)}_{c}(k)\phi^{\prime}(\hat{x}(k+1))\hat{w}^{(1)}_{c}(k)\right]^{T}\hat{G}(k-1)\right)_{p}\phi^{\prime}_{a,p}(k)$, $U(k)\in\mathbb{R}^{n\times 1}$ and its element $U_{p}(k)=(u_{p}(k)-\hat{u}_{p}(k+1))\phi^{\prime}_{a,p}(k)$.

The squared Frobenius norm of $\tilde{w}^{(2)}_a(k+1)$ is given by

\footnotesize
\begin{equation}
\begin{aligned}
&\operatorname{tr}\left[\left(\tilde w^{(2)}_a(k+1)\right)^{T}\tilde w^{(2)}_a(k+1)\right]\\ =&\operatorname{tr}\left[\left(\tilde w^{(2)}_a(k)-l_ae_{a}(k)C(k)\phi_{a}(k)^{T}-\lambda l_{a}e_{s}(k) U(k)\phi_{a}(k)^{T}\right)^{T}\left(\tilde w^{(2)}_a(k)
-l_ae_{a}(k)C(k)\phi_{a}(k)^{T}-\lambda l_{a}e_{s}(k) U(k)\phi_{a}(k)^{T}\right)\right]\\
=& \operatorname{tr}\left[\tilde w^{(2)}_a(k)^{T}\tilde w^{(2)}_a(k)\right]+\operatorname{tr}\left[\left(l_{a}e_{a}(k)C(k)\phi_{a}(k)^{T}+\lambda l_{a} e_{s}(k)U(k)\phi_{a}(k)^{T}\right)^{T}\left(l_{a}e_{a}(k)C(k)\phi_{a}(k)^{T}+\lambda l_{a}e_{s}(k)U(k)\phi_{a}(k)^{T}\right)\right]\\
&\qquad-l_ae_{a}(k)\operatorname{tr}\left[\tilde w^{(2)}_a(k)^{T}C(k)\phi_{a}(k)^{T}\right]- l_a e_{a}(k)\operatorname{tr}\left[\phi_{a}(k)C(k)^{T}\tilde w^{(2)}_a(k)\right]\\
&\qquad-\lambda l_ae_{s}(k)\operatorname{tr}\left[\tilde w^{(2)}_a(k)^{T}U(k)\phi_{a}(k)^{T}\right]-\lambda l_ae_{s}(k)\operatorname{tr}\left[\phi_{a}(k)U(k)^{T}\tilde w^{(2)}_a(k)\right]\\
=& \operatorname{tr}\left[\tilde w^{(2)}_a(k)^{T}\tilde w^{(2)}_a(k)\right]+\operatorname{tr}\left[l^{2}_{a}e^{2}_{a}(k)\phi_{a}(k)C(k)^{T}C(k)\phi_{a}(k)^{T}\right]+\operatorname{tr}\left[\lambda^{2}l^{2}_{a}e^{2}_{s}(k)\phi_{a}(k)U(k)^{T}U(k)\phi_{a}(k)^{T}\right]\\
&\quad+2\lambda l^{2}_{a}e_{a}(k)e_{s}(k)\operatorname{tr}\left[\phi_{a}(k) C(k)^{T}U(k)\phi_{a}(k)^{T}\right]-2l_{a}e_{a}(k)\operatorname{tr}\left[\phi_{a}(k)C(k)^{T}\tilde w^{(2)}_a(k)\right]-2\lambda l_ae_{s}(k)\operatorname{tr}\left[\phi_{a}(k)U(k)^{T}\tilde w^{(2)}_a(k)\right]\\
=& \operatorname{tr}\left[\tilde w^{(2)}_a(k)^{T}\tilde w^{(2)}_a(k)\right]+l^{2}_{a}e^{2}_{a}(k)\Vert C(k)\Vert^{2}\Vert \phi_{a}(k)\Vert^{2}+\lambda^{2}l^{2}_{a}e^{2}_{s}(k)\Vert U(k)\Vert^{2}\Vert \phi_{a}(k)\Vert^{2}\\
&\quad+\underbrace{2\lambda l^{2}_{a}e_{a}(k)e_{s}(k)\operatorname{tr}\left[\phi_{a}(k) C(k)^{T}U(k)\phi_{a}(k)^{T}\right]}_{\text{term 1}}\underbrace{-2l_ae_{a}(k)\operatorname{tr}\left[\phi_{a}(k)C(k)^{T}\tilde w^{(2)}_a(k)\right]}_{\text{term 2}}-\underbrace{2\lambda l_ae_{s}(k)\operatorname{tr}\left[\phi_{a}(k)U(k)^{T}\tilde w^{(2)}_a(k)\right]}_{\text{term 3}}
\label{}
\end{aligned}
\end{equation}

\normalsize
Use the cyclic property of the trace, $\operatorname{tr}(ABC)=\operatorname{tr}(BCA)$, and the Cauchy-Schwarz inequality \cite{HornJohnson2012MatrixAnalysis}, $|\operatorname{tr}(A^{T}B)|\leq\Vert A\Vert\Vert B\Vert$, in term 1:

\begin{equation}
\begin{aligned}
    \operatorname{tr}\left[\phi_{a}(k) C(k)^{T}U(k)\phi_{a}(k)^{T}\right]&=\operatorname{tr}\left[C(k)^{T}U(k)\left(\phi_{a}(k)^{T}\phi_{a}(k)\right)\right]\\  
    &=\operatorname{tr}\left[C(k)^{T}U(k)\right]\operatorname{tr}[\phi(k)^{T}\phi(k)]\\
    &\leq\big\vert\operatorname{tr}\left[C(k)^{T}U(k)\right]\big\vert\operatorname{tr}[\phi(k)^{T}\phi(k)]\\
    &\leq \Vert C(k)\Vert\Vert U(k)\Vert\Vert \phi(k)\Vert^{2}
\end{aligned}    
\end{equation}


Use $2ab\leq a^{2}+b^{2}$ in term 1:

\begin{equation}
\begin{aligned}
    &2\lambda l^{2}_{a}e_{a}(k)e_{s}(k)\operatorname{tr}\left[\phi_{a}(k) C(k)^{T}U(k)\phi_{a}(k)^{T}\right]\\
    &\leq 2\lambda l^{2}_{a}e_{a}(k)e_{s}(k) \Vert C(k)\Vert\Vert U(k)\Vert\Vert \phi(k)\Vert^{2}\\
    &\leq  l^{2}_{a}e^{2}_{a}(k)\Vert C(k) \Vert^{2}\Vert \phi(k) \Vert^{2}+\lambda^{2}l^{2}_{a}e^{2}_{s}(k)\Vert U(k) \Vert^{2}\Vert \phi(k)\Vert^{2}
\end{aligned}
\end{equation}

Consider the cross term 2, and use $\operatorname{tr}(ABC)=\operatorname{tr}(BCA)$:

\begin{equation}
\begin{aligned}
  &-2 l_a e_{a}(k)\operatorname{tr}\left(\phi_{a}(k)C(k)^{T}\tilde{w}^{(2)}_{a}(k)\right)\\
   &=-2 l_a e_{a}(k)\operatorname{tr}\left(C(k)^{T}\tilde{w}^{(2)}_{a}(k)\phi_{a}(k)\right)  \\
   &=-2 l_a e_{a}(k)C(k)^{T}\tilde{w}^{(2)}_{a}(k)\phi_{a}(k) \\
  &\leq l_{a}\left(\left\|e_{a}(k)-C(k)^{T}\tilde{w}^{(2)}_{a}(k)\phi_{a}(k)\right\|^{2}-\left\|C(k)^{T}\tilde{w}^{(2)}_{a}(k)\phi_{a}(k)\right\|^{2}-e^{2}_{a}(k)\right)\\
  &\leq l_{a}\left(\left\|e_{a}(k)-C(k)^{T}\zeta_a(k)\right\|^{2}-\left\|C(k)^{T}\zeta_a(k)\right\|^{2}-e^{2}_{a}(k)\right)
\end{aligned}
\end{equation}\\
where $\zeta_{a}(k)=\tilde{w}^{(2)}_{a}(k)\phi_{a}(k)$.

Similarly, term 3 becomes

\begin{equation}
\begin{aligned}
  -2 \lambda l_a e_{s}(k)\operatorname{tr}\left[\phi_{a}(k)U(k)^{T}\tilde{w}^{(2)}_{a}(k)\right]& =-2 \lambda l_a e_{s}(k)\operatorname{tr}\left[U(k)^{T}\tilde{w}^{(2)}_{a}(k)\phi_{a}(k)\right]  \\
  & =-2  \lambda l_a e_{s}(k)U(k)^{T}\tilde{w}^{(2)}_{a}(k)\phi_{a}(k) \\
  &\leq  \lambda l_{a} \left(2e^{2}_{s}(k)+\Vert U(k)^{T}\tilde{w}^{(2)}_{a}(k)\phi_{a}(k)\Vert^{2}-e^{2}_{s}(k)\right)\\
  &\leq  \lambda l_{a} \left(2e^{2}_{s}(k)+\Vert U(k)^{T}\zeta_{a}(k)\Vert^{2}-e^{2}_{s}(k)\right)
\end{aligned}
\end{equation}

Notice that $(a-b)^{2}-b^{2}\leq 2a^{2}+b^{2}$, and use $\Vert C(k)^{T}\zeta_{a}(k)\Vert^{2}\leq\Vert C(k)^{T}\Vert^{2}\Vert\zeta_{a}(k)\Vert^{2}$ by Cauchy–Schwarz inequality:

\begin{equation}
\begin{aligned}
&\left\|e_{a}(k)-C(k)^{T}\zeta_{a}(k)\right\|^{2}-\left\|C(k)^{T}\zeta_{a}(k)\right\|^{2}\\
&\quad\le2\left\|e_{a}(k)\right\|^{2}+\left\|C(k)^{T}\zeta_{a}(k)\right\|^{2}\\
&\quad\le2\left\|c(k)+\gamma\hat{V}(k+1)\right\|^{2}+\Vert C(k)^{T}\zeta_{a}(k)\Vert^{2}\\
&\quad\le2\left(\left\|
c(k)\right\|+\gamma\left\|\hat{V}(k+1)\right\|
\right)^{2}+\left\|C(k)^{T}\zeta_{a}(k)\right\|^{2}\\
&\quad\le4\|c(k)\|^{2}+4\gamma^{2}\left\|\hat{w}^{(2)}_{c}(k)\phi_{c}(k+1)\right\|^{2}+\left\|C(k)^{T}\zeta_{a}(k)
\right\|^{2}\\
&\quad\le4\|c(k)\|^{2}+4\gamma^{2}\left\|(\tilde{w}^{(2)}_{c}(k)+w^{(2)*}_{c})\phi_{c}(k+1)\right\|^{2}+\left\|C(k)^{T}\zeta_{a}(k)
\right\|^{2}\\
&\quad\le4\|c(k)\|^{2}+8\gamma^{2}\left\|\tilde{w}^{(2)}_{c}(k)\phi_{c}(k+1)\right\|^{2}+8\gamma^{2}\left\|w^{(2)*}_{c}\phi_{c}(k+1)\right\|^{2}+\left\|C(k)^{T}\zeta_{a}(k)\right\|^{2}\\
&\quad\le4\|c(k)\|^{2}+8\gamma^{2}\left\|\zeta_{c}(k+1)\right\|^{2}+8\gamma^{2}\left\|w^{(2)*}_{c}\phi_{c}(k+1)\right\|^{2}+\left\|C(k)^{T}\zeta_{a}(k)\right\|^{2}\\
&\quad \le4\|c(k)\|^{2}+8\gamma^{2}\left\|\zeta_{c}(k+1)\right\|^{2}+8\gamma^{2}\left\|w^{(2)*}_{c}\phi_{c}(k+1)\right\|^{2}+\left\|C(k)\Vert^{2}\Vert\zeta_{a}(k)\right\|^{2}
\end{aligned}
\end{equation}\\
where $\zeta_{c}(k+1)=\tilde{w}^{(2)}_{c}(k)\phi_{c}(k+1)$. \\

Therefore, the first difference is 
\begin{equation}
\begin{aligned}
&\Delta L_2(k)=\frac{1}{l_ak_{1}}\operatorname{tr}\left[
\tilde w^{(2)}_a(k+1)^{T}\tilde w^{(2)}_a(k+1)-\tilde w^{(2)}_a(k)^{T}\tilde w^{(2)}_a(k)\right]\\
&\leq\frac{1}{k_{1}}
\Big(2l_a e^{2}_{a}(k)\Vert C(k)\Vert^{2}\Vert \phi_a(k)\Vert^{2}+4\|c(k)\|^{2}+8\gamma^{2}\left\|\zeta_{c}(k+1)\right\|^{2}+8\gamma^{2}\left\|w^{(2)*}_{c}\phi_{c}(k+1)\right\|^{2}\\
&+\Vert C(k)\Vert^{2}\Vert\zeta_{a}(k)\Vert^{2}-e^{2}_{a}(k)\Big)\\
&+\frac{\lambda}{k_{1}}\left(2\lambda l_{a}e^{2}_{s}(k)\Vert U(k)\Vert^{2}\Vert \phi_{a}(k)\Vert^{2}+2 e^{2}_{s}(k)+\left\|U(k)^{T}\zeta_a(k)\right\|^{2}-e^{2}_{s}(k)\right)\\
\end{aligned}
\end{equation}
\end{myProof}

\begin{myRem}
Note that $\Delta L_{2}$ depends on $C(k)$, which incorporates the estimated control effectiveness $\hat{G}(k-1)$. Therefore, a system with stronger control effectiveness, i.e., a more sensitive system, may exhibit a larger bound, which is consistent with intuition.
\end{myRem}

\begin{myRem}
   $C(k)$ is defined differently from that in ADHDP \cite{Vamvoudakis2010878} because the objective functions used for policy improvement are different, namely, $c(k)+\hat{V}(k+1)$ in IHDP, while $\hat{Q}(x(k),u(k))$ in ADHDP.
\end{myRem}

\begin{myRem}
    During RLS identification, the estimated control effectiveness $\hat{G}(k-1)$ is initialized as a zero vector and may have signs opposite to those of $G(k-1)$ in the initial identification stage. These sign mismatches can lead to an opposite direction of the policy update, as shown in \cite{Tamimi2008TSMC38}. As a result, the learned policy may produce opposite responses to the same state when applied to the actual system. Therefore, we assume that RLS identification eventually yields $\hat{G}(k-1)$ with signs consistent with those of $G(k-1)$ after the online identification process, which ensures the converged policy performs well in the actual system.
\end{myRem}

\begin{myLem}\label{Lemma3}
Under Assumption \ref{assumption_1}, define $L_3(k) = \frac{1}{l_c k_{2}} \operatorname{tr}\left[\tilde w_c^{(1)}(k)^{T}\, \tilde w_c^{(1)}(k) \right]$, and its difference $\Delta L_{3}(k)=L_{3}(k+1)-L_{3}(k)$ is bounded by

\begin{equation}
\begin{aligned}
\Delta L_3(k)\leq\frac{1}{k_{2}}\Big(l_c e^{2}_{c}(k)\|a(k)\|^{2}\|y(k)\|^{2} +\Vert a(k)\Vert^{2}\Vert\tilde w^{(1)}_c(k)y(k)\Vert^{2}+ e^{2}_{c}(k)\Big)\label{eq:lemma3_bound}
\end{aligned}\\
\end{equation}
where $k_{2} > 0$ is a weighting factor and $a(k)\in\mathbb{R}^{N_{h_{c}}\times 1}$ is a vector with the element $
a_j(k) = \hat{w}^{(2)}_{c,j}(k) \phi^{\prime}_{j}(k)$.
\end{myLem}

\begin{myProof}
Consider the update rule \eqref{equation_critic_w1_update}, the weight error is 

\begin{equation}
\begin{aligned}
\tilde w^{(1)}_c(k+1)
= \hat w^{(1)}_c(k+1) - w_c^{*(1)}= \tilde w^{(1)}_c(k)- l_ce_{c}(k) B(k)
\label{eq:wtilde}
\end{aligned}
\end{equation}\\
where $B(k)\in\mathbb{R}^{N_{h_{c}}\times m}$ and its element $B_{ji}(k) =\hat{w}^{(2)}_{c,j}(k)\phi_{j}^{\prime}(k) y_i(k)$, then $B(k)=a(k)y(k)^{T}$.

The squared Frobenius norm of $\tilde w^{(1)}_c(k+1)$ is given by
\begin{equation}
\begin{aligned}
\operatorname{tr}\left[\tilde w^{(1)}_c(k+1)^{T}\tilde w^{(1)}_c(k+1)\right]&=\operatorname{tr}\left[\left(\tilde w^{(1)}_c(k)- l_ce_{c}(k)B(k)\right)^{T}\left(\tilde w^{(1)}_c(k)- l_ce_{c}(k)B(k)\right)\right]\\
&=\operatorname{tr}\left[\tilde w^{(1)}_c(k)^{T}\tilde w^{(1)}_c(k)\right]+\operatorname{tr}\left[l_c^{2} e^{2}_{c}(k)B(k)^{T}B(k)\right]\\
&\qquad-l_c e_{c}(k)\operatorname{tr}\left[\tilde w^{(1)}_c(k)^{T}B(k)\right]-l_c e_{c}(k)\operatorname{tr}\left[B(k)^{T}\tilde w^{(1)}_c(k)\right]\\
&=\operatorname{tr}\left[\tilde w^{(1)}_c(k)^{T}\tilde w^{(1)}_c(k)\right]+\operatorname{tr}\left[l_c^{2} e^{2}_{c}(k)B(k)^{T}B(k)\right]\\
&\qquad-2l_c e_{c}(k)\operatorname{tr}\left[B(k)^{T}\tilde w^{(1)}_c(k)\right]\\
&=\operatorname{tr}\left[\tilde w^{(1)}_c(k)^{T}\tilde w^{(1)}_c(k)\right]+\operatorname{tr}\left[l_c^{2} e^{2}_{c}(k)y(k)a(k)^{T}a(k)y(k)^{T}\right]\\
&\qquad-2l_c e_{c}(k)\operatorname{tr}\left[B(k)^{T}\tilde w^{(1)}_c(k)\right]\\
&=\operatorname{tr}\left[\tilde w^{(1)}_c(k)^{T}\tilde w^{(1)}_c(k)\right]+l_c^{2} e^{2}_{c}(k)\Vert a(k)\Vert^{2}\Vert y(k)\Vert^{2}\\
&\qquad-2l_c e_{c}(k)\operatorname{tr}\left[B(k)^{T}\tilde w^{(1)}_c(k)\right]
\label{eq:trace_expand}
\end{aligned}
\end{equation}

By $\operatorname{tr}(ABC)=\operatorname{tr}(BCA)$:
\begin{equation}
\begin{aligned}
    \operatorname{tr}\left[B(k)^{T}\tilde{w}^{(1)}_{c}(k)\right]  =&\operatorname{tr}\left[y(k)a(k)^{T}\tilde{w}^{(1)}_{c}(k)\right]\\
    =&\operatorname{tr}\left[a(k)^{T}\tilde w^{(1)}_c(k)y(k)\right]\\
    =&a(k)^{T}\tilde w^{(1)}_c(k)y(k) 
\end{aligned}
\end{equation}

Use inequality $-2ab\leq a^{2}+b^{2}$, the cross term becomes

\begin{equation}
\begin{aligned}
- 2 l_ce_{c}(k)a(k)^{T}\tilde w^{(1)}_c(k)y(k) &\leq l_{c}\left( \Vert a(k)^{T}\tilde w^{(1)}_c(k)y(k)\Vert^{2} + e^{2}_{c}(k)\right)\\
&\leq l_{c}\left( \Vert a(k)^{T}\Vert^{2} \Vert\tilde w^{(1)}_c(k)y(k)\Vert^{2} + e^{2}_{c}(k)\right)\\
&= l_{c}\left( \Vert a(k)\Vert^{2} \Vert\tilde w^{(1)}_c(k)y(k)\Vert^{2} + e^{2}_{c}(k)\right)
\label{eq:last_term}
\end{aligned}
\end{equation}


Therefore, the upper bound is given by

\begin{equation}
\begin{aligned}
\Delta L_3(k)\leq\frac{1}{k_{2}}\Big(l_c e^{2}_{c}(k)\|a(k)\|^{2}\|y(k)\|^{2} +\Vert a(k)\Vert^{2}\Vert\tilde w^{(1)}_c(k)y(k)\Vert^{2}+ e^{2}_{c}(k)\Big)
\end{aligned}
\end{equation}

\end{myProof}

\begin{myLem}\label{Lemma4}
Under Assumption \ref{assumption_1}, define $L_4(k)= \frac{1}{l_a k_{3}}\operatorname{tr}\!\left[\tilde w^{(1)}_a(k)^{T}\tilde w^{(1)}_a(k)\right]$, and its first difference  $\Delta L_{4}(k)=L_{4}(k+1)-L_{4}(k)$ is bounded by

\begin{equation}
\begin{aligned}
\Delta L_{4}(k)\le\frac{1}
{k_{3}}\Big(l_{a} e^{2}_{a}(k)\left\| C(k)^{T} D(k)\right\|^{2}\|x(k)\|^{2}+ e^{2}_{a}(k)+\left\| C(k)^{T} D(k)\right\|^{2}\left\|\tilde w^{(1)}_{a}(k)\, x(k)\right\|^{2}\Big)
\end{aligned}
\end{equation}
where $k_{3} > 0$ is a weighting factor, and $D(k)\in\mathbb{R}^{n\times N_{h_{a}}}$, and its element\\
$D_{pj}(k)=\hat w^{(2)}_{a,pj}(k)\phi^{\prime}_{a,pj}(x(k))$.
\end{myLem}

\begin{myProof}
According to Equations \eqref{equation_action_wa1_update}-\eqref{equation_action_e2_w1_update}, the weight errors are
\begin{equation}
\begin{aligned}
\tilde w^{(1)}_a(k+1)
&= \hat w^{(1)}_a(k+1) - w_a^{(1)*} \\
&= \tilde w^{(1)}_a(k)- l_a e_{a}(k)D(k)^{T} C(k) x(k)^{T}-\lambda l_{a}e_{s}(k)D(k)^{T}U(k)x(k)^{T}
\end{aligned}
\end{equation}

and the squared Frobenius norm:

\begin{equation}
\begin{aligned}
&\operatorname{tr}\left[\tilde w^{(1)}_a(k+1)^{T}\tilde w^{(1)}_a(k+1)\right]\\
=& \operatorname{tr}\Big[\left(\tilde w^{(1)}_a(k)- l_a e_{a}(k)D(k)^T C(k)x(k)^{T}-\lambda l_a e_{s}(k)D(k)^T U(k)x(k)^{T}\right)^{T}\\
&\qquad\left(\tilde w^{(1)}_a(k)- l_a e_{a}(k)D^T(k) C(k) x(k)^{T}- \lambda l_a e_{s}(k)D^T(k) U(k)x(k)^{T}\right)\Big]\\
=&\operatorname{tr}\left[\tilde w^{(1)}_a(k)^{T}\tilde w^{(1)}_a(k)\right]+\operatorname{tr}\Big[\left(l_{a}e_{a}(k)D(k)^{T}C(k)x(k)^{T}+\lambda l_{a}e_{s}(k)D(k)^{T}U(k)x(k)^{T}\right)^{T}\\
&\qquad \quad \qquad \qquad \qquad \qquad \qquad \left(l_{a}e_{a}(k)D(k)^{T}C(k)x(k)^{T}+\lambda l_{a}e_{s}(k)D(k)^{T}U(k)x(k)^{T}\right)\Big] \\
&\qquad-l_{a}\operatorname{tr}\left[\tilde w^{(1)}_a(k)^{T}\left(e_{a}(k)D(k)^{T}C(k)x(k)^{T}+\lambda e_{s}(k)D(k)^{T}U(k)x(k)^{T}\right)\right]\\
&\qquad-l_a \operatorname{tr}\left[\left(e_{a}(k)x(k)C(k)^{T}D(k)+\lambda e_{s}(k)x(k)U(k)^{T}D(k)\right)\tilde{w}^{(1)}_{a}(k)\right]\\
=&\operatorname{tr}\left[\tilde w^{(1)}_a(k)^{T}\tilde w^{(1)}_a(k)\right]+l^{2}_{a}e^{2}_{a}(k)\operatorname{tr}\left[ x(k)C(k)^{T}D(k)D(k)^T C(k) x(k)^{T}\right]\\
&\qquad+\lambda^{2}l^{2}_{a}e^{2}_{s}(k)\operatorname{tr}\left[ x(k)U(k)^{T}D(k)D(k)^{T} U(k) x(k)^{T}\right] \\
&\qquad+2\lambda l^{2}_{a}e_{s}(k)e_{a}(k)\operatorname{tr}\left[x(k)U(k)^{T}D(k)D(k)^{T}C(k)x(k)^{T}\right]-2l_a e_{a}(k)\operatorname{tr}\left[x(k)C(k)^{T}D(k)\tilde{w}^{(1)}_{a}(k)\right]\\
&\qquad -2\lambda l_a e_{s}(k)\operatorname{tr}\left[x(k)U(k)^{T}D(k)\tilde{w}^{(1)}_{a}(k)\right]\\
=&\operatorname{tr}\left[\tilde w^{(1)}_a(k)^{T}\tilde w^{(1)}_a(k)\right]+l^{2}_{a}e^{2}_{a}(k)\Vert C(k)^{T}D(k)\Vert^{2}\Vert x(k) \Vert^{2}+\lambda^{2}l^{2}_{a}e^{2}_{s}(k)\Vert U(k)^{T}D(k)\Vert^{2}\Vert x(k) \Vert^{2} \\
&\qquad+2\lambda l^{2}_{a}e_{s}(k)e_{a}(k)\operatorname{tr}\left[x(k)U(k)^{T}D(k)D(k)^{T}C(k)x(k)^{T}\right]-2l_{a} e_{a}(k)\operatorname{tr}\left[x(k)C(k)^{T}D(k)\tilde{w}^{(1)}_{a}(k)\right]\\
&\qquad-2\lambda l_a e_{s}(k)\operatorname{tr}\left[x(k)U(k)^{T}D(k)\tilde{w}^{(1)}_{a}(k)\right]
\end{aligned}
\end{equation}

Using the cyclic property of the trace $\operatorname{tr}(ABC)=\operatorname{tr}(BCA)$ and the fact that \\
$C(k)^{T}D(k)\tilde{w}^{(1)}_{a}(k)x(k)$ and $C(k)^{T}D(k)\tilde{w}^{(1)}_{a}(k)x(k)$ are scalars:

\begin{equation}
\begin{aligned}
    \operatorname{tr}\left[ x(k)C(k)^{T}D(k)\tilde{w}^{(1)}_{a}(k)\right]&=\operatorname{tr}\left[ C(k)^{T}D(k)\tilde{w}^{(1)}_{a}(k)x(k)\right]\\
    &=C(k)^{T}D(k)\tilde{w}^{(1)}_{a}(k)x(k)
\end{aligned}
\end{equation}

\begin{equation}
\begin{aligned}
    \operatorname{tr}\left[ x(k)U(k)^{T}D(k)\tilde{w}^{(1)}_{a}(k)\right]&=\operatorname{tr}\left[ U(k)^{T}D(k)\tilde{w}^{(1)}_{a}(k)x(k)\right]\\
    &=U(k)^{T}D(k)\tilde{w}^{(1)}_{a}(k)x(k)
\end{aligned}
\end{equation}

The cross terms can be transformed using $-2ab\leq a^{2}+b^{2}$ and Cauchy–Schwarz inequality:

\begin{equation}
\begin{aligned}
-2 l_a e_{a}(k)C(k)^{T}D(k)\tilde{w}^{(1)}_{a}(k)x(k)
&\leq l_a \left(e^{2}_{a}(k)+\left\| C(k)^{T} D(k)\tilde w^{(1)}_{a}(k) x(k)\right\|^{2}\right)\\
&\leq l_a \left(e^{2}_{a}(k)+\Vert C(k)^{T} D(k)\Vert^{2}\Vert\tilde w^{(1)}_{a}(k) x(k)\Vert^{2}\right)
\end{aligned}
\end{equation}

\begin{equation}
\begin{aligned}
-2 l_a e_{s}(k)U(k)^{T}D(k)\tilde{w}^{(1)}_{a}(k)x(k)
&\leq l_a \left(e^{2}_{s}(k)+\left\| U(k)^{T} D(k)\tilde w^{(1)}_{a}(k) x(k)\right\|^{2}\right)\\
&\leq l_a \left(e^{2}_{s}(k)+\Vert U(k)^{T} D(k)\Vert^{2}\Vert\tilde w^{(1)}_{a}(k) x(k)\Vert^{2}\right)
\end{aligned}
\end{equation}
And

\begin{equation}
\begin{aligned}
    \operatorname{tr}\left[x(k)U(k)^{T}D(k)D(k)^{T}C(k)x(k)^{T}\right]&=    \operatorname{tr}\left[U(k)^{T}D(k)D(k)^{T}C(k)x(k)^{T}x(k)\right]\\
    &=\operatorname{tr}\left[U(k)^{T}D(k)D(k)^{T}C(k)\right]\operatorname{tr}\left[x(k)^{T}x(k)\right]\\
    &=\operatorname{tr}\left[U(k)^{T}D(k)D(k)^{T}C(k)\right]\Vert x(k)\Vert^{2}\\
    &\leq\Vert U(k)^{T}D(k)\Vert \Vert D(k)^{T}C(k)\Vert \Vert x(k)\Vert^{2}\\
    &\leq\Vert U(k)^{T}D(k)\Vert \Vert C^{T}(k)D(k)\Vert \Vert x(k)\Vert^{2}
\end{aligned}
\end{equation}

which yields

\begin{equation}
\begin{aligned}
&2\lambda l^{2}_{a}e_{s}(k)e_{a}(k)\operatorname{tr}\left[x(k)U(k)^{T}D(k)D(k)^{T}C(k)x(k)^{T}\right]\\
\leq& 2\lambda l^{2}_{a}e_{s}(k)e_{a}(k) \Vert U(k)^{T}D(k)\Vert \Vert C^{T}(k)D(k)\Vert \Vert x(k)\Vert^{2}\\
\leq &l^{2}_{a}\left(e^{2}_{a}(k)\Vert C^{T}(k)D(k)\Vert^{2}\Vert x(k)\Vert^{2}+\lambda^{2}e^{2}_{s}(k)\Vert U(k)^{T}D(k)\Vert^{2}\Vert x(k)\Vert^{2}\right)
\end{aligned}  
\end{equation}

Finally, we can obtain the upper bound

\begin{equation}
\begin{aligned}
\Delta L_{4}(k)&\leq \frac{1}
{k_{3}}\Big(2l_{a} e^{2}_{a}(k)\left\| C(k)^{T} D(k)\right\|^{2}\|x(k)\|^{2}+ e^{2}_{a}(k)+\left\| C(k)^{T} D(k)\right\|^{2}\left\|\tilde w^{(1)}_{a}(k)\, x(k)\right\|^{2}\Big)\\
&+\frac{\lambda}{k_{3}}\Big(2\lambda l_{a} e^{2}_{s}(k)\left\| U(k)^{T} D(k)\right\|^{2}\|x(k)\|^{2}+ e^{2}_{s}(k)+\left\| U(k)^{T} D(k)\right\|^{2}\left\|\tilde w^{(1)}_{a}(k)\, x(k)\right\|^{2}\Big)
\end{aligned}
\end{equation}

\end{myProof}

\subsection{Convergence of the weight errors}

We introduce the Lyapunov candidate
\begin{equation}
L(k) = L_1(k) + L_2(k) + L_3(k) + L_4(k)    
\end{equation}

\begin{myTheo}[Main Theorem]
Let the weights of the critic and actor be updated by gradient descent algorithm, and assume that the one-step cost is a bounded semidefinite function. Under Assumption \ref{assumption_1}, the estimation errors $w_c^\ast - \hat w_c(k)$ and $w_a^\ast - \hat w_a(k)$ are uniformly ultimately bounded (UUB) if $k_{1}>\frac{8\overline{\phi}^{2}}{\kappa^{2}\underline{\phi}^{2}}\gamma^{2}$ and

\begin{equation}\label{lc_constraint}
l_{c} < \min_{k} \frac{k_{2}-1}{k_{2}\big(\|\phi_c(k)\|^{2}
+ \frac{1}{k_{2}}\|a(k)\|^{2}\|y(k)\|^{2}\big)}
\end{equation}

\begin{equation}\label{condition_la_1}
l_{a} < \min_{k} \frac{k_{3}-k_{1}}{k_{3}\| C(k)\|^{2}\|\phi_a(k)\|^{2}
+k_{1}\|C(k)^{T}D(k)\|^{2}\|x(k)\|^{2}}
\end{equation}

\begin{equation}\label{condition_la_2}
    l_{a} < \min_{k} \frac{1}{2\lambda}\frac{k_{3}-k_{1}}{k_{3}\| U(k)\|^{2}\|\phi_a(k)\|^{2}
+k_{1}\|U(k)^{T}D(k)\|^{2}\|x(k)\|^{2}}
\end{equation}

\end{myTheo}

\begin{myProof}
Collect all terms of $\Delta L(k)$ based on the results of Lemmas \ref{Lemma1}-\ref{Lemma4}, $\Delta L(k)$ is bounded by

\small
\begin{equation}
\begin{aligned}
&\Delta L(k) = \Delta L_{1}(k)+\Delta L_{2}(k)+\Delta L_{3}(k)+\Delta L_{4}(k)\\
&\leq  l_c e_{c}^2(k)\|\phi_c(k)\|^2 + \|e_{c}(k)-\tilde{w}^{(2)}_{c}(k)\phi_c(k)\|^2
- \Vert \zeta_{c}(k)\Vert^2-e^{2}_{c}(k)\\
&+\frac{1}{k_{1}}
\Big(2l_a e^{2}_{a}(k)\| C(k)\|^{2}\|\phi_a(k)\|^{2}+4\|c(k)\|^{2}+8\gamma^{2}\left\|\zeta_{c}(k+1)\right\|^{2}+8\gamma^{2}\left\|w^{(2)*}_{c}\phi_{c}(k+1)\right\|^{2}\\
&+\|C(k)\|^{2}\|\zeta_{a}(k)\|^{2} -e^{2}_{a}(k)\Big)+\frac{\lambda}{k_{1}}\Big(2\lambda l_{a}e^{2}_{s}(k)\Vert U(k)\Vert^{2}\Vert \phi_{a}(k)\Vert^{2}+2e^{2}_{s}(k)+ \left\|U(k)^{T}\zeta_a(k)\right\|^{2}-e^{2}_{s}(k)\Big)\\
&+\frac{1}{k_{2}}\Big(l_c e^{2}_{c}(k)\|a(k)\|^{2}\|y(k)\|^{2} +\Vert a(k)\Vert^{2}\Vert\tilde w^{(1)}_c(k)y(k)\Vert^{2}+e^{2}_{c}(k)\Big)\\
&+\frac{1}
{k_{3}}\Big(2l_{a} e^{2}_{a}(k)\left\| C(k)^{T} D(k)\right\|^{2}\|x(k)\|^{2}+ e^{2}_{a}(k)+\left\| C(k)^{T} D(k)\right\|^{2}\left\|\tilde w^{(1)}_{a}(k)\, x(k)\right\|^{2}\Big)\\
&+\frac{\lambda}{k_{3}}\Big(2\lambda l_{a} e^{2}_{s}(k)\left\| U(k)^{T} D(k)\right\|^{2}\|x(k)\|^{2}+ e^{2}_{s}(k)+\left\| U(k)^{T} D(k)\right\|^{2}\left\|\tilde w^{(1)}_{a}(k)\, x(k)\right\|^{2}\Big)
\end{aligned}
\end{equation}


\normalsize

Collecting terms associated with $\zeta_c(k)/\zeta_c(k+1)$, $e_c(k)$, $e_a(k)$, and $e_s(k)$, we obtain:


\begin{equation}\label{Delta_L_k}
\begin{aligned}
\Delta L(k) \leq&
\underbrace{-\Big(\|\zeta_c(k)\|^{2}-\frac{8\gamma^{2}}{k_{1}}\|\zeta_c(k+1)\|^{2}\Big)}_{\text{term} 1}
-\underbrace{\Big(1 - l_c\|\phi_c(k)\|^{2}-\frac{l_c}{k_{2}}\|a(k)\|^{2}\|y(k)\|^{2}
- \frac{1}{k_{2}}\Big) e^{2}_{c}(k)}_\text{{term} 2}\\
&-\underbrace{\Big(\frac{1}{k_{1}}- \frac{2l_a}{k_{1}}\|C(k)\|^{2}\|\phi_a(k)\|^{2}
- \frac{2l_a}{k_{3}}\| C(k)^{T}D(k)\|^{2}\|x(k)\|^{2}- \frac{1}{k_{3}}
\Big)e^{2}_{a}(k)}_{\text{term} 3}\\
&-\underbrace{\left(\frac{\lambda}{k_{1}}-\frac{2\lambda^{2}l_{a}}{k_{1}} \Vert U(k)\Vert^{2}\Vert \phi_{a}(k)\Vert^{2}-\frac{2\lambda^{2}l_{a}}{k_{3}}\Vert U(k)^{T} D(k) \Vert^{2}\Vert x(k)\Vert^{2}-\frac{\lambda}{k_{3}}\right)e^{2}_{s}(k)}_{\text{term} 4}\\
&+\underbrace{\big\|w^{(2)*}_{c}\phi_{c}(k)- c(k)-\gamma \hat{w}^{(2)}_{c}(k)\phi_{c}(k+1)\|^2}_{\text{term} 5}+4\Vert c(k) \Vert^{2}+\frac{8\gamma^{2}}{k_{1}}\|w_{c}^{(2)*}\phi_c(k+1)\|^{2}\\
&+\frac{1}{k_{1}}\|C(k)\|^{2}\|\zeta_a(k)\|^{2}\\
&+2\frac{\lambda}{k_{1}}e^{2}_{s}(k)+\frac{\lambda}{k_{1}}\Vert U(k)^{T}\zeta_{a}(k)\Vert^{2}+\frac{\lambda}{k_{3}}\Vert U(k)^{T}D(k)\Vert^{2}\Vert\tilde{w}^{(1)}_{a}x(k)\Vert^{2}\\
&+\frac{1}{k_{2}}\|a(k)\|^{2}\|\tilde w^{(1)}_c(k)y(k)\|^{2}\\
&+\frac{1}{k_{3}}\|C(k)^{T}D(k)\|^{2}\|\tilde w^{(1)}_a(k)x(k)\|^{2}
\end{aligned}
\end{equation}

\begin{myRem}
    The bound of temporal smoothness loss $e_{s}(k)$ depends on the sampling time $T$, which can be  small when adopting a fast sampling strategy.
\end{myRem}

To ensure the negativity of term 1:

\begin{equation}
\begin{aligned}
    \frac{k_{1}}{8\gamma^{2}}&>\frac{\|\zeta_c(k+1)\|^{2}}{\|\zeta_c(k)\|^{2}}\label{inequality_gamma_alpha}
\end{aligned}
\end{equation}

To find an upper bound of $\frac{\|\zeta_c(k+1)\|^{2}}{\|\zeta_c(k)\|^{2}}$, we introduce an assumption \cite{Jenkins201856SIAM}:

\begin{myAss}
The regressor vector $\phi_{c}(x)$ is persistently exciting and bounded away from zero, which satisfies a non-degeneracy condition such that there exists $\kappa\geq0$ for which
\begin{equation}
    \Vert \tilde{w}^{(2)}_{c}(k)\phi_{c}(x(k))\Vert\geq \kappa\Vert \tilde{w}^{(2)}_{c}(k)\Vert\Vert\phi_{c}(x(k))\Vert, \forall k \in \mathbb{N}_{+}
\end{equation}
\end{myAss}

\begin{myRem}
    This assumption is made to ensure a non-degenerate learning signal from the regressor, such that the parameter update does not vanish in any direction and sufficient excitation is guaranteed for stable and convergent weight adaptation. In practice, a sinusoidal reference signal is applied to the system, and a disturbance is injected into the actuator input to ensure that this condition is satisfied.
\end{myRem}

Therefore

\begin{equation}
\begin{aligned}
    \frac{\|\zeta_c(k+1)\|^{2}}{\|\zeta_c(k)\|^{2}}&=\frac{\Vert\tilde{w}^{(2)}_{c}(k)\phi_{c}(\hat{x}(k+1))\Vert^{2}}{\Vert\tilde{w}^{(2)}_{c}(k)\phi_{c}(x(k))\Vert^{2}}\\
    &\leq \frac{\Vert \tilde{w}^{(2)}_{c}(k)\Vert^{2}\Vert\phi_{c}(\hat{x}(k+1))\Vert^{2}}{\kappa^{2}\Vert \tilde{w}^{(2)}_{c}(k)\Vert^{2}\Vert\phi_{c}(x(k))\Vert^{2}}\\
    &\leq \frac{\Vert\phi_{c}(\hat{x}(k+1))\Vert^{2}}{\kappa^{2}\Vert\phi_{c}(x(k))\Vert^{2}}\\
    &\leq \frac{\overline{\phi}^{2}}{\kappa^{2}\underline{\phi}^{2}}
\end{aligned}
\end{equation}\\
where $\underline{\phi}^{2}\leq \Vert \phi_{c}(x(k)) \Vert^{2}\leq \overline{\phi}^{2}$, $\forall k \in \mathbb{N}_{+}$. 

Connecting back to \eqref{inequality_gamma_alpha}:

\begin{equation}
\begin{aligned}
   k_{1}&>\frac{8\overline{\phi}^{2}}{\kappa^{2}\underline{\phi}^{2}}\gamma^{2}
\end{aligned}
\end{equation}

To ensure negativity of the term 2:

\begin{equation}
1 - l_c\|\phi_c(k)\|^{2}- \frac{l_c}{k_{2}}\|a(k)\|^{2}\|y(k)\|^{2}- \frac{1}{k_{2}} > 0
\end{equation}

Therefore,
\begin{equation}\label{lrc}
l_{c} < \min_{k} \frac{k_{2}-1}{k_{2}\big(\|\phi_c(k)\|^{2}
+ \frac{1}{k_{2}}\|a(k)\|^{2}\|y(k)\|^{2}\big)}
\end{equation}
In particular, $k_{2}>1$.

Similarly, to ensure negativity of term 3:
\begin{equation}
\frac{1}{k_{1}}
- \frac{2l_a}{k_{1}}\| C(k)\|^{2}\|\phi_a(k)\|^{2}
- \frac{2l_a}{k_{3}}\|C(k)^{T}D(k)\|^{2}\|x(k)\|^{2}
- \frac{1}{k_{3}} > 0
\end{equation}

Therefore,

\begin{equation}\label{lra}
l_{a} < \min_{k} \frac{1}{2}\frac{k_{3}-k_{1}}{k_{3}\| C(k)\|^{2}\|\phi_a(k)\|^{2}+k_{1}\|C(k)^{T}D(k)\|^{2}\|x(k)\|^{2}}
\end{equation}

In particular, $k_{3}>k_{1}$.

To ensure negativity of term 4:

\begin{equation}
    \frac{\lambda}{k_{1}}-\frac{2\lambda^{2}l_{a}}{k_{1}} \Vert U(k)\Vert^{2}\Vert \phi_{a}(k)\Vert^{2}-\frac{2\lambda^{2}l_{a}}{k_{3}}\Vert U(k)^{T} D(k) \Vert^{2}\Vert x(k)\Vert^{2}-\frac{\lambda}{k_{3}}>0
\end{equation}

Therefore,

\begin{equation}\label{lra_ts}
l_{a} < \min_{k} \frac{1}{2\lambda}\frac{k_{3}-k_{1}}{k_{3}\| U(k)\|^{2}\|\phi_a(k)\|^{2}
+k_{1}\|U(k)^{T}D(k)\|^{2}\|x(k)\|^{2}}
\end{equation}

Term 5 can be bounded as
\begin{equation}
\begin{aligned}
\big\| w_{c}^{(2)*}\phi_c(k) - c(k)
-\gamma \hat w^{(2)}_c(k)\phi_c(k+1)
\big\|^{2}
&\leq 4\|w_{c}^{(2)*}\phi_c(k)\|^{2}
+ 4c^{2}(k) + 2\gamma^{2}\|\hat w^{(2)}_c(k)\phi_c(k+1)\|^{2}
\end{aligned}
\end{equation}

Let $\overline{C},\overline{U}, \overline{w}_{a,1}, \overline{w}_{a,2}, \overline{w}_{c,1}, \overline{\phi}_a,\overline{\phi}_c, \overline{y}, \overline{x}, \overline{a}, \overline{D}, \overline{e}_{s}$ be upper bounds of $C$, $U$, $\tilde{w}^{(1)}_{a}$, $\tilde{w}^{(2)}_{a}$, $\tilde{w}^{(1)}_{c}$, $\phi_a$, $\phi_c$, $y$, $x$, $a$, $D$, $e_{s}$, and let $\overline{w}_{c,2} = \max\{w_{c}^{(2)*},\, \hat{w}^{\text{max}}_{c,2}\}$, where $\hat{w}^{\text{max}}_{c,2}$ is the upper bound of $\hat w^{(2)}_c(k)$. Finally, we obtain the following bound from \eqref{Delta_L_k}:
\small
\begin{equation}\label{upper_bounds}
\begin{aligned}
&\big\| w_{c}^{(2)*}\phi_c(k)-c(k)-\gamma\hat w^{(2)}_c(k)\phi_c(k+1)\big\|^{2}+4\Vert c(k)\Vert^{2}+\frac{8\gamma^{2}}{k_{1}}\|w_{c}^{(2)*}\phi_c(k+1)\|^{2}
+ \frac{1}{k_{1}}\|C(k)\|^{2}\|\zeta_a(k)\|^{2}\\
&\quad + \frac{1}{k_{2}}\|a(k)\|^{2}\|\tilde w^{(1)}_c(k)y(k)\|^{2}+\frac{1}{k_{3}}\|C(k)^{T}D(k)\|^{2}\|\tilde w^{(1)}_a(k)x(k)\|^{2}\\
&\quad+2\frac{\lambda}{k_{1}}e^{2}_{s}+\frac{\lambda}{k_{1}}\Vert U(k)^{T}\zeta_{a}(k)\Vert^{2}+\frac{\lambda}{k_{3}}\Vert U(k)^{T}D(k)\Vert^{2}\Vert\tilde{w}^{(1)}_{a}(k)x(k)\Vert^{2}\\
\leq& \Big(\frac{8\gamma^{2}}{k_{1}}+4+2\gamma^{2}\Big)(\overline{w}_{c,2}\overline{\phi}_c)^{2}+ 8\overline{c}^{2}+ \frac{1}{k_{1}}(\overline{C}\overline{w}_{a,2}\overline{\phi}_a)^{2}+ \frac{1}{k_{2}}(\overline{w}_{c,1}\overline{y}\overline{a})^{2}
+ \frac{1}{k_{3}}(\overline{C}\overline{D} \overline{w}_{a,1}\overline{x})^{2}\\
&\quad +2\frac{\lambda}{k_{1}}\overline{e}^{2}_{s}+\frac{\lambda}{k_{1}}(\overline{U}\overline{w}_{a,2}\overline{\phi}_{a})^{2}+\frac{\lambda}{k_{3}}(\overline{U}\overline{D} \overline{w}_{a,1}\overline{x})^{2}
= M
\end{aligned}
\end{equation}
\normalsize
The conclusion is, if $k_{1}>\frac{8\overline{\phi}^{2}}{\kappa^{2}\underline{\phi}^{2}}\gamma^{2}$, and $l_c$, $l_a$ satisfying \eqref{lrc}, \eqref{lra}, \eqref{lra_ts} and $\|\zeta_c(k)\|^{2}-\frac{8\gamma^{2}}{k_{1}}\|\zeta_c(k+1)\|^{2}>M$, we obtain $\Delta L(k) < 0$.  
This implies that the estimation errors are ultimately uniformly bounded based on Theorem \ref{UUB_theorem}.
\end{myProof}

\begin{myRem}
    The constraints on the actor learning rate consist of two components: the first arises from the approximate Bellman loss, while the second is induced by the TS loss. When the TS loss is omitted, it results in relaxed learning-rate bounds. This relaxed bound can be regarded as the actor learning-rate constraint for IHDP required to ensure UUB convergence, although it has not been explicitly discussed in previous studies \cite{ZhouIAC2017,ZhouIMAV2016,ZHOUAST2020105}.
\end{myRem}

\begin{myRem}
    The numerical bounds $M$ and learning rates constraints \eqref{lc_constraint}\eqref{condition_la_1}\eqref{condition_la_2} are difficult to be verified numerically for a cascaded controller in the presence of a filter. This requires future investigation on simple nonlinear models, as used in \cite{Vamvoudakis2010878,Bhasin201382}.
\end{myRem}

\section{Action-Increment Constrained Policy Optimization}\label{section_primal_dual_optimization}




\textit{Motivation problem.} In Section~\ref{section_TS_IHDP}, we have incorporated a penalty on temporal action increment into the policy objective $E_{a}(k)$, an approach also adopted in CAPS \cite{Mysore2021ICRA}. However, in many previous studies, this penalty is empirical, and there is no well-defined metric for quantifying the required policy smoothness for a specific control task. Therefore, in this section, we consider using a prescribed smoothness criterion based on practical control rate constraints, and adaptively tune the penalty weight, i.e., $\lambda(k)$ adjusts according to $g(k)-\varepsilon$. 

To this end, we introduce an upper bounded action increment constraint into the constrained policy optimization (CPO) formulation \cite{PaternainTAC2023}. This leads to the primal optimization problem:





\begin{equation}\label{constrained_policy_optimization}
\begin{aligned}
P^{*}_{\hat{w}_{a}(k)}\triangleq\min_{\hat{w}_a(k)}\ &E_{a,1}(k) \triangleq \frac{1}{2}\Vert c(k) +\gamma\hat{V}(k+1)\Vert^{2}\\
    &\text{s.t.}\; g(k)\triangleq\underbrace{\frac{1}{8}\Vert u(k)-\hat{u}(k+1)\Vert^{4}\leq\varepsilon}_{\text{control smoothness constraint}}  
\end{aligned}
\end{equation}\\
where $g(k)$ is a function of temporal action increment, $\varepsilon>0$ is a prescribed smoothness threshold designed by actuator rate constraints. Section \ref{section_simulation} will specify the values of $\varepsilon$ based on the constraints on pitch-rate commands and control-surface deflections.

Similar to CPO \cite{PaternainTAC2023}, the action-increment-constrained policy optimization (AI-CPO) \eqref{constrained_policy_optimization} can be reformulated to an unconstrained optimization problem using the Lagrangian approach. The resulting Lagrangian is 

\begin{equation}
    \mathcal{L}(\hat{w}_{a}(k),\lambda(k))\triangleq E_{a,1}(k)+\lambda (k) (g(k)-\varepsilon)
\end{equation}\\
where $\lambda (k)$ is the multiplier updated at each $k$. Then the dual function to be optimized is defined as

\begin{equation}\label{dual_function}
    d_{\hat{w}_{a}(k)} \triangleq \min_{\hat{w}_{a}(k)} \mathcal{L}(\hat{w}_{a}(k),\lambda(k))
\end{equation}

An upper bound exists for the dual function \eqref{dual_function} for any $\lambda(k)$, i.e., $P^{*}_{\hat{w}_{a}(k)}\leq d_{\hat{w}_{a}(k)}(\lambda(k))$. The next problem is how to find the $\lambda(k)$ that enables the tightest upper bound, which defines the following dual problem:

\begin{equation}    
D^{*}_{\hat{w}_{a}(k)}\triangleq\max_{\lambda(k)>0}\min_{\hat{w}_{a}(k)}\mathcal{L}(\hat{w}_{a}(k),\lambda(k)) 
\end{equation}

The update of actor parameters $\hat{w}_{a}(k)$ aligns with the update laws in \eqref{equation_wa2_update_rule} and \eqref{equation_action_wa1_update}. The only difference is that, in this section, $\lambda(k)$ is allowed to vary online with the time step $k$. The update of  $\lambda(k)$ is by



\begin{equation}\label{update_law_lambda}    \lambda(k+1)=\left[\lambda(k)+\eta_{\lambda} (g(k)-\varepsilon)\right]_{+}
\end{equation}\\
where $\eta_{\lambda}>0$ is the update rate, $[\cdot]_{+}$ ensures positivity.

According to Equation \eqref{update_law_lambda}, the weight $\lambda(k)$ is increased to reinforce the penalty when the action increment exceeds the prescribed threshold $\varepsilon$, and is decreased to relax the penalty when the action increment remains below the threshold $\varepsilon$. This reflects a key principle of safe RL: using constraint-violation information to guide policy optimization toward constraint satisfaction.

\section{Online Learning Flight Control Design}\label{section_online_learning_flight_control_design}

In this section, online learning flight control is designed using TS-IHDP. A cascaded actor network is adopted, consisting of an outer-loop actor and an inner-loop actor for hierarchical tracking control tasks (see Figure \ref{Outer_loop_actor_training}). The outer-loop actor generates a pitch-rate reference, while the inner-loop actor performs actuator-level tracking control. A low-pass filter is applied to the pitch-rate reference to ensure smoothness, similar to command-filtered backstepping \cite{LunguAST2020105526,ZhangAST2024109552}.

\subsection{Dynamics model}
Consider a nonlinear longitudinal (pitch plane) dynamics model of the air vehicle \cite{Hull19951ACC}: 

\begin{equation} \label{aerialvehiclemodel}
\begin{aligned}
    \dot{\alpha} &= \left( \frac{fgQS}{WV}\right)\cos(\alpha)[\phi_{z}(\alpha)+b_{z}\delta] + q\\
    \dot{q} &= \left(\frac{fQSd}{I_{yy}}\right)[\phi_{m}(\alpha)+b_{m}\delta]
\end{aligned}
\end{equation}\\
where the state vector $[\alpha,q]^{T}$ consists of the angle of attack and the pitch rate, $\delta$ is the control-surface deflection. The aerodynamic coefficients $\phi_{z}(\alpha),\phi_{m}(\alpha),b_{z},b_{m}$ and physics coefficients can be found in Appendix \ref{appendix_aerodynamic_coefficients} and \cite{Hull19951ACC}. This generic dynamics model is commonly used for autopilot design and verification \cite{HullACC1997611921,KimCEP2004149}, making it suitable for a clear evaluation of the proposed learning-based controller. Assuming the measurements of states are noise-free, the tracking errors are defined as
$e_{1}=\alpha-\alpha_{\text{ref}}$, $e_{2}=q-q_{\text{ref}}$, where $\alpha_{\text{ref}}$, $q_{\text{ref}}$ are reference commands. 

The tracking error dynamics are obtained as

\begin{equation}\label{tracking_error_model}
\begin{aligned}   &\dot{e}_{1}=f_{1}(\alpha)+e_{2}+q_{\text{ref}}+d_{1}(\alpha,\delta)-\dot{\alpha}_{\text{ref}}\\
&\dot{e}_{2} = f_{2}(\alpha) + d_{2}(\alpha,\delta)-\dot{q}_{\text{ref}}
\end{aligned}
\end{equation}\\
where the nonlinear functions are defined as

\small
\begin{equation*}
\begin{aligned}
    f_{1}(\alpha) &=\left(\frac{fgQS}{WV}\right)\cos(\alpha)\phi_{z}(\alpha),\;
    d_{1}(\alpha,\delta) = \left(\frac{fgQS}{WV}\right)\cos(\alpha)b_{z}\delta\\
    f_{2}(\alpha) &=  \left(\frac{fQSd}{I_{yy}}\right)\phi_{m}(\alpha),\qquad \quad 
    d_{2}(\alpha,\delta) = \left(\frac{fQSd}{I_{yy}}\right)b_{m}\delta
\end{aligned}
\end{equation*}
\normalsize

\subsection{Incremental model}

The incremental models of \eqref{aerialvehiclemodel} are obtained by linearization using Taylor expansion:


\begin{equation}\label{incremental_models_vehicle}
\begin{aligned}
    \Delta \alpha(k+1) =&\Delta \alpha(k)+ T\Big(F_{1,1}(k-1)\Delta \alpha(k)\\
    &+F_{1,2}(k-1)\Delta q(k)+G_{1}(k-1)\Delta \delta(k)\Big)\\
    \Delta q(k+1) =& \Delta q(k)+ T\Big(F_{2}(k-1)\Delta \alpha(k)\\
    &+G_{2}(k-1)\Delta \delta(k)\Big)
\end{aligned}
\end{equation}\\
where $\Delta \alpha(k)= \alpha(k)- \alpha(k-1)$, $\Delta q(k)= q(k)- q(k-1)$, $\Delta \delta(k)= \delta(k)- \delta(k-1)$. The model parameters correspond to the Jacobian of the nonlinear model evaluated at $(\alpha,\delta,q)=(\alpha(k-1),\delta(k-1),q(k-1))$, i.e. $F_{1,1}(k-1)=\frac{\partial (f_{1}+d_{1})}{\partial \alpha}$, $F_{1,2}(k-1)=\frac{\partial (f_{1}+d_{1})}{\partial q}=1$, $G_{1}(k-1)=\frac{\partial (f_{1}+ d_{1})}{\partial \delta}$, $F_{2}(k-1)=\frac{\partial (f_{2}+d_{2})}{\partial \alpha}$, $G_{2}(k-1)=\frac{\partial (f_{2}+d_{2})}{\partial \delta}$.




\subsection{Flight control design}

\subsubsection{Cascaded control structure}

\begin{figure}
    \centering
   \includegraphics[width=0.6\linewidth]{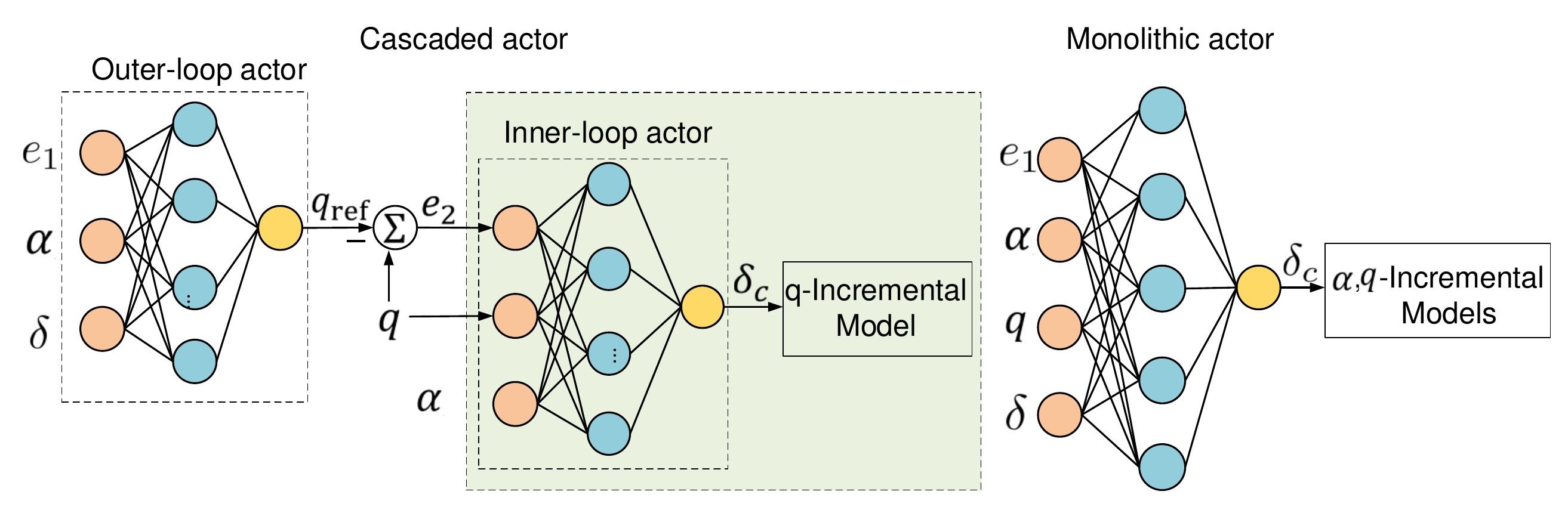}
    \caption{Cascaded actor versus monolithic actor. The former employs an outer-loop actor to generate a pitch-rate reference, which is then passed to an inner-loop actor that produces the control-surface command, providing an explicit mechanism for learning and tracking the pitch-rate reference. The latter is a fully connected MLP that approximates an end-to-end control policy and does not explicitly learn a pitch-rate reference; it is commonly used for ADP-based nonlinear systems stabilization \cite{Sokolov20159} and tracking control \cite{ModaresAuto20141780}.}\label{Outer_loop_actor_training}
\end{figure}

The outer-loop and inner-loop control laws are parameterized by $q_{\text{ref}} = W_{1}(e_{1},\alpha,\delta) $ and $\delta_{c} = W_{2}(e_{2},q,\alpha)$, respectively. The nested control law is 

\begin{equation}\label{nested_control_law}
    \delta_{c}(e_{2},q,\alpha) = \underbrace{W_{2}\big(q-\underbrace{W_{1}(e_{1},\alpha,\delta)}_{\text{outer-loop actor}},q,\alpha\big)}_{\text{inner-loop actor}} 
\end{equation}\\
where $W_{1}(\cdot;\hat{w}_{a_{1}}), W_{2}(\cdot;\hat{w}_{a_{2}})$ are neural networks with trainable weights $\hat{w}_{a_{1}},\hat{w}_{a_{2}}$. $\delta_{c}$ denotes the command for control-surface deflection. The cascaded actor is illustrated in Figure \ref{Outer_loop_actor_training}, along with a comparison to the monolithic actor. 

\begin{figure}
    \centering
    \includegraphics[width=0.7\linewidth]{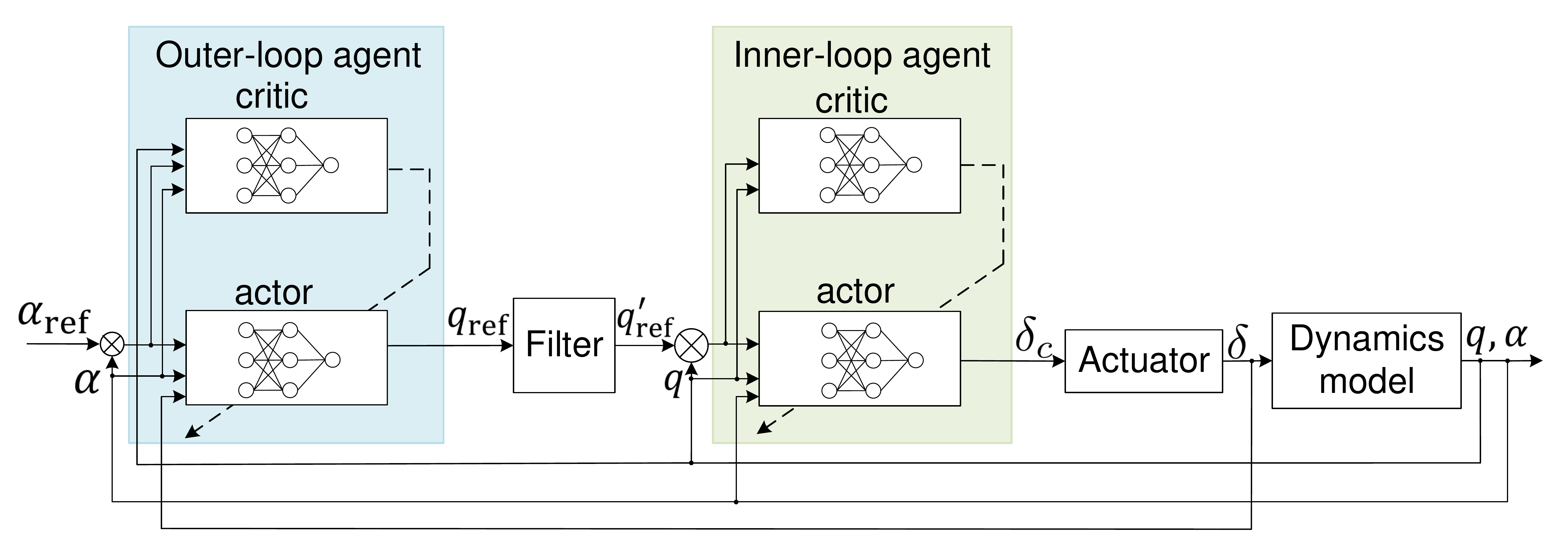}
    \caption{Diagram of command-filtered cascaded online learning flight control.} \label{cascaded_online_learning_structure} 
\end{figure}

\subsubsection{Outer-loop agent}
A pitch rate reference policy is learned. The critic input takes $[e_{1}(k),\alpha(k)]^{T}$ and the output is the estimated state value $\hat{V}_{1}(k)$. The actor input is $[e_{1}(k),\alpha(k),\delta(k)]^{T}$ and the output is $q_{\text{ref}}(k)$. The critic update is via temporal-difference (TD) learning \cite{Sutton2018RL} to minimize $E^{C_{1}}(k) = \frac{1}{2}e_{c,1}^{2}(k)$, where the TD error $e_{c,1}(k) = \hat{V}_{1}(k)-c_{1}(k)- \gamma \hat{V}^{\prime}_{1}(k+1)$. The one-step cost is $c_{1}(k)=e_{1}^{2}(k)+aq_{\text{ref}}^{2}(k)$, where $a>0$ trades off between tracking error and control effort. $\hat{V}^{\prime}_{1}(\cdot)$ denotes the target critic, which employs delayed weight updates to stabilize learning \cite{Li2026AST112516}.

The multi-objective policy optimization is formulated as 


\begin{equation}
\hat{w}^{*}_{a1}(k)
= \arg\min_{\hat{w}_{a1}}
\Big(\frac{1}{2}e^{2}_{a,1}(k)
+ \frac{1}{2}\lambda_{1}(k)e^{2}_{s,1}(k)\Big)
\end{equation}\\
where $e_{a,1}(k)=c_{1}(k)+ \gamma \hat{V}^{\prime}_{1}(k+1)$ and the temporal smoothness loss $e_{s,1}(k) = \frac{1}{2}\left(q_{\text{ref}}(k)-\hat{q}_{\text{ref}}(k+1)\right)^{2}$. 

\subsubsection{Inner-loop agent}

A control-surface deflection command policy is learned. An issue lies in that the non-minimum-phase dynamics associated with $\delta$ may inadvertently affect the $\alpha$-dynamics, as presented in \eqref{aerialvehiclemodel}. This motivates the adoption of a cascaded control structure, which isolates this effect from the inner-loop agent’s training. As a result, the outer-loop agent is prevented from attempting to control the angle-of-attack directly by using the control-surface deflection.

The critic input and output are $[e_{2}(k),q(k)]^{T}$, $\hat{V}_{2}(k)$. The actor input and output are $[e_{2}(k),q(k),\alpha(k)]^{T}$, $\delta_{c}(k)$. The critic is trained to minimize $E^{C_{2}}(k) = \frac{1}{2}e_{c,2}^{2}(k)$, where $e_{c,2}(k) = \hat{V}_{2}(k)-c_{2}(k)- \gamma \hat{V}^{\prime}_{2}(k+1)$ is the TD error. The one-step cost is $c_{2}(k)=e_{2}^{2}(k)+b\delta_{c}^{2}(k)$, $b>0$. $\hat{V}^{\prime}_{2}$ is the target critic. 

The multi-objective policy optimization is formulated as

\begin{equation}\label{policy_optimization2}
\hat{w}^{*}_{a2}(k)
= \arg\min_{\hat{w}_{a2}}
\Big(\frac{1}{2}e^{2}_{a,2}(k)
+ \frac{1}{2}\lambda_{2}(k)e^{2}_{s,2}(k)\Big)
\end{equation}\\
where $e_{a,2}(k)=c_{2}(k)+ \gamma \hat{V}^{\prime}_{2}(k+1)$ and the temporal smoothness loss $e_{s,2}(k) = \frac{1}{2}\left(\delta_{c}(k)-\hat{\delta}_{c}(k+1)\right)^{2}$. $\hat{\delta}_{c}(k+1)$ is the output of the actor given the predicted states as input.

\subsubsection{Training}


For the outer-loop actor, it interacts with the $\alpha$-incremental model for critic and actor training. The output action $q_{\text{ref}}$ is passed to the inner-loop actor as an input, which then generates the control signal $\delta$ to interact with the $q$-incremental model. Since these two incremental models are independent of each other, the two agents are also trained separately. Figure~\ref{Outer_loop_actor_training} illustrates the structure of the cascaded actor.


\subsubsection{Command filter}

A second-order low-pass filter is used to smooth the pitch rate reference: 

\begin{equation}\label{low_pass_filter}
\begin{aligned}
    \dot{d}_{1} &= d_{2}\\
    \dot{d}_{2}&= -2\zeta\omega_{n} d_{2}-\omega^{2}_{n}(d_{1}-q_{\text{ref}})
\end{aligned}
\end{equation}\\
where $d_{1},d_{2}$ are filtered signal and its differentiation, $\omega_{n}$ is the natural frequency, $\zeta$ is the damping factor. To integrate the filter into the cascaded flight control design, it takes $d_{1}=q^{\prime}_{\text{ref}}$ and $d_{2}=\dot{q}^{\prime}_{\text{ref}}$ as filter outputs, while only $d_{1}$ is fed to the inner-loop actor, as shown in Figure \ref{cascaded_online_learning_structure}. Here, $q^{\prime}_{\text{ref}}$ denotes the filtered pitch rate reference.

\begin{myRem}
   What is the low-pass filter's non-Markov effect in our online learning method?
   
   The dynamics in \eqref{aerialvehiclemodel} are assumed to satisfy the Markov property, i.e., the one-step-ahead state depends only on the current state and action \cite{Sutton2018RL}. However, the second-order low-pass filter dynamics in \eqref{low_pass_filter} make the one-step-ahead reference $q^{\prime}_{\text{ref}}(k+1)$ depend on its values at the two preceding time steps, making $q^{\prime}_{\text{ref}}$ a non-Markovian variable. As a result, the overall dynamical system is non-Markovian, and the Markov property can be recovered by augmenting the state fed to NNs with filter states \cite{WangICLR2024}, which would increase the computation cost and learning complexity. In practice, this effect is mitigated by the short sampling time (0.001s), since the filtered reference evolves slowly relative to the sampling rate. In our method, the agents interact with two incremental models \eqref{incremental_models_vehicle} decoupled via the cascaded actor structure, instead of the true system dynamics, thereby preserving the Markov property in both the incremental models and the identification process. For the outer-loop agent, the non-Markovian variable $q^{\prime}_{\text{ref}}$ is not involved. For the inner-loop agent, the tracking error $e_{2}=q-q^{\prime}_{\text{ref}}$ is non-Markovian, as its transition depends on the values of $q^{\prime}_{\text{ref}}$  from the previous two time steps. Owing to the hierarchical control structure, in which the $\alpha$-to-$q_{\text{ref}}$ mapping is generated by the outer-loop actor and the $q_{\text{ref}}$-to-$q^{\prime}_{\text{ref}}$ mapping is generated by the known filter dynamics (see Figure \ref{Outer_loop_actor_training}) and are decoupled from the training of the inner-loop agent, $q^{\prime}_{\text{ref}}$ serves as an external signal to the inner-loop agent. Consequently, no gradient is propagated through $q^{\prime}_{\text{ref}}$ to $\delta$, and its non-Markovian property has no effect on the training of either the inner-loop critic or the inner-loop actor. Finally, the same principle applies to both first-order and second-order filters. 
   
   To summarize, although \cite{Mysore2021ICRA} shows that filtering a post-training policy may cause nonvanishing oscillations, our method integrates a filter during both the training and operation phases, which avoids this disadvantage. Moreover, unlike the filter-state augmentation method in \cite{WangICLR2024}, which increases implementation complexity, our training method does not require augmenting the state with filter variables, owing to the hierarchical structure of the cascaded learning tasks.
\end{myRem}

\section{Simulations}\label{section_simulation}

\subsection{Setup}\label{section_simulation_setup}
The air vehicle dynamics model \eqref{aerialvehiclemodel} is numerically simulated through the fourth-order Runge Kutta method with time step 0.01s. The actuator dynamics are modeled as a first-order system $\tau \dot{\delta}(t)+\delta(t)=\delta_{c}(t), \tau =0.005$. To constrain actions, the outer-loop and inner-loop actors are scaled to $[-20^{\circ},20^{\circ}]$ and $[-20^{\circ}/\text{s},20^{\circ}/\text{s}]$. The reference signal is set as $\alpha_{\text{ref}}(t) = 10^{\circ}\sin(\frac{2\pi}{T_{\text{ref}}}t), T_{\text{ref}}=10\text{s}$. The network weights are randomly initialized using a uniform distribution $\mathcal{U}(-0.01,0.01)$, ensuring an initially low-gain control law. Also, this initialization prevents excessive sensitivity to inaccurate policy gradients during the early stages of training. All RL agents start with same initial NN weights to ensure a fair comparison. Under Assumption \ref{Assumption_persistent_excitation}, sufficient excitation of the system inputs is essential for both incremental model identification and NN training. Although the time-varying reference $\alpha_{\text{ref}}(t)$ provides a significant level of excitation, we use a complementary band-limited multi-sine excitation signal $\delta_{\text{exc}}(t)=\omega(t)\sum^{3}_{i=1}A_{i}\sin(2\pi f_{i}t+\phi_{i})$, $(A_{1},A_{2},A_{3})=(0.2^{\circ},0.1^{\circ},0.05^{\circ})$, $(f_{1},f_{2},f_{3})=(1,3,5)$ Hz, $(\phi_{1},\phi_{2},\phi_{3})=(0, 0.3,1.1)$ rad in the control surface deflection to further excite the system dynamics \cite{KleinMorelli2006}. The Hann-type window $\omega(t)=\frac{1}{2}(1+\cos(\frac{\pi t}{10})),0\leq t\leq 10$, and $\omega(t)=0,t>10$ is used to smoothly attenuate the excitation signal to zero in 10s \cite{HarrisIEEE1978}. The excitation signal $\delta_{\text{exc}}(t)$ degrades the temporal smoothness of the actions by disturbing state transitions and perturbing $e_{s,1},e_{s,2}$. However, the resulting smoothness degradation gradually diminishes as the excitation signal decays over time. The baseline control systems are considered as: (1) the TS-IHDP-trained monolithic actor, with the actor learning rate $l_{a}=0.5$, the weight of the smoothness loss $\lambda=0.0002$. This baseline system is compared with the cascaded actor to demonstrate its advantages; (2) the IHDP-trained cascaded actor \cite{ZhouIAC2017}, i.e. $\lambda_{1}=\lambda_{2}=0$. For our developed TS-IHDP for cascaded control system, we select three values of the update rate $\eta_{\lambda,2}$ for the weight $\lambda_{2}$ in the dual update \eqref{update_law_lambda}, denoted as TS-IHDP 1, TS-IHDP 2, and TS-IHDP 3. The rest of hyperparameters are listed in Table \ref{Hyperparameters_of_RL_agents_3.4.1}. The code and datasets used in this study are publicly available in \cite{Li2026FlightControlData}.

\setlength{\tabcolsep}{1pt}
\begin{table}[htbp]
  \centering
  \caption{Hyperparameters of RL agents}
  \begin{threeparttable}
    \begin{tabular}{l|l|l}
    \hline
    Parameter & Outer-loop  & Inner-loop \\
    \hline
    critic learning rate $l_{c1}$, $l_{c2}$ & 0.1 & 0.1\\
    actor learning rates $l_{a1}$, $l_{a2}$  & $1.5\times 10^{-6}$ & $8.5\times 10^{-7}$\\
    discount factor $\gamma$ & 0.6 & 0.6\\
    forgetting factor $\rho$   & 0.9 & 0.9\\
    hidden layer size\tnote{a} & 7 & 7 \\
    critic hidden-layer & tanh & tanh \\
    actor activation function & tanh & tanh \\
    control effort weights $a,b$ & $5\times 10^{-6}$  & $1\times10^{-5}$\\ 
    target critic delay factor & 0.9&0.9\\
    Update rates $\eta_{\lambda,1}, \eta_{\lambda,2}$ & 3000 & 0.0044,0.00448,0.0045\\
    \hline 
    \end{tabular}
    \begin{tablenotes}
    \footnotesize
    \item[a] A small number of hidden-layer neurons (=7) in the actor network is verified through simulations to provide sufficient control accuracy for $\alpha$-tracking and $q$-tracking.
    \end{tablenotes}
 \end{threeparttable}
 \label{Hyperparameters_of_RL_agents_3.4.1}
\end{table}

\subsection{Online training phase}

\subsubsection{Incremental model identification}\label{section_online_identification}

\begin{figure}
    \centering 
    \includegraphics[width=0.6\linewidth]{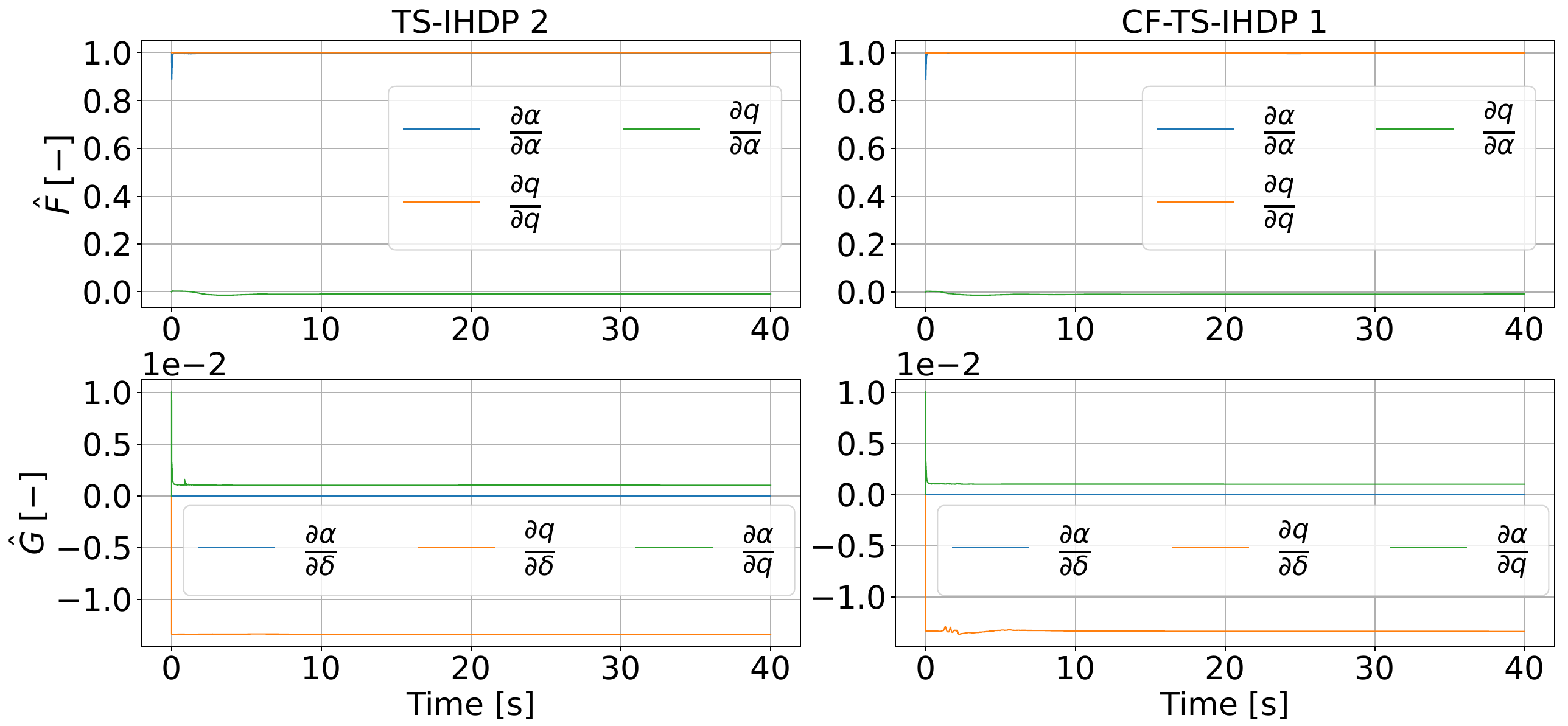}
    \caption{Incremental model parameters.} \label{Fig_incremental_model}
\end{figure}

Figure \ref{Fig_incremental_model} shows that the parameters of the angle-of-attack and pitch-rate incremental models \eqref{incremental_models_vehicle} are identified online within approximately 3s, using TS-IHDP 2 and CF-TS-IHDP 1 as representative examples. As discussed in Section \ref{section_TS_IHDP}, the gradient equations \eqref{equation_action_e1_w2_update} and \eqref{equation_action_e1_w1_update} use these identified parameters to construct policy gradients. Therefore, the identified parameters are expected to have not only the correct signs but also reasonably accurate numerical values to provide meaningful NN updates. It is observed that the identified value of $\frac{\partial q}{\partial \delta}$ is negative, whereas that of $\frac{\partial \alpha}{\partial q}$ is positive. These signs are consistent with the actual system dynamics and therefore contribute to obtaining the correct policy-gradient directions, as reflected in the gradient propagation equations \eqref{equation_action_e1_w2_update} and \eqref{equation_action_e1_w1_update}. Meanwhile, the use of a low-pass filter does not affect the identification results.

\subsubsection{Tracking control performance}

Figure \ref{fig:baselines} shows the TS-IHDP-trained monolithic actor fails to learn an effective and smooth pitch-rate policy, which leads to unsuccessful AoA tracking. The actuator responses aggressively even if the temporal smoothness loss is applied to penalize action variations. This phenomenon can be attributed to the absence of an explicit pitch-rate reference in a monolithic actor network (see Figure \ref{Outer_loop_actor_training}), that makes learning an effective pitch-rate-control policy difficult. Second, the IHDP-trained cascaded actor learns tracking the AoA reference with a large tracking errors (see Figure \ref{fig:TS-IHDP-1-2-3}). Both the outer-loop and inner-loop actors respond aggressively to tracking errors through high-gain and saturated control policies (see Figure \ref{Fig_landscape}). This action saturation phenomenon has also been observed in online flight control using IGDHP for F-16 nonlinear model \cite{SunJGCD2021} and in offline PPO-trained quadrotor flight control \cite{Mysore2021ICRA}. Reference \cite{SunJGCD2021} migrates system oscillations by scaling the actor output within a tight bound $\pm 5^{\circ}$. However, the aggressive learning process can still drive the actor into saturation, resulting in a bang–bang control law.

In contrast, our developed TS-IHDP methods regularize the actor network smoothness in temporal scale and learn control policies that naturally generate actuator actions within $\pm 6^{\circ}$ after sufficient training (within 15s), thereby preventing actor saturation, as shown in Figure \ref{fig_saturation}. This demonstrates that temporal policy regularization moderates policy updates and helps prevent actor saturation compared with the results in \cite{SunJGCD2021}. Specifically, TS-IHDP 2,3 enable smooth actor outputs, while TS-IHDP 1 produces switching and noisy actions, because the update of $\lambda_{2}$ is slower. Since the policy evolution of online IHDP has not been sufficiently presented in  \cite{ZHOUAST2020105}, we re-evaluate the method and present the resulting policy curves in Figure \ref{fig:baselines} for comparison. For convenience, “TS-IHDP” hereafter refers to the cascaded actor trained by TS-IHDP, as the monolithic actor is beyond the scope of the following discussion.

\subsubsection{Primal-dual method for smoothness weight tuning}

In the dual update law \eqref{update_law_lambda}, the initial weights are set as $\lambda_{1}(0)=\lambda_{2}(0)=0$ for outer-loop and inner-loop agents. The thresholds can be selected based on the actuator rate constraints in specific control tasks. In this paper, we set $\varepsilon_{1}=\varepsilon_{2}=1.25\times10^{-9}$, which indicate expected constraints on command increments are $\vert q_{\text{ref}}(k)-\hat{q}_{\text{ref}}(k+1)\vert\leq0.01^{\circ}/\text{s}$ and $\vert \delta_{c}(k)- \hat{\delta}_{c}(k+1)\vert\leq 0.01^{\circ}$. To reduce the influence of severe constraint violations during the exploration phase, which may lead to uninformative updates of $\lambda_{1}(k),\lambda_{2}(k)$, these weights are updated only when the constraint violations remain within a small and informative range, i.e., $g_{1}(k)\leq 0.0001$ and $g_{2}(k)\leq 0.01$.

\begin{figure}[tbp]
    \centering 
\includegraphics[width=0.6\linewidth]{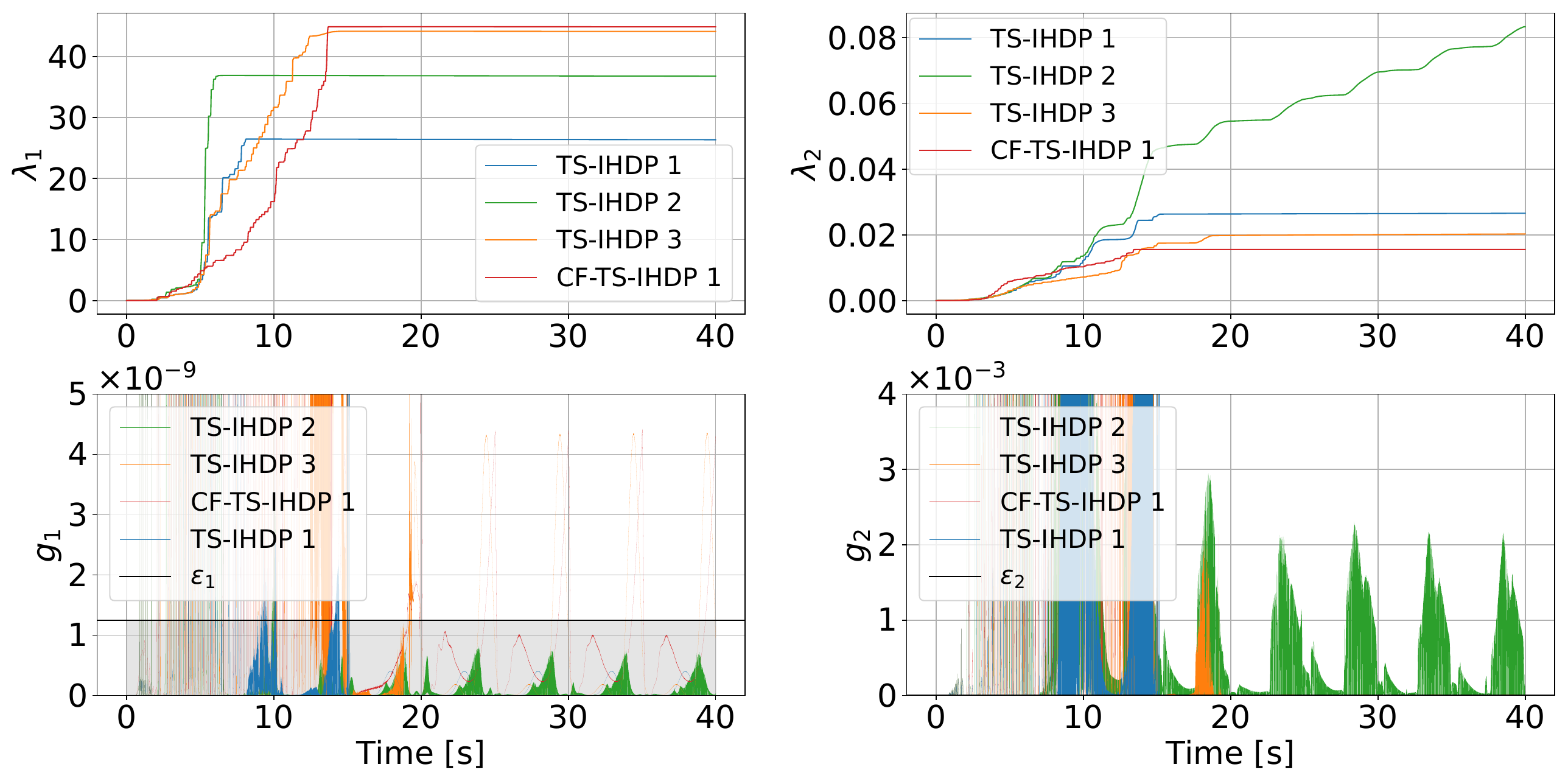}
    \caption{Evolution of smoothness weights. The short peaks after 15s are caused by tracking errors switching near zero. } \label{adaptive_weight}
\end{figure}

Figure \ref{adaptive_weight} shows that $\lambda_{1}(k),\lambda_{2}(k)$ increase during the initial exploration phase (approximately the first 15 s), and remain near constant once $g_{1}(k),g_{2}(k)$ are close to the thresholds $\varepsilon_{1},\varepsilon_{2}$. First, we observe the smallest $\lambda_{1}$ after sufficient growth by TS-IHDP 1, indicating the mildest penalty on $\vert q_{\text{ref}}(k)-\hat{q}_{\text{ref}}(k+1)\vert$, which leads to unsmooth and noisy $q_{\text{ref}}(k)$. Second, we observe a consistent increase in the inner-loop weight $\lambda_{2}(k)$ after 15s by TS-IHDP 2 due to the periodic peaks of $g_{2}(k)$ exceeding $\varepsilon_{2}$. These peaks are the result of $e_{2}$ switching near zero that causes jumping actor outputs. This issue can be mitigated by setting a higher threshold $\varepsilon_{2}$, such that the detected jumpings no longer trigger dual updates, or using a smaller update rate $\eta_{\lambda_{2}}$, as in TS-IHDP 1. Despite these jumpings-based weight updates are uninformative, and $\lambda_{2}(k)$ does not remain near constant after exploration, we still present TS-IHDP 2 as an illustrative case of excessive smoothness regularization of the inner-loop actor. The consequence is that the actor behaviors are overly restricted and hence large tracking errors appear near the peaks of $\alpha_{\text{ref}}$, as shown in Figure~\ref{fig:TS-IHDP-1-2-3-zoom}. This further causes rapidly changing actions in response of tracking errors, and the observed peaks of $g_{1}(k),g_{2}(k)$ in Figure \ref{adaptive_weight}. This suggests that an overly large smoothness penalty can stagnate policy behaviors and paradoxically degrade action smoothness.

\begin{myRem}

What are the advantages of the update law \eqref{update_law_lambda} for a dual variable $\lambda(k)$ over simply using a monotonically increasing or constant one?

The proposed dual update \eqref{update_law_lambda} slows down as the action increment approaches the prescribed threshold. Consequently, $\lambda(k)$ may converge to a value that yields action smoothness close to the desired level. In contrast, a purely increasing update law, i.e., $\varepsilon=0$, has no stopping criterion and may result in inconsistent levels of smoothness depending on the update rate $\eta_{\lambda}$. Likewise, selecting a constant $\lambda$ provides no explicit criterion for evaluating or achieving the desired level of smoothness \cite{Mysore2021ICRA}.  
\end{myRem}

\begin{figure}
    \centering
    \includegraphics[width=0.6\linewidth]{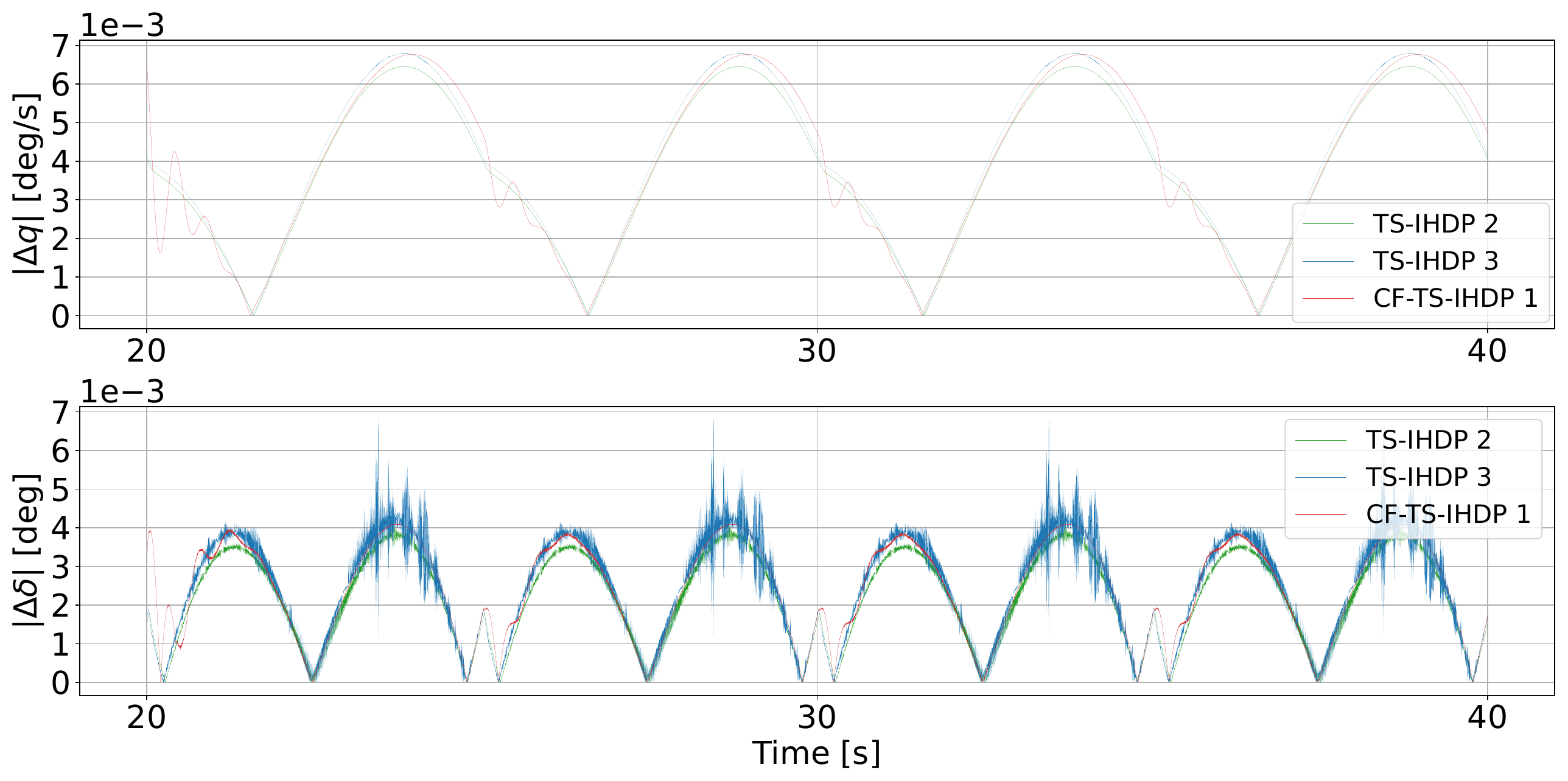}
    \caption{Increments of actions. The presented methods satisfy the constraints $\Delta q(k)\leq0.01^{\circ}/\text{s}$ and $\Delta \delta(k)\leq 0.01^{\circ}$.} \label{action_incement}
\end{figure}

\begin{myRem}
    
What if the command increment does not finally go below the threshold?

The values of thresholds $\varepsilon_{1}$ and $\varepsilon_{2}$ are set significantly smaller than the pitch rate and actuator rate limits, such that they do not substantially increase the actuator burden when command increments slightly exceed these thresholds (Figure \ref{adaptive_weight}). According to the update law \eqref{update_law_lambda}, the smoothness weight $\lambda_{1}(k),\lambda_{2}(k)$ do not vanish when the action smoothness is close to the threshold, since the corresponding weight updates become slow in this situation. The proposed update law cannot enforce the consistent satisfaction of the policy smoothness requirement, because the policy continues to adapt in response to tracking errors and may generate periodic and short action variations that exceed the threshold, as shown in Figure~\ref{adaptive_weight}. Therefore, the weights $\lambda_{1},\lambda_{2}$ provide persistent and dynamical regulation of policy smoothness.    

On the other hand, the command increments generated from actor networks are different from the actual action increments due to actuator and system dynamics. In Figure \ref{adaptive_weight}, although periodic peaks of $g_{1},g_{2}$ exceed threshold $\varepsilon_{1},\varepsilon_{2}$, the variants of actual $\Delta q(k),\Delta \delta(k)$ in Figure \ref{action_incement} remain within the corresponding thresholds. Therefore, the policy smoothness may slightly exceed the prescribed thresholds while the actual action increments remain within their corresponding limits.
\end{myRem}

\subsubsection{Low-pass filter}
\begin{figure}
    \centering
    \begin{subfigure}{\linewidth}
        \centering
        \includegraphics[width=0.6\linewidth]{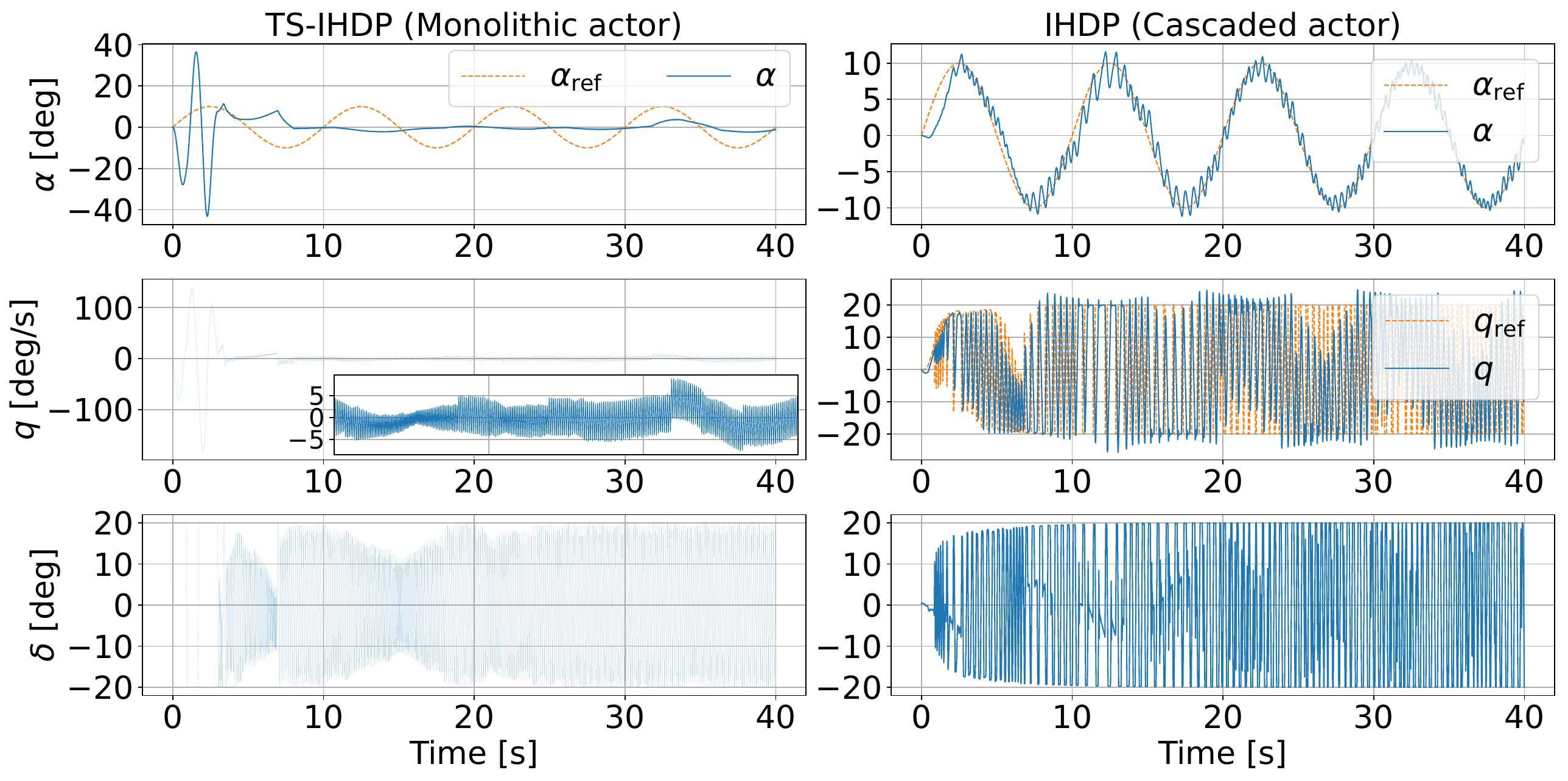}
        \caption{Unsuccessful baselines.}
        \label{fig:baselines}
    \end{subfigure}

    \begin{subfigure}{\linewidth}
        \centering
        \includegraphics[width=0.6\linewidth]{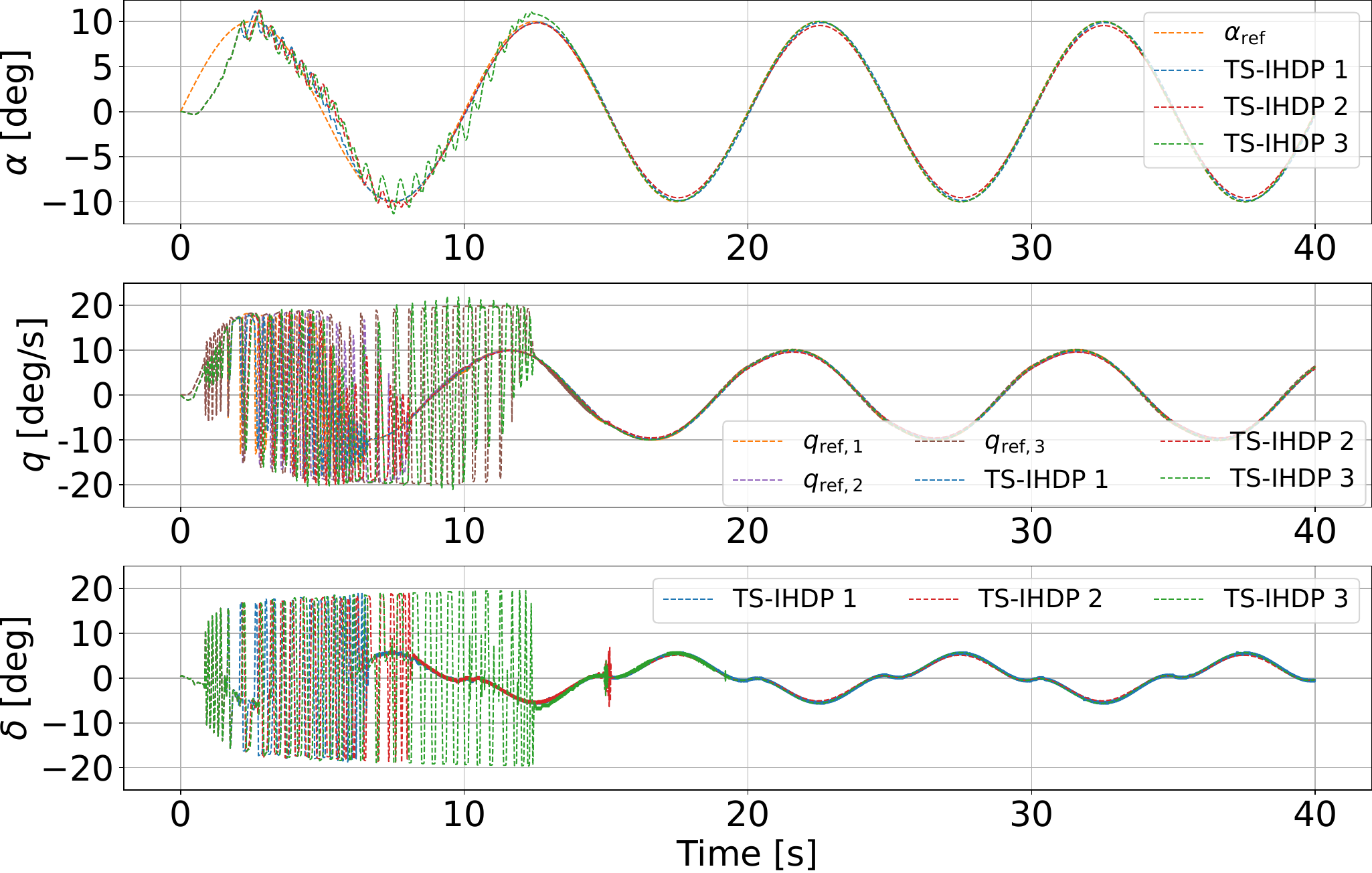}
        \caption{Online learning by TS-IHDP with different update rates $\eta_{\lambda_{2}}$.}
        \label{fig:TS-IHDP-1-2-3}
    \end{subfigure}

    \vspace{2mm}

    \begin{subfigure}{\linewidth}
        \centering
        \includegraphics[width=0.6\linewidth]{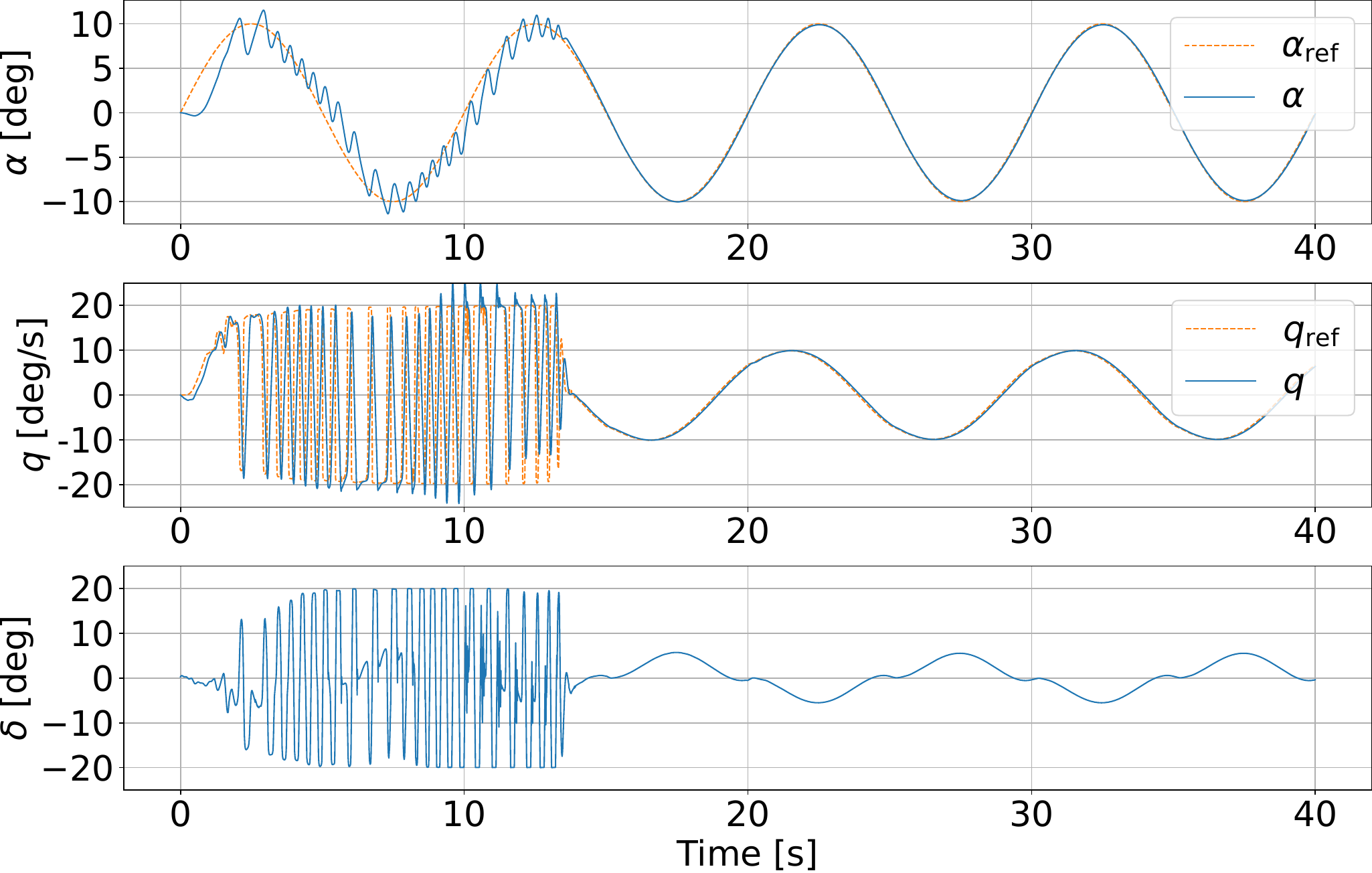}
        \caption{Online learning by CF-TS-IHDP 1.}
        \label{fig:CF-TS-IHDP-1}
    \end{subfigure}
    \caption{Comparison of different online learning strategies.}
\end{figure}

\begin{figure}
        \centering
        \includegraphics[width=0.6\linewidth]{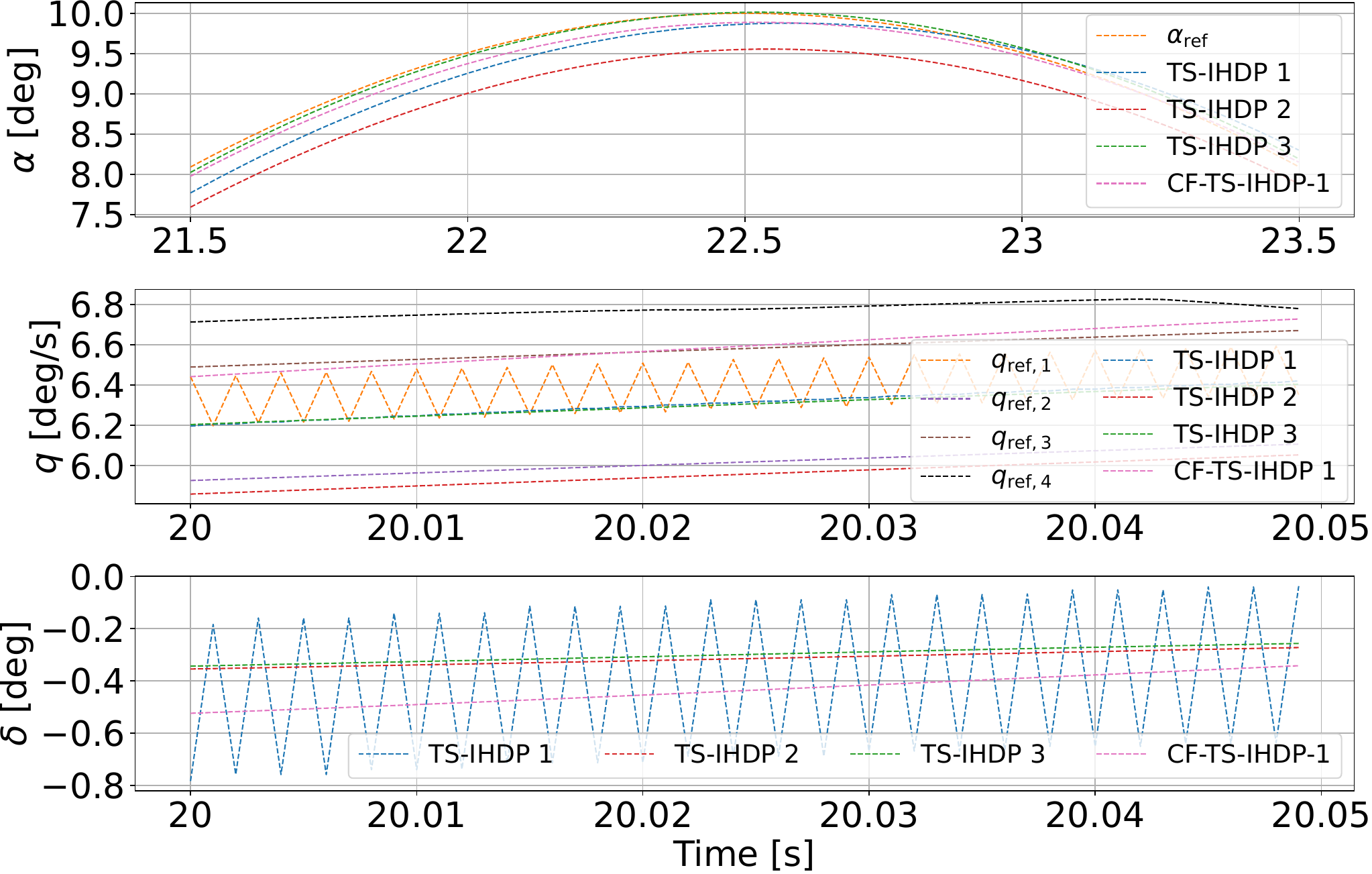}
        \caption{Enlarged view of the results in Figs.~\ref{fig:TS-IHDP-1-2-3} and \ref{fig:CF-TS-IHDP-1}.}
        \label{fig:TS-IHDP-1-2-3-zoom}
\end{figure}

We introduce a low-pass filter ($\omega_{n}=20 \text{rad/s}$, $\zeta=0.7$) into the cascaded actor for pitch rate command filtering (see Figure \ref{cascaded_online_learning_structure}). The comparisons between TS-IHDP 1 and Command-Filtered (CF-) TS-IHDP 1 in Figures \ref{fig:TS-IHDP-1-2-3} and \ref{fig:CF-TS-IHDP-1} show the filter generates a smoother pitch-rate reference and further leads to smoother control-surface deflection.

\begin{figure}
    \centering    \includegraphics[width=0.6\linewidth]{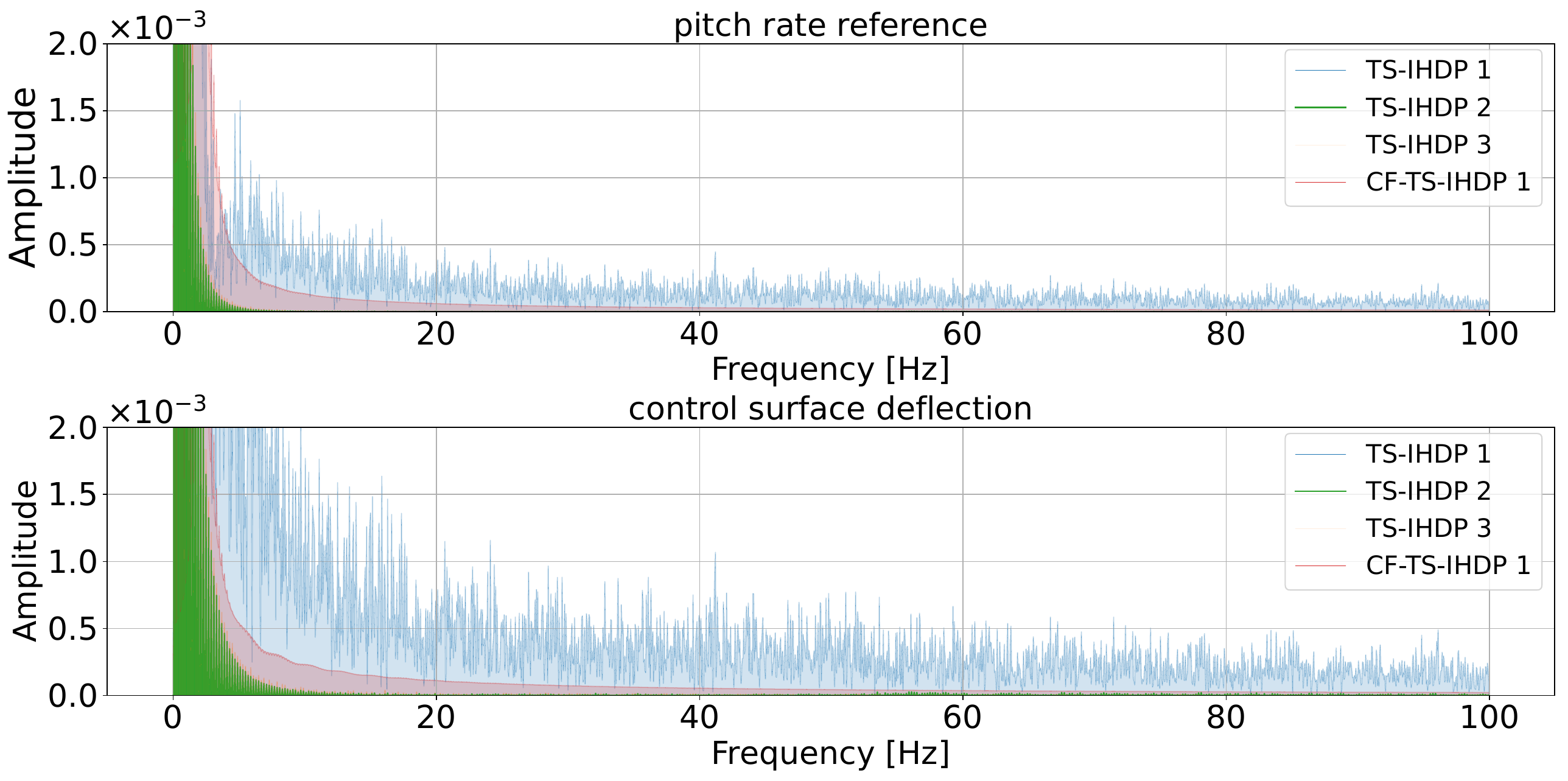}
    \caption{Spectrum of actions. } \label{fig_spectrum_actions}
\end{figure}

\begin{myRem}
Why a filter is needed for TS-IHDP? 

An issue of TS-IHDP 1 lies in that it imposes the mildest penalty on outer-loop policy smoothness, which causes frequently switching command $q_{\text{ref}}(k)$ and additional actuator burden. This issue can be addressed either by increasing the weight update rate, or by introducing a filter during online learning. We will show in the next subsection that the filter improves temporal smoothness to a level similar to that of TS-IHDP 3.
\end{myRem}  

\FloatBarrier

\subsubsection{Control smoothness}
We evaluate the control smoothness in the frequency domain using the fast Fourier transform (FFT) of the action sequence over the time interval [20,40]s after policy convergence. The FFT efficiently computes the discrete Fourier transform (DFT), which transforms the action sequence from the time domain into the frequency representation, defined as $U(i) = \sum_{l=0}^{N-1} u(l) \, e^{-j \frac{2\pi}{N} in}, i = 0,1,\ldots,N-1$, where $l$ is the index of $N$ samples, $U(i)$ is $i$-th frequency component \cite{OppenheimSchafer2009}. Based on the results of FFT, the action smoothness is quantified by $\text{Sm} = \frac{2}{(N-1)f_{s}}\sum^{N-1}_{i=0}M_{i}f_{i}$, where $M_{i}$, $f_{i}$ are the amplitude and frequency of $i$-th frequency component, respectively, and $f_{s}$ is the sampling frequency \cite{Mysore2021ICRA}. Sm  measures the mean weighted normalized frequency, which simultaneously accounts for the frequencies and amplitudes of action signals. The large values of Sm indicate more high-frequency components in the action signal and, consequently, a greater potential for costly actuation, whereas lower values of Sm indicate smoother responses and lower actuation. The spectra in Figure~\ref{fig_spectrum_actions} show that the high-frequency components of $q_{\text{ref}}$ and $\delta$ below 100 Hz in TS-IHDP 1 are attenuated by the low-pass filter. TS-IHDP 2 and 3 exhibit reduced control signal amplitudes compared to TS-IHDP 1. These observations are supported by the statistics of Sm in Table \ref{tab:smoothness_measures}. TS-IHDP 2 exhibits an Sm value in the inner loop that is nearly half that of TS-IHDP 3 due to the further increase in $\lambda_{2}$ after 15s (Figure \ref{adaptive_weight}). This imposes a stricter smoothness requirement on the inner-loop policy, which may lead to sluggish control command responses and is a primary reason for the reduced tracking accuracy observed in Figure \ref{figure_model_uncertainty}.

To quantify action smoothness in the time domain, the Mean Control Increment (MCI) is adopted and defined as $\overline{\vert\Delta u\vert}=\frac{1}{N}\sum^{N}_{k=1}\vert u(k)-u(k-1) \vert$, similar to the action fluctuation ratio for a stochastic policy \cite{Song2023PMLR}. Table \ref{tab:smoothness_measures} shows that TS-IHDP 1 exhibits the largest degree of temporal unsmoothness in the time domain over all the methods, including IHDP. However, using a low-pass filter (CF-TS-IHDP 1) achieves an improved level of smoothness comparable to that of TS-IHDP 2 and 3. Notably, although CF-TS-IHDP 1 and TS-IHDP 2/3 exhibit similar control signals in the time domain, the former yields a slightly larger value of Sm. This is because the frequency-domain metric Sm is more sensitive to small differences in high-frequency spectral components, that are not visually apparent in the time-domain representation by MCI.

Figure \ref{action_incement} shows the action increments defined as $\Delta q(k) = \left|q(k) - q(k-1)\right| $, $\Delta \delta(k) = \left|\delta(k) - \delta(k-1) \right|$. Under the first-order actuator model in Section \ref{section_simulation_setup}, the maximum allowable change is bounded by $\vert \Delta \delta(k) \vert_{\max} = 0.6^{\circ}/\text{s}$ over a time step of 0.001s. IHDP and TS-IHDP 1 are not presented because their action increments almost reach the actuator limit and therefore lie outside the range shown in this figure. Other methods are shown to achieve small action increments after the policies converge, i.e, $\vert\Delta q(k)\vert\leq 0.007^{\circ}/\text{s}$ and $|\Delta\delta(k)|\leq0.007^{\circ}$. Since actuator load is closely related to the magnitude and rate of control-surface deflection, these reductions compared to IHDP and TS-IHDP 1 imply a lower actuation demand and improved actuator friendliness.

\begin{table}[htbp]
\centering
\caption{Smoothness statistics.}
\label{tab:smoothness_measures}
\begin{threeparttable}
\begin{tabular}{lcc|cc}
\toprule
& \multicolumn{2}{c}{Outer-loop} & \multicolumn{2}{c}{Inner-loop} \\
\cmidrule(lr){2-3} \cmidrule(lr){4-5}
Method & Sm$\cdot 10^{-7}$  & MCI$\cdot 10^{-3}$ &Sm$\cdot 10^{-7}$  & MCI$\cdot 10^{-3}$ \\
\hline
IHDP & 20357.663 & 218.136 & 8563.684 & 281.160 \\
TS-IHDP-1 & 187.117 & 233.639 & 384.250 & 597.576 \\
TS-IHDP-2 & \underline{2.432} & \underline{3.858}  & \underline{3.624}  & \underline{2.208} \\
TS-IHDP-3 & 2.796 &  4.081 & 7.403 & 2.442 \\
CF-TS-IHDP-1 & 29.843 & 3.950 & 52.832  & 2.435 \\
\bottomrule
\end{tabular}
\begin{tablenotes}
\footnotesize
\item The best result in each metric is underlined.
\end{tablenotes}
 \end{threeparttable}
\end{table}

\subsubsection{Network Convergence}

\begin{figure}
    \centering 
    \includegraphics[width=0.6\linewidth]{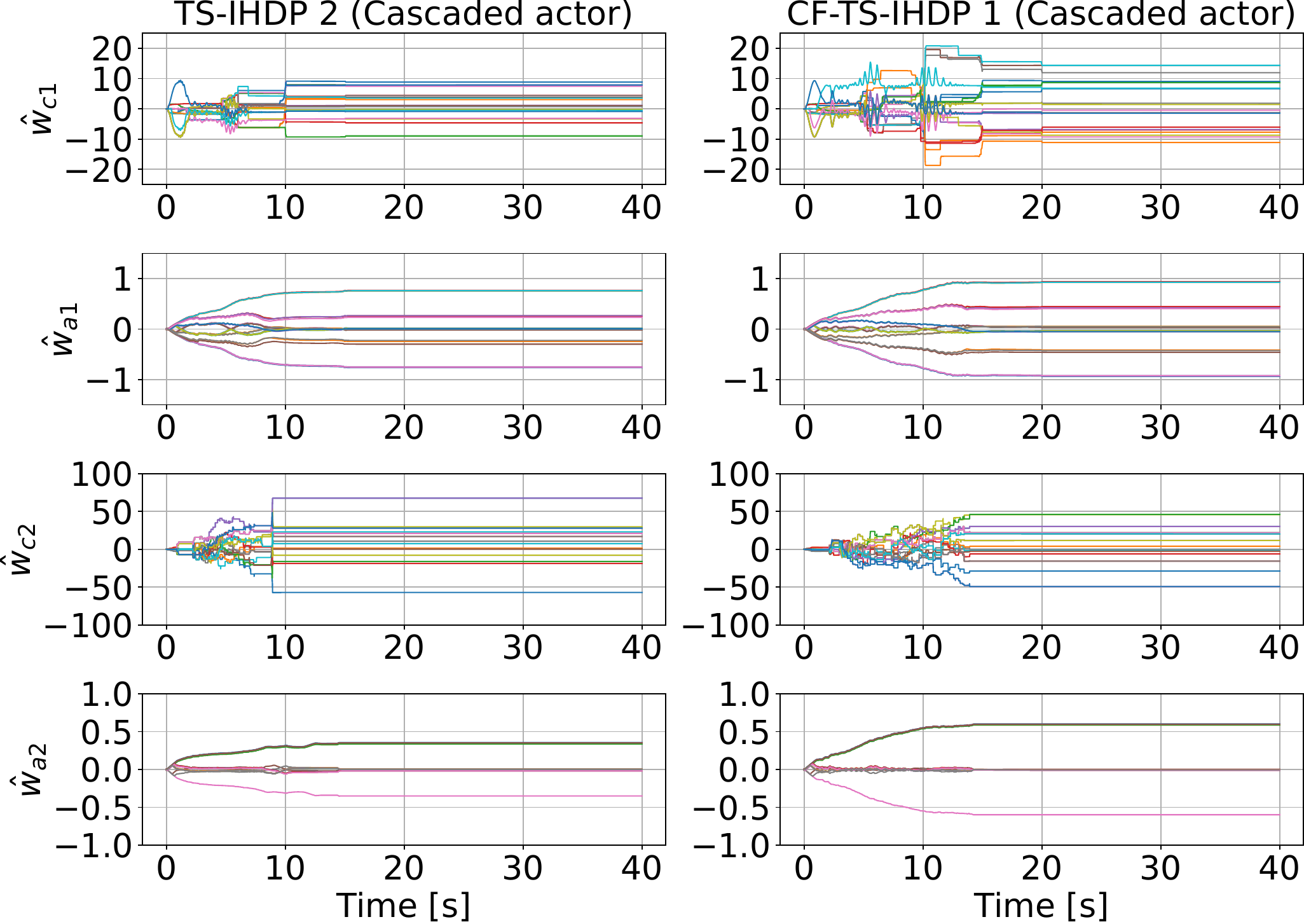}
    \caption{Critic and actor weights. $\hat{w}_{c1}$ and $\hat{w}_{c2}$ denote the critic weights of the outer-loop and inner-loop actors, respectively, while $\hat{w}_{a1}$ and $\hat{w}_{a2}$ denote the corresponding actor weights. Different lines represent the weights in both hidden and output layers.} \label{network_weights_TS}
\end{figure}

Figure \ref{network_weights_TS} shows the critic weights  converge in 15s. By TS-IHDP 2, the outer-loop actor weights associated with the input $e_{1}$ converge to absolute values of approximately 0.75, whereas those by CF-TS-IHDP 1 converge approximately to 1.0. This difference reflects the distinct control policies learned for tracking the same reference $\alpha_{\text{ref}}$: CF-TS-IHDP 1 learns a pitch rate reference with higher tracking-error feedback gains than that TS-IHDP 2 does after a longer exploration, which enables smaller AoA tracking error, especially observed in tracking the peaks of $\alpha_{\text{ref}}$ in Figure \ref{fig:TS-IHDP-1-2-3-zoom}. Similar observations hold for the inner-loop actors.

\begin{figure*}
    \centering 
    \includegraphics[width=0.8\linewidth]{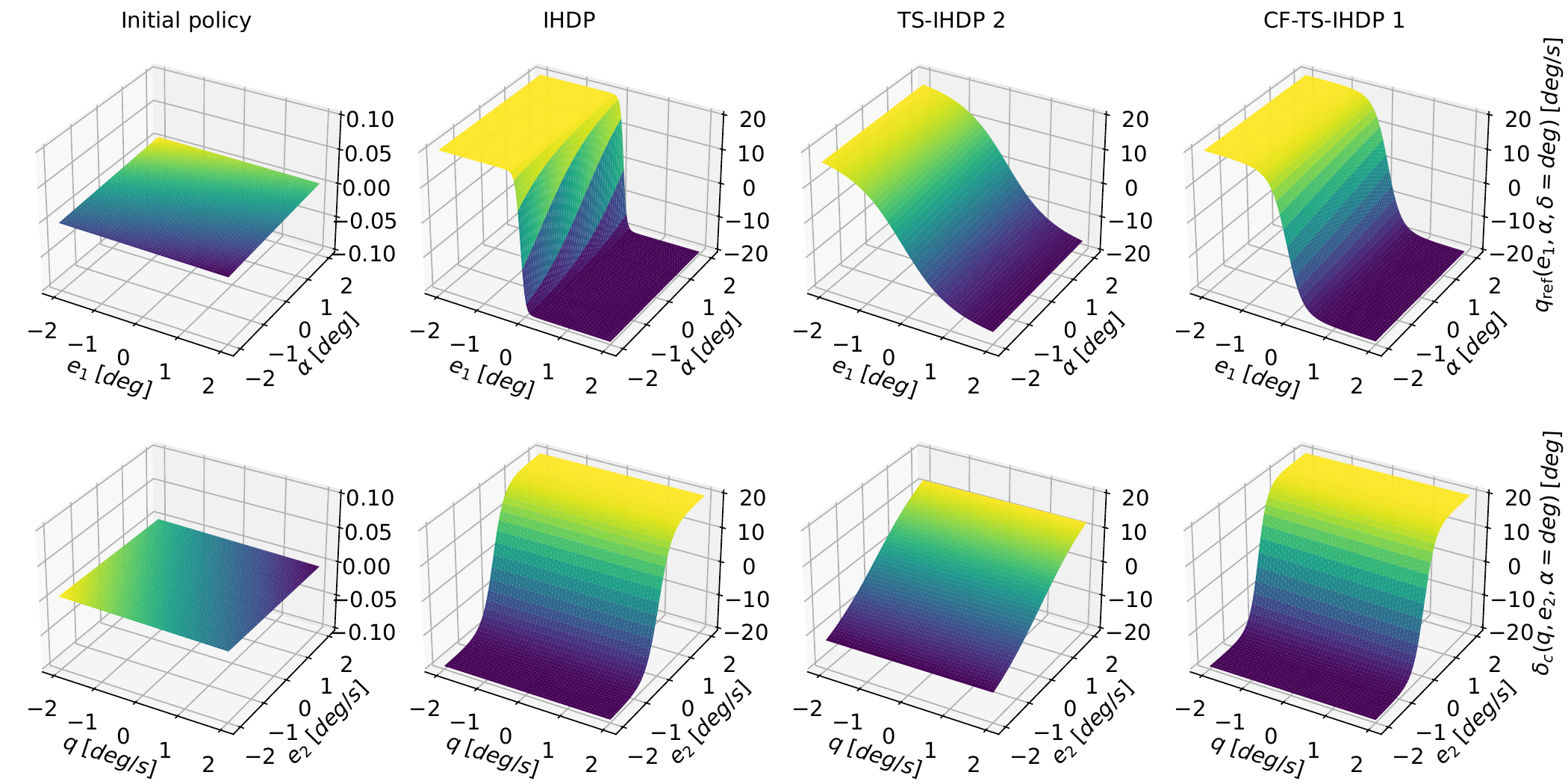}
    \caption{Policy network surface over the local state space. The upper four plots show the outer-loop actor evaluated at $\delta=0^{\circ}$, while the lower four plots show the inner-loop actor evaluated at $\alpha=0^{\circ}$.} \label{Fig_landscape}
\end{figure*}

Figure \ref{Fig_landscape} shows actor network surfaces in local state spaces, with $e_{1}\in[-2^{\circ},2^{\circ}]$, $\alpha\in[-2^{\circ},2^{\circ}]$, $e_{2}\in[-2^{\circ}/\text{s},2^{\circ}/\text{s}]$, $q\in[-2^{\circ}/\text{s},2^{\circ}/\text{s}]$. The initial policies exhibit nearly flat surfaces and the outputs remain close to zero. The IHDP policies produce steepest policy surfaces, indicating highest-gain control laws with respect to tracking errors, and reach saturation for approximately $|e_{1}|>0.1^{\circ}$, $|e_{2}|>0.1^{\circ}/\text{s}$, respectively. The TS-IHDP 2 and CF-TS-IHDP 1 policies, however, yield smoother surfaces, suggesting milder policies. Notably, the inner-loop actor
of the CF-TS-IHDP 1 exhibits a stronger control policy due
to a longer exploration, that provides better tracking accuracy
as shown in Figure \ref{fig:TS-IHDP-1-2-3-zoom}. These observations are consistent with
the differences in the actor weights shown in Figure~\ref{network_weights_TS}.

\begin{figure}
    \centering 
    \includegraphics[width=0.6\linewidth]{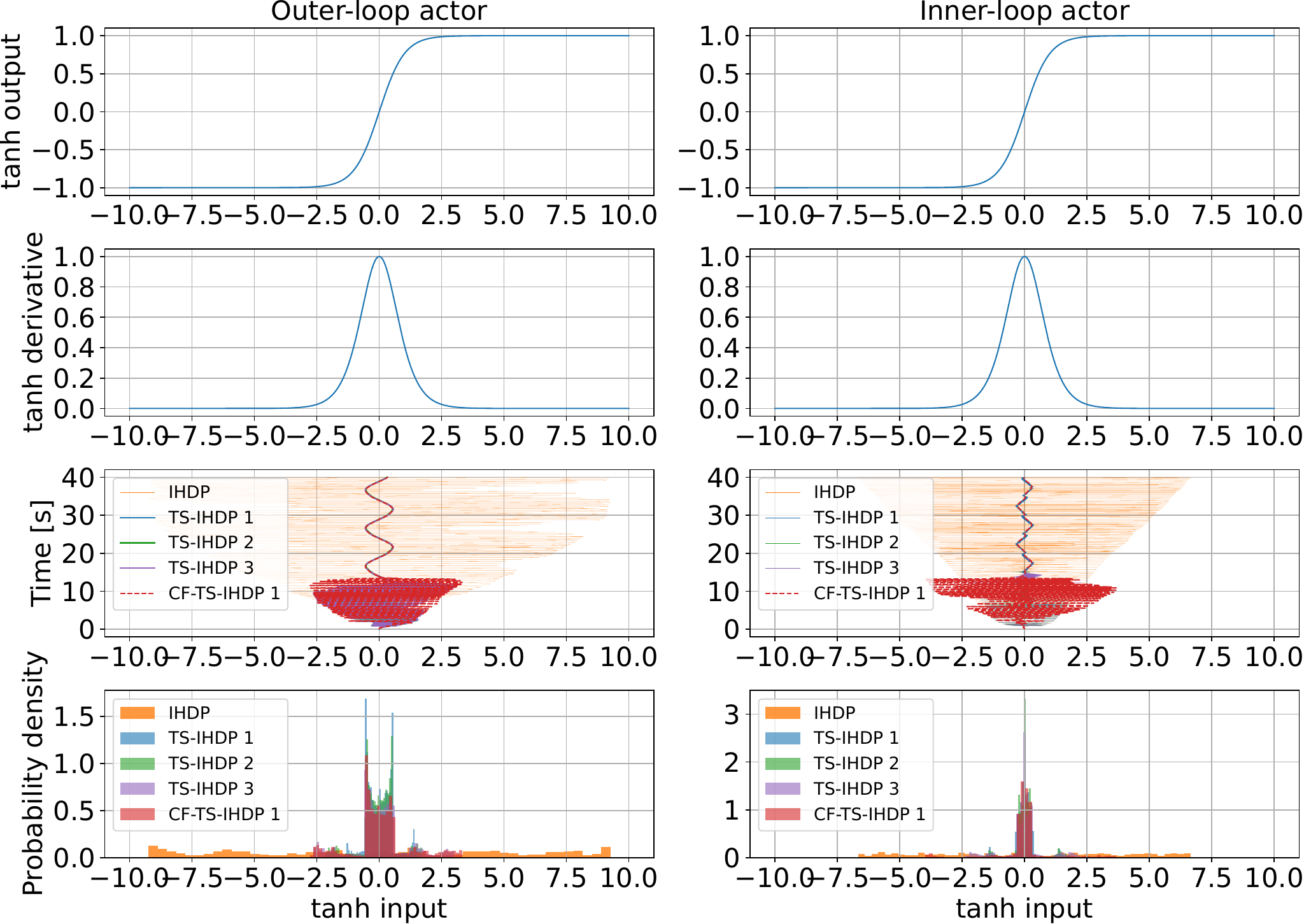}

    \caption{$\tanh(\cdot)$ in output layers of actors.} \label{fig_saturation}
\end{figure} 

\begin{figure}
    \centering 
    \includegraphics[width=0.6\linewidth]{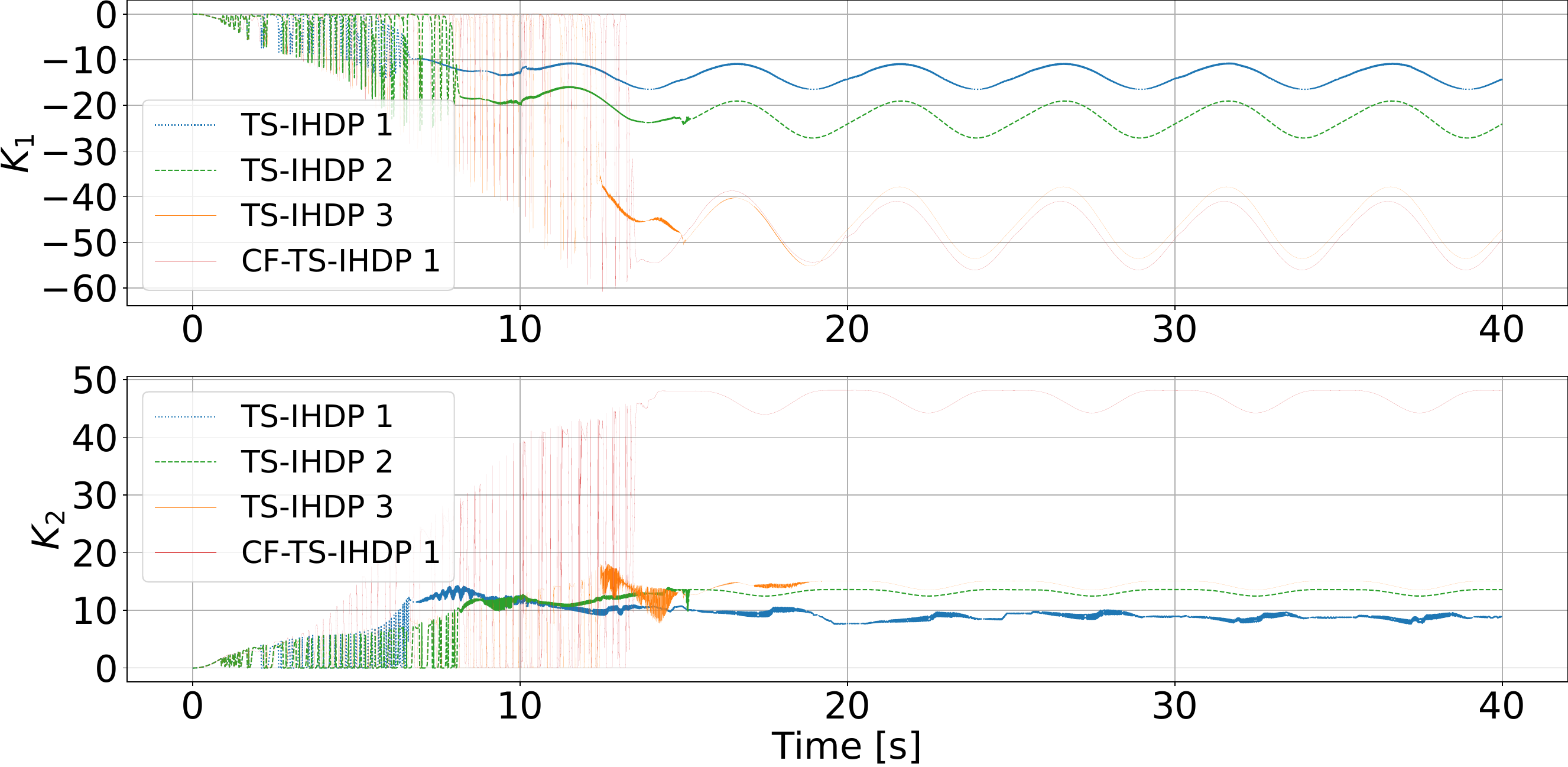}

    \caption{Sensitivity measures.} \label{fig_feedback_gain_filter}
\end{figure} 

\subsubsection{Activation function}

With large inputs, the actor's activation function ($\tanh$) outputs values near its limits, and its derivative ($\tanh^{\prime}$) vanishes (see Figure~\ref{fig_saturation}). Consequently, policy gradient vanishes and hence policy update becomes slow. Another issue with the saturated actor is sensitive and produces aggressive control actions in response to small tracking errors, which increases oscillations and degrades system stability.

Actor saturation is usually caused by an aggressive exploration process, which occurs for two main reasons: (1) using a large learning rate that would update the actor weights rapidly, pushing the weights into into saturation regions of $\tanh$. This process is irreversible due to vanishing gradients in these regions; (2) lack of policy smoothness regularization. The policy may generate aggressive actions during the exploration phase, which causes system oscillations and further amplify policy update, increasing the possibility of actor saturation. Meantime, the policy would keep aggressive and continues to degrade tracking performance even as the tracking error decreases, because policy update is slow when the actor becomes near saturated. In Figure \ref{fig_saturation}, switching actions occur even though $\tanh$ is not saturated during 10–20s, indicating that an unsaturated actor can also generate oscillatory actions. This necessitates using policy smoothness regularization to restrict the policy update. Finally, the cascaded actor exhibits an interaction issue: the time-varying dynamics and reference signal $\alpha_{\text{ref}}$ give rise to switching tracking errors $e_{1}$. These errors propagate through the outer-loop actor, producing a switching $q_{\text{ref}}$, which generates switching errors $e_{2}$ and induces switching $\delta$ through the inner-loop actor.

In Figure \ref{fig_saturation}, IHDP makes the $\tanh$ inputs within intervals [7.5,10] (the outer-loop actor) and intervals [5,7.5] (the inner-loop actor). In these intervals, the derivative $\tanh^{\prime}(\cdot)\leq0.001$ and attenuates the overall policy gradient. As a comparison, TS-IHDP and CF-TS-IHDP prevent reaching to the saturation regions. The inputs are kept within 'unsaturated' intervals [-0.6,0.6] (the outer-loop actor) and [-0.4,0.4] (the inner-loop actor) after policies converge. The derivatives satisfy $\tanh^{\prime}(\cdot)\geq 0.6$ for the outer-loop actor and $\tanh^{\prime}(\cdot)\geq 0.85$ for the inner-loop actor. Probability density distribution histograms over $4\times10^{4}$ states show more concentrated distributions of $\tanh$ input around zero by TS-IHDP methods and CF-TS-IHDP 1, leading to improved learning sensitivity. This demonstrates temporal smoothness regularization and its combination with a low-pass filter effectively mitigate actor saturation.

\subsubsection{Actor sensitivity}

We define the local sensitivity measures of the outer-loop and inner-loop policies to the tracking errors as

\begin{subequations}
\begin{align}
     K_{1}(k) &= \frac{\partial q_{\text{ref}}(k)}{ \partial e_{1}(k)}\\
     K_{2}(k) &= \frac{\partial \delta_{c}(k)}{\partial e_{2}(k)}
\end{align}
\end{subequations} \\
which are local Jacobian entries of the policy networks outputs with respect to individual input components (i.e., the tracking errors), and can also be interpreted as element-wise local Lipschitz constants of policy networks \cite{Song2023PMLR}.

Figure \ref{fig_feedback_gain_filter} shows that applying a low-pass filter in CF-TS-IHDP 1 promotes increases of $K_{1}$ and $K_{2}$ by prolonging the exploration process. The longer exploration time in TS-IHDP 3 promotes an increase in $K_{1}$. Consequently, these two methods exhibit greater sensitivity, and hence higher local feedback gains to tracking errors. This increased sensitivity leads to improved tracking accuracy, as demonstrated in Table \ref{tab:tracking_error_metric}.

\subsection{Online operation phase}

The robustness of the NN policies to model uncertainties is evaluated in this subsection. The weights of post-training actors are obtained from the converged values at 40s in online learning phase and kept fixed during the online operation phase. The system dynamics \eqref{aerialvehiclemodel} are considered to be uncertain, by perturbating the aerodynamic coefficients within $\pm 30\%$ of their nominal values, including $\Delta Q$, $\Delta \phi_{z}$, $\Delta \phi_{m}$, $\Delta b_{z}$, and $\Delta b_{m}$, as considered in \cite{Hull19951ACC}. The reference trajectory is a sequence of piecewise-constant AoA commands with both increasing and decreasing segments, representing typical maneuvering conditions. 

\begin{figure}
    \centering
    \includegraphics[width=0.6\linewidth]{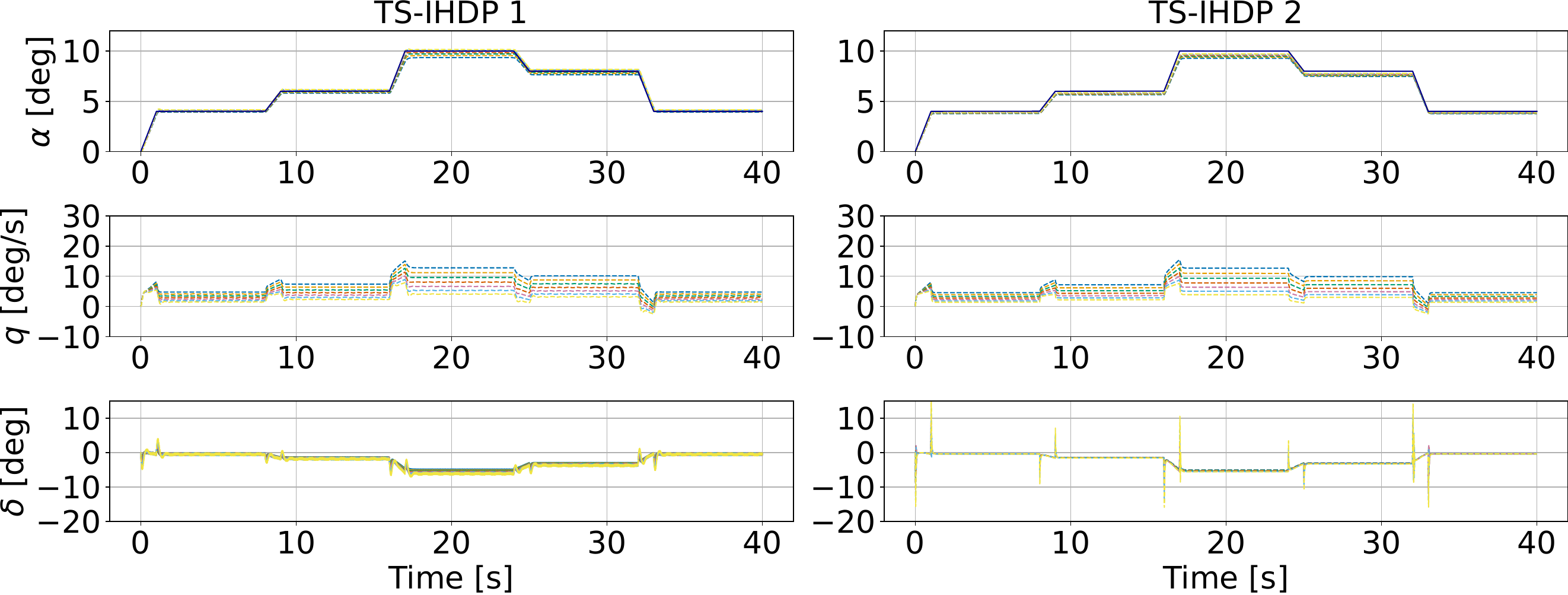}
    \vspace{2mm}
    
    \includegraphics[width=0.6\linewidth]{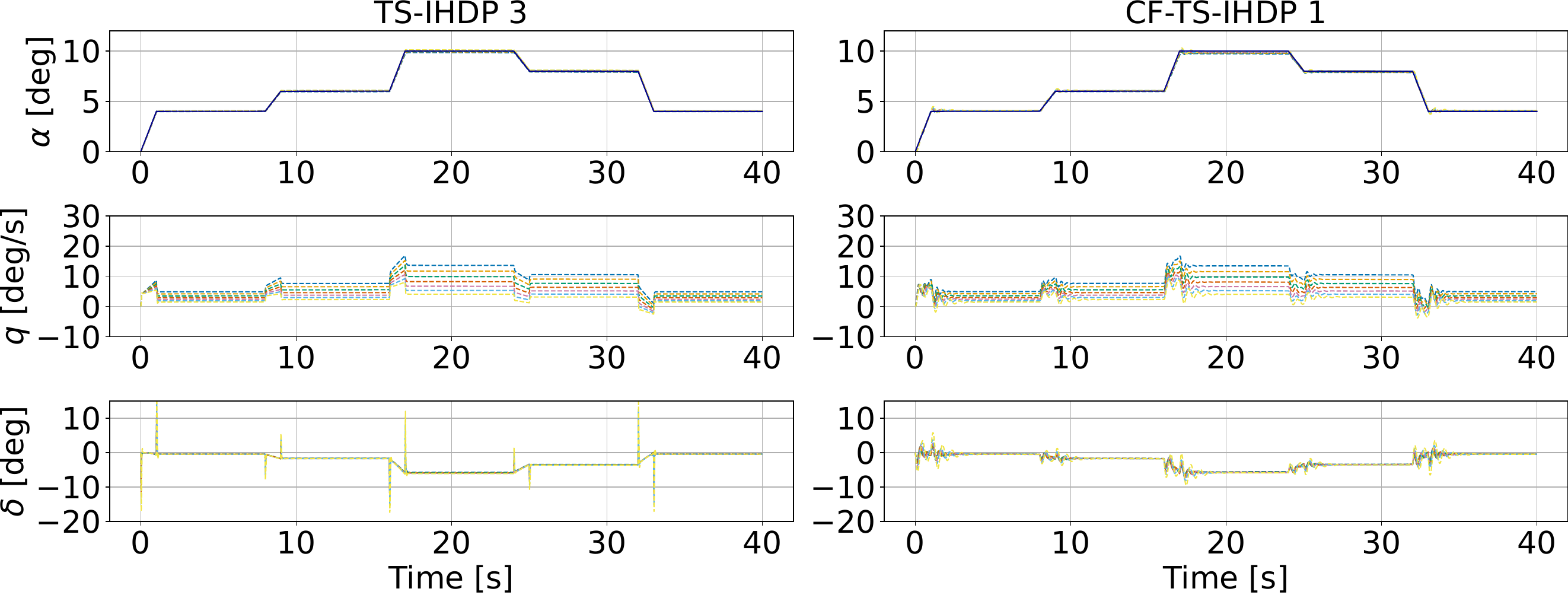}
    \includegraphics[width=0.6\linewidth]{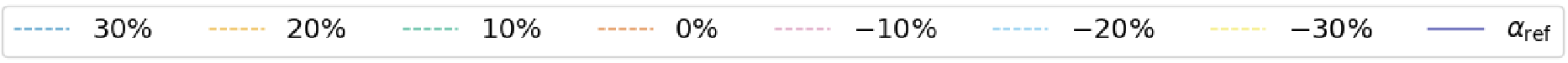}
    \caption{Tracking control under aerodynamic coefficients perturbations.}
    \label{figure_model_uncertainty}
\end{figure}

\FloatBarrier

\begin{figure}
    \centering
       
    \includegraphics[width=0.4\linewidth]{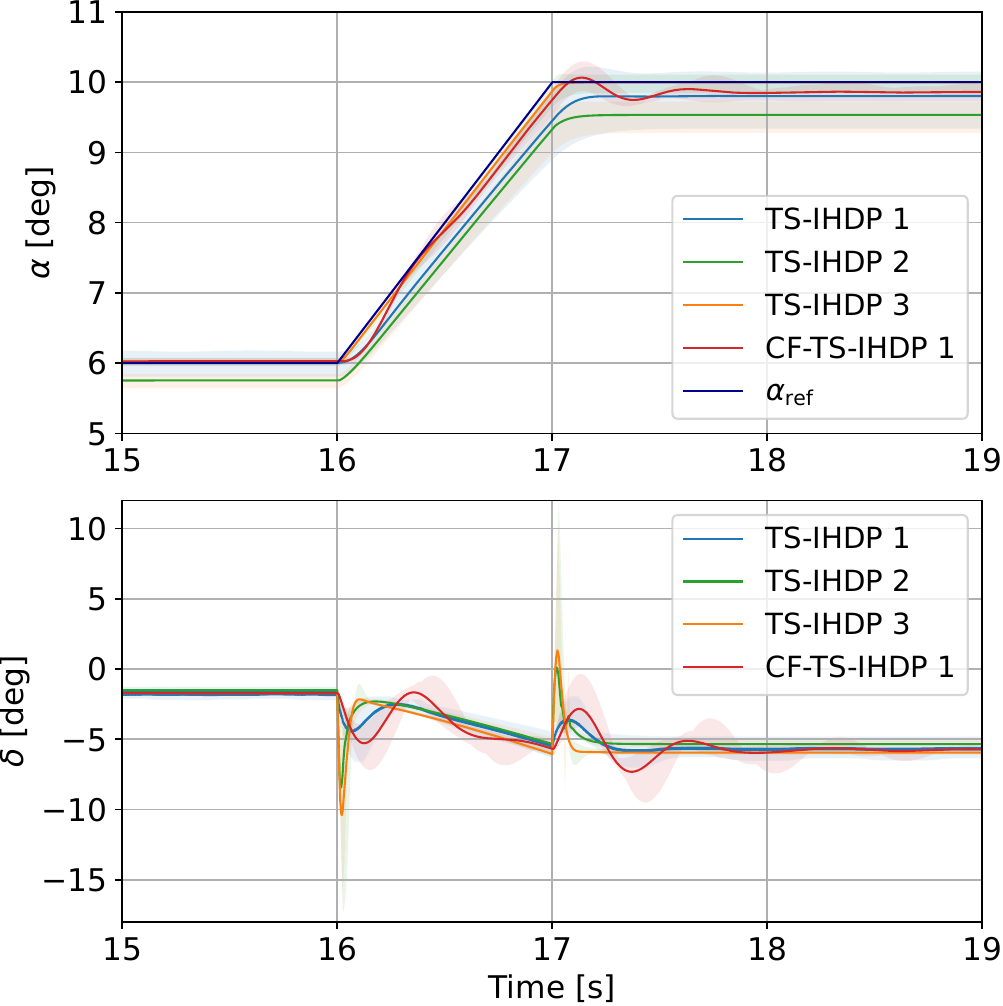}

    \caption{Enlarged view of $\alpha,\delta$ histories. The solid line represents the mean trajectory across seven uncertain scenarios, while the shaded region indicates the range between the maximum and minimum bounds. The filter introduces oscillations due to the additional filter dynamics and phase lag.}
    \label{fig:model_uncertainty_zoom}
\end{figure}

Figure~\ref{figure_model_uncertainty} shows that all proposed methods successfully track $\alpha_{\text{ref}}$ under considered model uncertainties. TS-IHDP 1 continues to exhibit switching control actions, consistent with its behavior during the online learning phase. TS-IHDP 2,3 exhibit higher action peaks during the maneuvering phases of $\alpha_{\text{ref}}$. For one state-trajectory under model uncertainty, the AoA tracking error is quantified using the Mean Absolute Error (MAE), defined as $\overline{\vert e^{i}_{1}\vert}=\frac{1}{N}\sum^{N}_{k=1}\vert e^{i}_{1}(k)\vert$, where $N$ is the number of time samples, $i\in\{1,2,\cdots,7\}$ is the index of a trajectory. Over the presented uncertain scenarios, the mean of MAE is computed with $\overline{\vert e_{1}\vert}=\frac{1}{7}\sum^{7}_{i=1}\overline{\vert e^{i}_{1}\vert}$. This MAE mean reflects the average tracking accuracy under different model uncertainties, and a small MAE mean indicates better average accuracy. The variance of MAE is $\text{Variance}=\frac{1}{7}\sum^{7}_{i=1}(\overline{\vert e^{i}_{1}\vert}-\overline{\vert e_{1}\vert})^{2}$, which characterizes the consistency of the MAE across uncertain scenarios. A smaller variance indicates that the performance is less sensitive to variations in model parameters, implying stronger robustness to model uncertainties. Table \ref{tab:tracking_error_metric} shows that CF-TS-IHDP 1 achieves lower mean and variance of MAE than those of TS-IHDP 1, indicating that using a low-pass filter during online training improves the robustness of tracking control policy under model uncertainty. Among unfiltered methods, TS-IHDP 3 achieves the best tracking accuracy and robustness, and TS-IHDP 1 outperforms TS-IHDP 2. TS-IHDP 2 exhibits stricter inner-loop policy smoothness and reduced command responsiveness due to the additional increase in $\lambda_{2}$ after 15 s (Figure~\ref{adaptive_weight}), which leads to less accurate tracking during the maneuvering phases of the command $\alpha_{\text{ref}}$ (Figure \ref{fig:model_uncertainty_zoom}). This result suggests that improved command smoothness alone does not necessarily correspond to better tracking accuracy and robustness. On the contrary, overly strict policy smoothness requirement may lead to sluggish control command responses, a primary cause of the degraded tracking accuracy, as demonstrated by TS-IHDP 2.


\setlength{\tabcolsep}{1.0pt}
\begin{table}[htbp]
\centering
\caption{MAE across uncertain scenarios.}
\label{tab:tracking_error_metric}
\begin{threeparttable}
{\fontsize{8}{9}\selectfont
\begin{tabular}{c|c|c|c|c}

\hline
 &  TS-IHDP 1 &  TS-IHDP 2  &  TS-IHDP 3 & CF-TS-IHDP 1\\
\hline
Mean$\cdot 10^{-3}$ &146.771  & 277.355  &\underline{46.977} &77.240\\
Variance$\cdot 10^{-3}$&4.424 & 6.094 & 0.383&\underline{0.122} \\

\hline
\end{tabular}
}
\begin{tablenotes}
\footnotesize
\item The best result in each metric is underlined.
\end{tablenotes}
 \end{threeparttable}
\end{table}

%

\section{Conclusion}\label{section_conclusion}

This paper proposes temporally smoothed incremental heuristic dynamic programming (TS-IHDP), which regularizes temporal policy smoothness during online learning with slight additional computational cost. AoA tracking control simulations demonstrate the benefits of TS-IHDP in improving control smoothness and tracking accuracy, reducing system oscillations, and avoiding actor NN saturation. Next, the primal-dual method is shown to adaptively adjust the smoothness weight according to a prescribed smoothness criterion, thereby simplifying the process of manual tuning. However, the achieved policy smoothness remains sensitive to the update rate $\eta_{\lambda}$, which should be considered when applying this method. Furthermore, a low-pass filter used in the TS-IHDP-based control system improves residual policy unsmoothness, at the inherent expense of phase delay and signal bias. A notable finding is that the hierarchical structure of the control tasks enables the Markov property of the states to be preserved during online learning. Finally, post-training simulations under model uncertainties show that command-filtered TS-IHDP 1 achieves (1) improved robustness compared with its unfiltered counterpart (TS-IHDP 1) and (2) robustness comparable to TS-IHDP 3 with the best-optimized smoothness. It is shown that an overly strict policy smoothness penalty improves smoothness at the expense of control accuracy and policy robustness, which can be interpreted as over-constraining the actuator responsiveness.


\textit{Limitations and future work:} (1) The time-varying $\lambda(k)$ would changes the convergence results in Section \ref{section_convergence_analysis} and can be further studied. (2) TS-IHDP is model-free and uses a random initial policy, which can lead to oscillatory system behavior during online exploration and hence limit its practical applicability. Future work should consider initializing the policy with an offline-trained safe policy to improve online system stability while preserving sufficient exploration. (3) This paper considers a longitudinal dynamics model to verify TS-IHDP, its performance with full six-degree-of-freedom (6-DOF) models warrants further investigation. (4) To clearly evaluate TS-IHDP through simulation, output feedback, measurement noise, and partial observability \cite{ZHOUAST2020105} are not considered in this paper. These practical factors may adversely affect NNs' training and tracking control performance, which require further investigation.


\appendix
\section{Aerodynamic coefficients}\label{appendix_aerodynamic_coefficients}

The aerodynamic coefficients are taken from \cite{Hull19951ACC} and are approximated by $b_{z}=-0.034, b_{m} = -0.206$, and 

\begin{equation}
\begin{aligned}
    &\phi_{z}(\alpha) = 0.000103\alpha^{3}-0.00945\alpha|\alpha|-0.170\alpha\\
    &\phi_{m}(\alpha) = 0.000215\alpha^{3} - 0.0195\alpha|\alpha| - 0.051\alpha
\end{aligned}
\end{equation}

These approximations of $b_{z},b_{m},\phi_{z}(\alpha),\phi_{m}(\alpha)$ hold for $\alpha\in[- 20^{\circ}, 20^{\circ}]$. The physical coefficients are provided in Table \ref{physcial_coefficient_aerial_vehicle}. In addition, the actuator dynamics are considered as a first-order inertial model with the time delay $0.005$s. The control-surface deflection is constrained to $\pm 20^{\circ}$, with a maximum deflection rate of 600$^{\circ}$/s.

\begin{table}[htbp]
  \centering
  \caption{Physical parameters (adapted from \cite{Hull19951ACC})}
  \begin{threeparttable}
    \begin{tabular}{lll}
    \hline
    Notations & Definition & Value\\
    \hline
    $g$ & acceleration of gravity & 9.815 m/s$^{2}$\\
    $W$ & weight & 204.3 kg\\
    $V$ & speed & 947.715 m/s\\
    $I_{yy}$ & pitch moment of inertia & 247.438 kg$\cdot$ m$^{2}$\\
   $f$ & radians to degrees & 180/$\pi$\\
   $Q$ & dynamic pressure & 29969.861 kg/m$^{2}$\\
   $S$ & reference area & 0.041 m$^{2}$\\
    $d$ & reference diameter & 0.229 m\\
    \hline
    \end{tabular}
 \end{threeparttable}
 \label{physcial_coefficient_aerial_vehicle}
\end{table}

\printcredits

\section*{Acknowledgement}

The first author acknowledges the financial support provided by the China Scholarship Council (CSC) for his Ph.D. study under Grant No. 202006290022. 
\section*{Declaration of Competing Interest}

The authors declare that they have no known competing financial interests or personal relationships that could have appeared to influence the work reported in this paper.





\bibliographystyle{unsrtnat}

\bibliography{cas-refs}

@ARTICLE{ZHOUCEP201813,
author = {Ye Zhou and Erik-Jan {van Kampen} and Qi Ping Chu},
title = {Incremental model based online dual heuristic programming for nonlinear adaptive control},
journal = {Control Engineering Practice},
volume = {73},
pages = {13-25},
year = {2018},
doi = {https://doi.org/10.1016/j.conengprac.2017.12.011},

}

@CONFERENCE{ZhouIMAV2016,
author = {Ye Zhou and Erik-Jan van Kampen and Qi Ping Chu},
title = {Incremental Model Based Heuristic Dynamic Programming for Nonlinear Adaptive Flight Control},
booktitle = {Proceedings of the International Micro Air Vehicles Conference and Competition},
chapter = {},
pages = {},
doi = {},
URL = {https://www.imavs.org/papers/2016/25.pdf},
eprint = {},
year = {2016},
}

@ARTICLE{ZHOUAST2020105,
title = {Incremental model based online heuristic dynamic programming for nonlinear adaptive tracking control with partial observability},
journal = {Aerospace Science and Technology},
volume = {105},
pages = {106013},
year = {2020},
doi = {https://doi.org/10.1016/j.ast.2020.106013},
author = {Ye Zhou and Erik-Jan {van Kampen} and Qiping Chu},
}

@article{Sun2020103425,
title = {Incremental model-based global dual heuristic programming with explicit analytical calculations applied to flight control},
journal = {Engineering Applications of Artificial Intelligence},
volume = {89},
pages = {103425},
year = {2020},
doi = {https://doi.org/10.1016/j.engappai.2019.103425},
author = {Bo Sun and Erik-Jan {van Kampen}},
}

@ARTICLE{SunJGCD2021,
author = {Bo Sun and Erik-Jan van Kampen},
title = {Reinforcement-Learning-Based Adaptive Optimal Flight Control with Output Feedback and Input Constraints},
journal = {Journal of Guidance, Control, and Dynamics},
volume = {44},
number = {9},
pages = {1685-1691},
year = {2021},
doi = {https://doi.org/10.2514/1.G005715},
}

@CONFERENCE{ZhouIAC2017,
  author = {Ye Zhou and Erik-Jan van Kampen and QiPing Chu},
  booktitle={68th International Astronautical Congress}, 
  title={Launch vehicle adaptive flight control with incremental model based heuristic dynamic programming}, 
  year={2017},
  volume={},
  number={},
  pages={7154-7164},
  doi={},
}

@article{Vamvoudakis2010878,
title = {Online actor–critic algorithm to solve the continuous-time infinite horizon optimal control problem},
journal = {Automatica},
volume = {46},
number = {5},
pages = {878-888},
year = {2010},
doi = {https://doi.org/10.1016/j.automatica.2010.02.018},
author = {Kyriakos G. Vamvoudakis and Frank L. Lewis},
}

@ARTICLE{Sokolov20159,
title = {Complete stability analysis of a heuristic approximate dynamic programming control design},
journal = {Automatica},
volume = {59},
pages = {9-18},
year = {2015},
doi = {https://doi.org/10.1016/j.automatica.2015.06.001},
author = {Yury Sokolov and Robert Kozma and Ludmilla D. Werbos and Paul J. Werbos},
}

@ARTICLE{Bhasin201382,
title = {A novel actor–critic–identifier architecture for approximate optimal control of uncertain nonlinear systems},
journal = {Automatica},
volume = {49},
number = {1},
pages = {82-92},
year = {2013},
doi = {https://doi.org/10.1016/j.automatica.2012.09.019},
author = {S. Bhasin and R. Kamalapurkar and M. Johnson and K.G. Vamvoudakis and F.L. Lewis and W.E. Dixon},
}

@ARTICLE{Lin201738,
author = {Hanquan Lin and Qinglai Wei and Derong Liu},
title = {Online identifier–actor–critic algorithm for optimal control of nonlinear systems},
journal = {Optimal Control Applications and Methods},
volume = {38},
number = {3},
pages = {317-335},
doi = {https://doi.org/10.1002/oca.2259},
year = {2017}
}

@INPROCEEDINGS{Mysore2021ICRA,
  author={Siddharth Mysore and Bassel Mabsout and Renato Mancuso and Kate Saenko},
  booktitle={2021 IEEE International Conference on Robotics and Automation (ICRA)}, 
  title={Regularizing Action Policies for Smooth Control with Reinforcement Learning}, 
  year={2021},
  volume={},
  number={},
  pages={1810-1816},
  doi={https://doi.org/10.1109/ICRA48506.2021.9561138},
}

@article{Gavra202447JGCD,
author = {Vlad Gavra and Erik-Jan van Kampen},
title = {Evolutionary Reinforcement Learning: Hybrid Approach for Safety-Informed Fault-Tolerant Flight Control},
journal = {Journal of Guidance, Control, and Dynamics},
volume = {47},
number = {5},
pages = {887-900},
year = {2024},
doi = {https://doi.org/10.2514/1.G008112},

}

@CONFERENCE{Homola2025AIAA,
author = {Marek Homola and Yifei Li and Erik-Jan van Kampen},
title = {Uncertainty-Driven Distributional Reinforcement Learning for Flight Control},
booktitle = {AIAA SCITECH 2025 Forum},
chapter = {},
pages = {},
doi = {https://doi.org/10.2514/6.2025-2793},
year = {2025},
}

@ARTICLE{Niederlinski1995TAC,
  author={A. Niederlinski},
  journal={IEEE Transactions on Automatic Control}, 
  title={An upper bound for the recursive least squares estimation error}, 
  year={1995},
  volume={40},
  number={9},
  pages={1655-1660},
  doi={https://doi.org/10.1109/9.412640},
}

@BOOK{Haykin2009NeuralNetworks,
  author    = {Haykin, S.},
  title     = {Neural Networks and Learning Machines},
  edition   = {3},
  publisher = {Pearson Prentice Hall},
  address   = {Upper Saddle River, NJ, USA},
  year      = {2009},
  isbn      = {978-0-13-147139-9}
}

@BOOK{Khalil2002NonlinearSystems,
  author    = {Khalil, H. K.},
  title     = {Nonlinear Systems},
  edition   = {3},
  publisher = {Prentice Hall},
  address   = {Upper Saddle River, NJ, USA},
  year      = {2002},
  isbn      = {978-0130673893}
}

@BOOK{HornJohnson2012MatrixAnalysis,
  author    = {Horn, R. A. and Johnson, C. R.},
  title     = {Matrix Analysis},
  edition   = {2},
  publisher = {Cambridge University Press},
  address   = {Cambridge, UK},
  year      = {2012},
  doi       = {https://doi.org/10.1017/CBO9780511810817}
}

@CONFERENCE{Hull19951ACC,
  author={Hull, R.A. and Schumacher, D. and Zhihua Qu},
  booktitle={Proceedings of 1995 American Control Conference}, 
  title={Design and evaluation of robust nonlinear missile autopilots from a performance perspective}, 
  year={1995},
  volume={1},
  number={},
  pages={189-193},
  doi={https://doi.org/10.1109/ACC.1995.529235},
  }

@article{Jenkins201856SIAM,
author = {Jenkins, Benjamin M. and Annaswamy, Anuradha M. and Lavretsky, Eugene and Gibson, Travis E.},
title = {Convergence Properties of Adaptive Systems and the Definition of Exponential Stability},
journal = {SIAM Journal on Control and Optimization},
volume = {56},
number = {4},
pages = {2463-2484},
year = {2018},
doi = {https://doi.org/10.1137/15M1047805},

}

@BOOK{Blakelock1991AC,
  author    = {Blakelock, J. H.},
  title     = {Automatic Control of Aircraft and Missiles},
  edition   = {2},
  publisher = {Wiley-Interscience},
  address   = {New York, NY, USA},
  year      = {1991}
}

@ARTICLE{Lewis20090903IEEEMagazine,
  author={Lewis, Frank L. and Vrabie, Draguna},
  journal={IEEE Circuits and Systems Magazine}, 
  title={Reinforcement learning and adaptive dynamic programming for feedback control}, 
  year={2009},
  volume={9},
  number={3},
  pages={32-50},
  doi={https://doi.org/10.1109/MCAS.2009.933854}}

@article{Ferrari20004JGCD,
author = {Ferrari, Silvia and Stengel, Robert F.},
title = {Online Adaptive Critic Flight Control},
journal = {Journal of Guidance, Control, and Dynamics},
volume = {27},
number = {5},
pages = {777-786},
year = {2004},
doi = {https://doi.org/10.2514/1.12597},
}

@InProceedings{Song2023PMLR,
  title = 	 {LipsNet: A Smooth and Robust Neural Network with Adaptive Lipschitz Constant for High Accuracy Optimal Control},
  author =       {Song, Xujie and Duan, Jingliang and Wang, Wenxuan and Li, Shengbo Eben and Chen, Chen and Cheng, Bo and Zhang, Bo and Wei, Junqing and Wang, Xiaoming Simon},
  booktitle = {Proceedings of the 40th International Conference on Machine Learning},
  pages = {32253--32272},
  year =  {2023},
  volume = {202},
  series = {Proceedings of Machine Learning Research},
  month =  {23--29 Jul},
  publisher = {PMLR},
  url = 	 {https://proceedings.mlr.press/v202/song23b.html},
  
}

@book{Sutton2018RL,
  author    = {Richard S. Sutton and Andrew G. Barto},
  title     = {Reinforcement Learning: An Introduction},
  edition   = {2nd},
  year      = {2018},
  publisher = {MIT Press},
  address   = {Cambridge, MA}
}

@ARTICLE{Tamimi2008TSMC38,
  author={Al-Tamimi, Asma and Lewis, Frank L. and Abu-Khalaf, Murad},
  journal={IEEE Transactions on Systems, Man, and Cybernetics, Part B (Cybernetics)}, 
  title={Discrete-Time Nonlinear HJB Solution Using Approximate Dynamic Programming: Convergence Proof}, 
  year={2008},
  volume={38},
  number={4},
  pages={943-949},
  doi={https://doi.org/10.1109/TSMCB.2008.926614}
}

@ARTICLE{Werbos1977advanced,
  title={Advanced Forecasting Methods for Global Crisis Warning and Models of Intelligence},
  author={Werbos, Paul J.},
  journal={General Systems Year-book},
  year={1977},
  note={Early foundational work on backpropagation, dynamic programming, and adaptive systems}
}

@ARTICLE{Prokhorov1997TINN08,
  author={Prokhorov, D.V. and Wunsch, D.C.},
  journal={IEEE Transactions on Neural Networks}, 
  title={Adaptive critic designs}, 
  year={1997},
  volume={8},
  number={5},
  pages={997-1007},
  doi={https://doi.org/10.1109/72.623201}}

@book{OppenheimSchafer2009,
author = {Alan V. Oppenheim and Ronald W. Schafer},
title = {Discrete-Time Signal Processing},
edition = {3},
publisher = {Pearson},
year = {2009},
address = {Upper Saddle River, NJ, USA},
}

@book{KleinMorelli2006,
  author    = {Vladislav Klein and Eugene A. Morelli},
  title     = {Aircraft System Identification: Theory and Practice},
  publisher = {American Institute of Aeronautics and Astronautics (AIAA)},
  year      = {2006},
  edition   = {1},
  isbn      = {9781563478321},
  address   = {Reston, VA}
}

@article{HarrisIEEE1978,
title = {On the use of windows for harmonic analysis with the discrete Fourier transform},
journal = {Proceedings of the IEEE},
volume = {66},
number = {1},
pages = {51-83},
year = {1978},
doi = {https://doi.org/10.1109/PROC.1978.10837},
author = {Fredric J. Harris},
}

@article{Li2026AST112516,
  author = {Yifei Li and Erik-Jan van Kampen},
  title = {Deep deterministic policy gradient with symmetric data augmentation for lateral attitude tracking control of a fixed-wing aircraft},
  journal = {Aerospace Science and Technology},
  volume = {176},
  pages = {112516},
  year = {2026},
  doi = {https://doi.org/10.1016/j.ast.2026.112516},
}

@inbook{KampenAIAA2006,
author = {Erik-Jan van Kampen and Q.P. Chu and J.A. Mulder},
title = {Continuous Adaptive Critic Flight Control Aided with Approximated Plant Dynamics},
booktitle = {AIAA Guidance, Navigation, and Control Conference and Exhibit},
chapter = {},
pages = {},
doi = {https://doi.org/10.2514/6.2006-6429},
year = {2006},
}

@ARTICLE{ZhouTNNL34,
  author={Ye Zhou},
  journal={IEEE Transactions on Neural Networks and Learning Systems}, 
  title={Efficient Online Globalized Dual Heuristic Programming With an Associated Dual Network}, 
  year={2023},
  volume={34},
  number={12},
  pages={10079-10090},
  doi={https://doi.org/10.1109/TNNLS.2022.3164727}}

@article{SinghAST2022107653,
title = {Robust backstepping sliding mode aircraft attitude and altitude control based on adaptive neural network using symmetric BLF},
journal = {Aerospace Science and Technology},
volume = {126},
pages = {107653},
year = {2022},
issn = {1270-9638},
doi = {https://doi.org/10.1016/j.ast.2022.107653},
author = {Prabhjeet Singh and Dipak K. Giri and Ajoy K. Ghosh},
}

@article{YuAST2020106244,
title = {Adaptive backstepping control for air-breathing hypersonic vehicle subject to mismatched uncertainties},
journal = {Aerospace Science and Technology},
volume = {107},
pages = {106244},
year = {2020},
doi = {https://doi.org/10.1016/j.ast.2020.106244},
author = {Chaojun Yu and Ju Jiang and Ziyang Zhen and Ajeet Kumar Bhatia and Shuo Wang},
}

@article{Du2024AST109474,
author = {Desong Du and Yanfang Liu and Ouyang Zhang and Naiming Qi and Weiran Yao and Wei Pan},
title = {Bayesian quadrature policy optimization for spacecraft proximity maneuvers and docking},
journal = {Aerospace Science and Technology},
volume = {154},
pages = {109474},
year = {2024},
doi = {https://doi.org/10.1016/j.ast.2024.109474}
}

@INPROCEEDINGS{KobayashiIROS2022,
  author={Kobayashi, Taisuke},
  booktitle={2022 IEEE/RSJ International Conference on Intelligent Robots and Systems (IROS)}, 
  title={L2C2: Locally Lipschitz Continuous Constraint towards Stable and Smooth Reinforcement Learning}, 
  year={2022},
  volume={},
  number={},
  pages={4032-4039},
  doi={10.1109/IROS47612.2022.9981812}}

@article{DudaAST1998265,
title = {Flight control system design considering rate saturation},
journal = {Aerospace Science and Technology},
volume = {2},
number = {4},
pages = {265-275},
year = {1998},
issn = {1270-9638},
doi = {https://doi.org/10.1016/S1270-9638(98)80004-7},
author = {Holger Duda}
}

@article{Hong2024PMLR,
  title={A Primal-Dual-Critic Algorithm for Offline Constrained Reinforcement Learning},
  author={Kihyuk Hong and Yuhang Li and Ambuj Tewari},
  journal={Processding to machine learning research},
  volume={238},
  year={2024}
}

@article{MoghimiAST2021,
title = {Pitch autopilot design for an autonomous aerial vehicle in the presence of amplitude and rate saturation},
journal = {Aerospace Science and Technology},
volume = {108},
pages = {106371},
year = {2021},
doi = {https://doi.org/10.1016/j.ast.2020.106371},
author = {Saber {Moghimi Rad} and Alireza Khosravi and Pouria Sarhadi},
}

@ARTICLE{PaternainTAC2023,
  author={Santiago Paternain and Miguel Calvo-Fullana and Luiz F. O. Chamon and Alejandro Ribeiro},
  journal={IEEE Transactions on Automatic Control}, 
  title={Safe Policies for Reinforcement Learning via Primal-Dual Methods}, 
  year={2023},
  volume={68},
  number={3},
  pages={1321-1336},
  doi={https://doi.org/10.1109/TAC.2022.3152724}}

@article{Wei2025Auto112029,
title = {Primal–dual adaptive dynamic programming for finite-horizon optimal control of nonlinear systems with isoperimetric constraints},
journal = {Automatica},
volume = {173},
pages = {112029},
year = {2025},
doi = {https://doi.org/10.1016/j.automatica.2024.112029},
author = {Qinglai Wei and Tao Li and Jie Zhang and Fei-Yue Wang},
}

@article{ModaresAuto20141780,
author={Modares, Mamidrezia and Lewis, Frank. L.},
title = {Optimal tracking control of nonlinear partially-unknown constrained-input systems using integral reinforcement learning},
journal = {Automatica},
volume = {50},
number = {7},
pages = {1780-1792},
year = {2014},
doi = {https://doi.org/10.1016/j.automatica.2014.05.011},
}

@dataset{Li2026FlightControlData,
  author       = {Li, Yifei},
  title        = {{Python code and data supporting chapter 3 of the Phd thesis: temporally smoothed cascaded online learning flight control}},
  year         = {2026},
  month        = aug,
  day          = {27},
  publisher    = {4TU.ResearchData},
  doi          = {10.4121/22A294E7-3550-4E6A-B415-12107E3F159C.V1},
  url          = {https://doi.org/10.4121/22A294E7-3550-4E6A-B415-12107E3F159C.V1}
}

@INPROCEEDINGS{HullACC1997611921,
  author={Hull, R.A. and Zhihua Qu},
  booktitle={Proceedings of the 1997 American Control Conference (Cat. No.97CH36041)}, 
  title={Dynamic robust recursive control design and its application to a nonlinear missile autopilot}, 
  year={1997},
  volume={1},
  number={},
  pages={833-837 vol.1},
  doi={10.1109/ACC.1997.611921}}

@article{KimCEP2004149,
title = {A robust adaptive nonlinear control approach to missile autopilot design},
journal = {Control Engineering Practice},
volume = {12},
number = {2},
pages = {149-154},
year = {2004},
doi = {https://doi.org/10.1016/S0967-0661(03)00016-9},
author = {Seung-Hwan Kim and Yoon-Sik Kim and Chanho Song}
}

@article{ZhangAST2024109552,
title = {Command-filtered incremental backstepping attitude control of spacecraft with predefined-time stability},
journal = {Aerospace Science and Technology},
volume = {155},
pages = {109552},
year = {2024},
doi = {https://doi.org/10.1016/j.ast.2024.109552},
author = {Haichao Zhang and Haowei Huang and Bing Xiao and Kejun Dong}
}

@article{LunguAST2020105526,
title = {Backstepping and dynamic inversion combined controller for auto-landing of fixed wing UAVs},
journal = {Aerospace Science and Technology},
volume = {96},
pages = {105526},
year = {2020},
doi = {https://doi.org/10.1016/j.ast.2019.105526},
author = {Mihai Lungu}
}

@inproceedings{WangICLR2024,
  title     = {Addressing Signal Delay in Deep Reinforcement Learning},
  author    = {Wang, Wei and Han, Dongqi and Luo, Xufang and Li, Dongsheng},
  booktitle = {The Twelfth International Conference on Learning Representations},
  year      = {2024},
  url       = {https://openreview.net/forum?id=Z8UfDs4J46}
}



\end{document}